\documentclass[
headinclude=true, 
open=right,
twoside=true,
BCOR=5mm, 
    fontsize=11pt,
paper=17cm:24cm, 
pagesize, 
DIV=13, 
headings=normal, 
appendixprefix=true,
numbers=noendperiod, 
toc=bibliography, 
parskip=false, 
version=last 
]{scrbook}

\makeatletter
\renewcommand{\@chapapp}{}

\makeatother

\setcapindent{0pt}

\usepackage{fancyhdr}

\usepackage{graphicx}

\usepackage[T1]{fontenc}
\usepackage{titlesec, blindtext, color}
\titleformat{\chapter}[display]
{\normalfont\LARGE\bfseries\scshape\raggedright}{
\thechapter}
   {25pt}{\LARGE}
\titleformat{\section}
   {\normalfont\Large\bfseries\raggedright}{\thesection}{1em}{}
\titleformat{\subsection}
   {\normalfont\large\bfseries\raggedright}{\thesubsection}{1em}{}
\titleformat{\paragraph}[runin]
   {\normalfont\bfseries}{\theparagraph}{1em}{}

\usepackage{amsmath}
\usepackage{multirow}
\usepackage{amsfonts}
\usepackage{amssymb,graphics,psfrag,float}
\usepackage{array,epsfig,multirow,graphicx}
\usepackage{comment}
\usepackage{slashed}
\usepackage{hyperref}
\usepackage{tensor}
\usepackage{booktabs}
\usepackage{colortbl}
\usepackage{hhline}
\usepackage{mathtools}
\usepackage{enumitem}
\usepackage{xfrac}
\usepackage{tikz}
\usetikzlibrary{matrix}
\usepackage[nosort]{cite}
\usetikzlibrary{decorations.markings}
\usetikzlibrary{shapes,arrows}
\usetikzlibrary{decorations.pathreplacing}
\usetikzlibrary{arrows,positioning}
\usepackage[toc]{appendix}

\numberwithin{equation}{chapter}
\numberwithin{table}{chapter}

\newcommand{\beq}{\begin{equation}}
\newcommand{\eeq}{\end{equation}}
\newcommand{\be}{\begin{equation}}
\newcommand{\ee}{\end{equation}}
\newcommand{\bea}{\begin{eqnarray}}
\newcommand{\eea}{\end{eqnarray}}
\newcommand{\ben}{\begin{eqnarray*}}
\newcommand{\een}{\end{eqnarray*}}               
\newcommand{\ba}{\begin{aligned}}
\newcommand{\ea}{\end{aligned}}
\newcommand{\bt}{\begin{tabular}}
\newcommand{\et}{\end{tabular}}
\newcommand{\bc}{\begin{center}}
\newcommand{\ec}{\end{center}}

\newcommand{\cO}{\mathcal{O}}
\newcommand{\cT}{\mathcal{T}}

\newcommand{\cC}{\mathcal{C}}
\newcommand{\cD}{\mathcal{D}}
\newcommand{\cL}{\mathcal{L}}
\newcommand{\cS}{\mathcal{S}}
\newcommand{\cK}{\mathcal{K}}
\newcommand{\cN}{\mathcal{N}}

\newcommand{\cA}{\mathcal{A}}

\newcommand{\cB}{\mathcal{B}}
\newcommand{\cF}{\mathcal{F}}

\newcommand{\cJ}{\mathcal{J}}
\newcommand{\cR}{\mathcal{R}}

\newcommand{\cV}{\mathcal{V}}

\newcommand{\cM}{\mathcal M}

\newcommand{\I}{\text{Im}}
\newcommand{\R}{\text{Re}}

\newcommand{\bbC}{\mathbb{C}}
\newcommand{\bbP}{\mathbb{P}}

\newcommand{\nn}{\nonumber}

\newcommand{\cref}{{\textbf  [check ref]}}

\definecolor{mppgreen}{RGB}{17,102,86}
\definecolor{mppgray}{RGB}{221,222,214}

\def\blfootnote{\xdef\@thefnmark{}\@footnotetext}
\long\def\symbolfootnote[#1]#2{\begingroup%
\def\thefootnote{\fnsymbol{footnote}}\footnote[#1]{#2}\endgroup}

\begin{document}

\pagenumbering{gobble}
\baselineskip=15pt

\vspace*{3cm}

\begin{center}
{
{\large \scshape Aspects of Calabi-Yau Fourfold Compactifications}

\vspace*{0.7cm}

{\large Sebastian Walter Greiner}
}
\end{center}

%

\newpage

\vspace*{15.5 cm}

\noindent 
ISBN: 978-90-393-7038-4 \\
Cover design by 123FreeVectors \\
Printed by Gildeprint

\newpage

\vspace*{1.5cm}
\begin{center}
 {\LARGE Aspects of Calabi-Yau Fourfold Compactifications} \\
 \vspace*{2 cm}
 {\Large Aspekten van Compactificaties op Calabi-Yau Vari\"eteiten van Dimensie Vier} \\
 \vspace*{0.4cm} 
 (met een samenvatting in het Nederlands) \\
 \vspace*{1.5cm}
 {\large Proefschrift} \\
 \vspace*{0.7cm}
 \end{center}
 ter verkrijging van de graad van doctor aan de Universiteit Utrecht op gezag van de
rector magnificus, prof. dr. H.R.B.M. Kummeling, ingevolge het besluit van het
college voor promoties in het openbaar te verdedigen op maandag 24 september
2018 des middags te 6.00 uur \\
 \begin{center}
 \vspace*{0.7cm}
 door \\
 \vspace*{0.7cm}
 {\Large Sebastian Walter Greiner} \\
 \vspace*{0.2cm}
 geboren op 9 september 1990 te G\"unzburg, Duitsland

\end{center}

\newpage
\noindent Promotor: Prof. dr. S.J.G. Vandoren \\
Copromotor: Dr. T.W. Grimm

\newpage

\tableofcontents

\newpage

\newpage\null\thispagestyle{empty}\newpage

\pagenumbering{arabic}
    
\chapter{Introduction} \label{Introduction}

The two cornerstones of modern fundamental physics are quantum field theory and general relativity. Einstein's general relativity as an extension of Newton's theory of gravity is based on the idea that matter and energy densities curve its surrounding space-time as dictated by Einstein's famous equations. This accounts for phenomena on astrophysical scales like the gravitational red-shift, gravitational lensing or gravitational waves as detected in 2016 by the LIGO experiment \cite{PhysRevLett.116.061102}. The framework of quantum field theory is tested to be realized in nature as the Standard Model of particle physics. This theory elevates classical matter like electrons to quantum fields that interact via mediating fields, for instance photons describing light in quantum electrodynamics. Quantum electrodynamics is up to date the most well-tested theory as its predictions were observed to an astonishing accuracy beyond any other physical theory. The great success of quantum field theory in particle physics was crowned in 2012 with the discovery of the Higgs particle at LHC \cite{Aad:2012tfa,201230} as suggested by theorists in the 60s. This proves the realization of the Higgs mechanism in nature accounting for the generation of masses of particles.

Although both theories provided great achievements for theoretical physics, unifying both into a single consistent theory of nature proves troublesome. The difficulties arise from the fact that general relativity loses is predictive power in the regime of high energies where a classical description of nature breaks down and we enter the realm of quantum field theory. In nature this occurs for instance in the vicinity of black holes where the extraordinary mass density within the horizon makes a high energy description of gravity necessary. Here the size of the black hole can theoretically be smaller than its Compton wave-length requiring a full quantum treatment. In the quantum description of gravity the gravitational force is mediated by gravitons, quantized particles whose interactions with matter model the attractive forces between the latter. Due to the positive mass-dimensions of the graviton self-coupling in Einstein gravity, the graviton propagator is plagued by divergences at high energies in the standard quantum field theory approach to gravity. Therefore a field theoretic description of general relativity can only be an effective theory at large distances which means that it only accounts for phenomena up to a certain energy scale. At this scale new physics is needed for a consistent quantum description of gravity dubbed quantum gravity. This new physics will contain additional degrees of freedom that are too massive to be observed at low energies and their interactions with the large distance spectrum is hoped to render the full theory finite. Although many alternative approaches to realize this new physics exist, the most promising candidate for a well-behaved quantum gravity is string theory \cite{Green:2012oqa,Green:2012pqa,Polchinski:1998rq,Polchinski:1998rr,Becker:2007zj,Blumenhagen:2013fgp}.

\section{String theory as quantum gravity}

String theory is the concept of modelling the fundamental constituents of nature not via point-like, but extended objects, one-dimensional strings. The extended nature of the string allows for rich dynamics that can be split into center-of-mass movements as well as vibrational modes, similar to a violin string vibrating at a specific eigenmode. The kind of the eigenmode of the string determines its quantum numbers and hence the sort of particle it represents to an observer agnostic about the strings' extended nature. Among this infinite number of eigenmodes is also a massless symmetric two-tensor mode which behaves as a graviton in space-time. Furthermore, there are modes corresponding to matter and non-abelian gauge fields. In this sense string theory is a quantum theory of gravity, as it contains a gravitational sector that is coupled to fermions or gauge-bosons. The infinite number of massive eigenmodes of the string provides the additional degrees of freedom for the new physics and resolves the high energy divergences. So far no inconsistencies were found in the string theory framework, therefore it is conjectured that string theory provides a unified framework for a quantum description of the interactions between matter and gravity.

Common to all string theories is a massless real scalar field in their spectrum called the dilaton which can be viewed as a dynamical coupling constant of the string interactions. For a weak string coupling, i.e. a small vacuum expectation value of the dilaton, we can recover a perturbative description of string theory. The interactions in perturbative quantum field theory of point particles following their worldlines are described by one dimensional graphs called Feynman diagrams. These are replaced in perturbative string theory by two-dimensional Riemann-surfaces describing the worldsheet of a string, the two-dimensional analogue of the worldlines of point particles. In this sense perturbative string theories whose effective theories might describe our world enjoy a twofold perturbative expansion in two scales, the string length and the string coupling. The perturbative expansion of scattering amplitudes of strings, perturbative in string coupling, can be viewed as summing over all possible worldsheet topologies where the contribution from each topology is suppressed by a power of string coupling corresponding to the number of holes (and handles) of the two-dimensional worldsheet. This expansion is the analogue of regular Feynman diagrams smearing out the localized interaction points of point particles to the worldsheet of interacting strings, hereby removing divergencies in the scattering amplitudes. The loops of Feynman-diagrams correspond here to the holes and handles of the worldsheet of the string.

These remarkable properties put, however, strong constraints on the consistent string theories. To avoid tachyons rendering a possible vacuum unstable and to contain fermions in its spectrum, a string theory needs to be supersymmetric. A quantum field theory endowed with supersymmetry shows a symmetry between its bosonic and fermionic degrees of freedom. The generators of these symmetries are called supercharges and map bosons to fermions and vice versa enhancing the regular symmetry group of space-time containing translations and rotations. The advantage of unbroken supersymmetry in a theory is an improved stability of the underlying quantum theory, as certain quantum corrections of fermions and bosons cancel due to the restrictions imposed by their additional symmetry. In such theories bosons and fermions combine into multiplets, irreducible representations of the enhanced Lorentz group. The number of supercharges is usually denoted by $ \cN $-supersymmetry with $ \cN $ counting the number of irreducible representations in which the supercharges occur. In four space-time dimensions we have minimal or $ \cN = 1 $ supersymmetry, if there are four supercharges in the theory transforming as one real Majorana fermion or two chiral Weyl fermions. 

Further necessary constraints on such a superstring theory are the requirement of conformal symmetry on the world-sheet and the string propagating in a ten dimensional space-time to avoid anomalies in the two-dimensional conformal quantum field theory. The conformal symmetry is in part responsible for the well-behaved nature of the string as it forbids the theory to have any scale allowing for the same description at all energy scales. The scale set by the string coupling only emerges by setting a vacuum expectation value for the dilaton which we choose small to find a perturbative description of the theory. The only scale left is the string length $ \alpha^\prime $ that serves as an over-all normalization of the world-sheet action. It is usually assumed to be of the same order as the Planck length.

After imposing the constraints only five distinct perturbative superstring theories are left: Type IIA and Type IIB string theory, the heterotic string theory with gauge group either $ E_8 \times E_8 $ or $ SO(32) $ and Type I string theory. As proposed by Witten in the mid 90s these five string theories are all related by dualities and can be viewed as different limits of an eleven-dimensional theory called M-theory \cite{Witten:1995em,Schwarz:1995jq,Horava:1995qa,Dasgupta:1995zm}. The crucial insight that led to this development was the realization that string theory contains $ (p+1) $-dimensional extended objects called $ p $-branes, see \cite{Polchinski:1995mt,Polchinski:1996fm,Polchinski:1996na} and \cite{Blumenhagen:2006ci} for a modern review, that are not included in the perturbative spectrum of the theory. The branes are non-perturbative as their mass scales inversely with the string coupling. For certain branes, the D-branes of Type II string theories, their fluctuations around a background configuration are, however, described by open strings ending on the D-branes that can be viewed as solitonic solutions of the underlying theory. The massless open string degrees of freedom describe a supersymmetric gauge theory, possibly non-Abelian, confined to the worldvolume of the brane. As the masses of the the open strings are determined by their tension and the distance between branes, we can find massless states at the intersection of branes. These intersections of branes can lead to chiral fermions and Yukawa couplings in their massless spectrum localized on the intersection of several branes where the open strings become massless. If we stack several branes, we obtain non-Abelian gauge theories on their common world-volume. This is a convenient way to realize the standard model on the world-volume of a D-brane which is known in the literature as brane-world scenario.

At large length scales or low energies when the extended nature of the string can not be resolved by an observer, we recover point-particle physics. In this limit the tension of the strings becomes infinitely high and therefore its higher excitations infinitely massive. The states that survive this limit fit into spectra of supergravities coupled to gauge fields, a unique ten-dimensional theory for each string theory and one unique eleven-dimensional supergravity as the low-energy limit of M-theory. Supergravity is the gauged version of supersymmetry and due to the fact that the (anti-)commutator of two supercharges is proportional to the momentum operator of the underlying space-time, gauging supersymmetry automatically implies diffeomorphism-invariance of the entire theory and hence it necessarily includes gravity. These supergravities are therefore effective low-energy descriptions of a UV-complete theory, provided by string theory. In this sense we recovered an effective description of the low-energy physics of a consistent theory of quantum gravity.  

Branes appear in this context as singularities of the space-time extending the notion of a black hole to higher dimensional objects. The extended nature of a brane breaks the translation invariance of the underlying space-time and therefore in general also some supersymmetry. Their perturbative degrees of freedom are captured by the so called Dirac-Born-Infeld (DBI) action, coupling gravity to the world volume of the brane, and topological Chern-Simons terms that model the interactions of the gauge-potentials with the background. In the weak coupling description this gives rise to gauge potentials coupled minimally to supergravity to leading order on the brane. The DBI and Chern-Simons actions of a brane are just added to the supergravity theory, in this sense we put branes into the supergravity theory by hand. The higher order interactions of the DBI theory are very hard to compute and unknown, but necessary to obtain a consistent theory. These corrections provide the interactions of the bulk or closed string degrees of freedom with the brane or open string degrees of freedom. The emerging backreaction of the branes onto the ambient geometry, however, is important to phenomenological applications as they might be responsible for conjectured no-go theorems constraining the validity of the perturbative low-energy description of string theory. A possible solution to this problem is F-theory, a non-perturbative description of type IIB string theory with seven-branes.

\section{Dimensional reduction in string theory}

Wait, we don't live in ten space-time dimensions! So how does that fit with the observation of a four dimensional space-time? One answer is compactification, the idea that the additional dimensions predicted from string theory are curled up in tiny compact dimensions that hence avoid detection. The compact space is called the internal space, as we do not observe it, whereas the extended dimensions are called external space of the compactification. Truncating the theory to its massless content we recover an effective low-energy description in four dimensional space-time, which is nicely described in \cite{ibáñez_uranga_2012}. This process to obtain a low-energy effective theory from a higher dimensional theory with small extra dimensions is called dimensional reduction.

We will illustrate this concept by considering a theory of a free massless scalar $ \phi(x^0, \ldots, x^4) $ in five-dimensional space-time $ M_{4,1} $ with one compact direction
\begin{equation}
 S_{4,1} = \int_{M_4,1} \partial_M \phi \, \partial^M \phi \,  , \quad M_{4,1} = M_{3,1} \times S^1 \, .
\end{equation}
The compact direction parametrizes a circle $ S^1 $ of radius $ R $ with periodic coordinate $ y $. Expanding $ \phi(x^0, \ldots, x^3, y) $ into its Fourier modes $ \phi_k (x^0, \ldots, x^3) $ as
\begin{equation} \label{KKmodes}
 \phi(x^0, \ldots, x^4) = \sum_{k \in \mathbb{Z}} e^{i k y / R }  \phi_k (x^0, \ldots, x^3) \, ,
\end{equation}
we obtain an infinite number of four-dimensional scalars $ \phi_k (x^0, \ldots, x^3) $ with mass-squares
\begin{equation}
 m^2 _k = \Big( \frac{k}{2 \pi R} \Big)^2 \, .
\end{equation}
These fields $ \phi_k $ are often called the tower of Kaluza-Klein (KK) modes, because their masses (the levels of the tower) increase with the internal momentum $ k \in \mathbb{Z} $ along the circle. In an effective theory where the additional dimension avoids detection, at energies $ E \ll 1/R $ only the lowest component of the $ KK $-tower $ \phi_0 $ can be excited and we find an effective four-dimensional theory of a free scalar
\begin{equation}
 S_{3,1,eff} = 2 \pi R \int_{M_3,1} \partial_\mu \phi_0 \, \partial^\mu \phi_0 \,  .
\end{equation}
This illustrates dimensional or Kaluza-Klein reduction that allows to derive an effective theory for small additional dimensions. This effective theory is a perturbation theory around a vacuum which is a solution of the full equations of motion of the higher dimensional theory. In the above example the vacuum solution is given by the space $ M_{3,1} \times S^1 $ where the field $ \phi $ has a zero vacuum expectation value.

The motivation of Kaluza \cite{Kaluza1921} and Klein \cite{Klein1926} for their original work was somewhat different than our modern approach coming from string theory. Their goal was to unify Einstein's general relativity with Maxwell's theory of electro-magnetism or in today's language a $ U(1) $-gauge-field theory. Starting from a five-dimensional theory of pure gravity, they compactified it on a circle and derived the effective theory around a flat vacuum. In this construction the five-dimensional metric $ g^{(5)} $ splits into the four-dimensional metric $ g^{(4)} $, a massless four-dimensional vector $ A $ and a massless real scalar $ \phi $. 
\begin{align}
 g^{(5)} = 
 \begin{pmatrix}
  g^{(4)} & A \\
  A^T & \phi
 \end{pmatrix} \, .
\end{align}
Remarkably, the five-dimensional diffeomorphism invariance descends to a $ U(1) $ gauge-symmetry of the four-dimensional vector which can be traced back to the periodicity of the circle. The real scalar is basically the radius or size of the circle and controls the coupling of the four-dimensional gauge-theory. In broader terms the size of the circle is a modulus of the internal geometry, a field with no potential. This expectation value of the modulus, here the radius of our background, determines the couplings in our effective theory. We find in particular that the geometry of the internal space has a crucial influence on the physics of the effective theory. 

Although the reduction as proposed by Kaluza and Klein unifies indeed gravity and electro-magnetism, it only does so on the classical level as the five-dimensional quantum theory already suffered from inconsistencies. The modern approach is to take as a starting point ten- or eleven-dimensional supergravity which we argued to be UV-completed by string theory or M-theory and perform the dimensional reduction on a six- or seven-dimensional internal space.

In this work we will consider only compactifications to flat Minkowski-space and the background values of the field strengths of the gauge-potentials will be set to zero. Furthermore do we only consider classical space-times as vacua where also the fermions have vanishing vacuum expectations value. For the lower dimensional theory to retain a certain amount of supersymmetry restricts the internal space to have a so called Killing spinor, i.e.~ the internal geometry allows for fermionic symmetries. This restricts the holonomy of the internal $ D $-dimensional Riemannian manifold which we assume to allow for spinors. In this case is the holonomy group generated by the transformations of a spinor when parallel transported along closed curves in the manifold and is in general $ Spin(D) $. To preserve supersymmetry we need to restrict to subgroups of the holonomy group. In $ D=6 $ dimensions Calabi-Yau manifolds serve as convenient backgrounds for string compactifications \cite{Strominger:1985it,Witten:1985bz,Strominger:1986uh,Greene:1996cy}, as they have $ SU(3) \subset Spin(6) \simeq SU(4) $ holonomy allowing for Killing spinors. They not only preserve supersymmetry, but are also Ricci-flat and therefore solve Einstein's equation for space-times without matter. The six-dimensional Calabi-Yau manifolds which are complex manifolds are called Calabi-Yau threefolds as they have three complex dimensions. 

The effective low-energy descriptions of string theory in four flat space-time dimensions $ \mathbb{M}_{3,1} $ preserving supersymmetry are therefore chosen to be perturbations around a supergravity background of the form
\begin{equation} \label{threefoldVacuum}
 M_{9,1} = \mathbb{M}_{3,1} \times Y_3 \, , \quad e^{\langle \phi \rangle} = g_s \, , 
\end{equation}
where $ Y_3 $ is a Calabi-Yau threefold of a very small length scale. The vacuum expectation of the dilaton $ \phi $, the real scalar common to a all superstring theories determines the string coupling $ g_s $ of the perturbation theory. It can be thought of as the radius in the compactification of M-theory on a circle. Giving the remaining fields background values breaks supersymmetry in general spontaneously and generates a potential for the fields rendering them massive. This fixes the fields at the minimum of the potential, i.e.~ stabilizes them at their vacuum expectation values. Non-zero vacuum expectation values of the various $ p $-form field strengths are called fluxes. Compactifications on Calabi-Yau backgrounds usually leads to hundreds of massless fields without potential, the moduli of the effective field theory. These moduli need to be stabilized at high mass scales as they are not observed in the four dimensional space-time. A convenient scenario to do so are called flux compactifications, \cite{Silverstein:2004id,Grana:2005jc,Douglas:2006es,Denef:2007pq,Samtleben:2008pe}, where the non-zero field strength values of the background generate a potential for the moduli. So far no dynamical mechanism to determine the correct vacuum for the effective low-energy theory is known and therefore we need to construct these backgrounds by hand.

A convenient framework for these constructions are the Type II theories, as a background of the form \eqref{threefoldVacuum} preserves eight supercharges, $ \cN=2 $ supersymmetry, which is then spontaneously broken by fluxes in the internal space. This approach provides a certain amount of stability to corrections. The downside is that these backgrounds do not allow for non-Abelian gauge theories or chiral matter in the four-dimensional external space. This can be cured by introducing space-time filling D-branes into the background. The corresponding D-branes need to wrap non-trivial cycles of the internal space to account for their tension and create a stable background. Since they break translational invariance of the internal space, they also break part of the supersymmetry.

The tension of the D-branes and their charges sourcing the gauge fields provide a non-trivial energy-momentum density of the background and hence it can no longer be Ricci-flat to solve Einstein's equations. There are two solutions to this problem, either we consider a complex internal manifold $ B_3 $ with positive curvature which is no longer Calabi-Yau or we introduce negative tension objects called orientifold planes or O-planes on the Calabi-Yau space $ Y_3 $. Both solutions are related by a so called orientifold-involution $ \sigma $, mapping $ Y_3 $ holomorphically to itself with fixed point loci the O-planes. The quotient of $ Y_3 $ by the action of $ \sigma $ is precisely the positive curvature background $ B_3 $ which contains only D-branes. The four-dimensional effective theories and their properties were derived in a series of papers \cite{Louis:2002ny,Grana:2003ek,Grimm:2004uq,Grimm:2004ua,Grimm:2005fa}.

Constructing these so called Type II orientifold flux backgrounds is a tedious task as we need to pay attention to various consistency conditions, like tadpole constraints, anomaly cancellation and preserving a weak coupling description. An elegant geometric solution to these complications is given by F-theory.

\section{F-theory on Calabi-Yau fourfolds}

The F-theory framework as proposed by Vafa in 1996, see \cite{Vafa:1996xn}, is the best understood description of string theory as it allows to include certain non-perturbative effects not accessible by other approaches. Starting point is Type IIB string theory whose massless bosonic spectrum contains a complex scalar, called the axio-dilaton $ \tau $, defined as the combination
\begin{equation}
 \tau = C_0 + i e^{-\phi} \, , \quad e^{\phi} = g_s \, .
\end{equation}
Here $ C_0 $ is the zero-form gauge potential of Type IIB and $ \phi $ the dilaton, whose vacuum expectation value determines the string coupling $ g_s $. Type IIB string theory and its low-energy supergravity obey a self-duality called S-duality under which the axio-dilaton transforms as
\begin{equation} \label{Sduality}
 \tau \, \rightarrow \, \frac{a \tau + b}{c \tau + d} \, , \quad 
 \begin{pmatrix} 
 a & b \\
 c & d 
 \end{pmatrix}
 \, \in \, SL(2, \mathbb{R}) \, ,
\end{equation}
which is further broken to $ SL(2,\mathbb{Z}) $ in the full string theory due to quantum effects.
This duality contains in particular the mapping
\begin{equation}
 \tau \, \rightarrow -\frac{1}{\tau} \, ,
\end{equation}
which allows for a strongly coupled region in the ten-dimensional space-time to have a dual weakly coupled description. The discrete duality in \eqref{Sduality} is the reparametrization invariance of the complex structure modulus $ \tau $ of a torus
\begin{equation} \label{TorusTau}
 T^2 = \mathbb{C} / \mathbb{Z} + \tau \mathbb{Z} \, .
\end{equation}
Therefore, we can interpret the ten-dimensional Type IIB string theory as a compactification of a putative twelve-dimensional F-theory on a torus $ T^2 $ with complex structure modulus $ \tau $ varying over the ten-dimensional space-time. This geometry is called a torus fibration over the space-time. The problems with this picture are that there is no interpretation for the volume of the torus, which should appear in a compactification as a modulus in the effective theory, and the fact that there is no twelve-dimensional supergravity with signature $ (1,11) $.

The more practical approach is to understand F-theory as a certain limit of M-theory. Here M-theory is compactified on a torus fibration with the torus complex structure modulus $ \tau $, as for example described in \cite{Denef:2008wq}. We call the two cycles of the torus A-cycle and B-cycle which are both circles. M-theory on a circle can be interpreted as Type IIA String Theory with coupling the radius of the circle. As we go to strong coupling in Type IIA we decompactify the space-time and recover the eleven-dimensional M-theory description. This can also be seen as the definition of M-theory. The second duality we will exploit is the duality between Type IIA String Theory on a cirlce and Type IIB String Theory on a circle with inverse radius. This duality is known as T-duality and interchanges states with momentum along the circle direction with stringy states that wind around the circle.

To obtain the desired F-theory vacuum, we first compactify M-theory on the A-cycle of the torus leading to Type IIA String Theory and then compactify further on the B-cycle and use T-duality to obtain a Type IIB vacuum with axio-dilaton $ \tau $ in nine dimensions. In the following we will restrict to a nine-dimensional space-time of the form $ \mathbb{M}_{2,1} \times B_3 $ with $ B_3 $ a three complex dimensional manifold which is K\"ahler. We further restrict to elliptic fibrations for which $ \tau $ depends on the coordinates of $ B_3 $ holomorphically, preserving a certain amount of supersymmetry in the effective theory. An elliptic curve is a torus with a marked point, for example the origin in \eqref{TorusTau}. It can be shown in particular that due to the holomorphiciity of the fibration the volume of the elliptic fiber is constant over $ B_3 $. Sending then this volume modulus of the torus to zero, the B-cycle radius becomes infinite, the vacuum decompactifies and we recover an effectively four-dimensional Lorentz-invariant Type IIB vacuum with axio-dilaton $ \tau $. To preserve minimal supersymmetry in four dimensions the total space of the elliptic fibration over $ B_3 $ needs to be Calabi-Yau. The resulting space has four complex dimensions and is hence called an elliptically fibered Calabi-Yau fourfold $ Y_4 $ whose geometry and the effective physics of its string theory compactifications will be the central topic of this thesis. They were first studied in \cite{Klemm:1996ts,Brunner:1996pk,Brunner:1996bu,Becker:1996ay,Mayr:1996sh,Becker:1996gj} and their effective theories were discussed in \cite{Haack:1999zv,Gates:2000fj,Haack:2000di,Berkovits:2001tg,Gukov:2002iq} for Type II string theories, in \cite{Haack:2001jz,Berg:2002es} for M-theory and in \cite{Grimm:2010ks} for F-theory.

The advantage of this construction is that the elliptic fibration can degenerate, i.e. a cycle along the torus fiber may shrink to zero size over complex codimension one loci. At these points $ \tau $ has a pole indicating the presence of a charged object which is called a space-time filling seven-brane. These are eight dimensional objects in $ \mathbb{M}_{3,1} \times B_3 $ that wrap four-cycles in $ B_3 $ and are charged magnetically under $ C_0 $. If there is more then one seven-brane at the same locus in the base $ B_3 $, for example they stack or intersect, the elliptic fiber degenerates further and the Calabi-Yau fourfold $ Y_4 $ becomes singular. At these degeneration loci in $ B_3 $ we find non-Abelian gauge groups for stacks of seven-branes where the number of branes in the stack corresponds to the rank of the gauge group. At the intersection of two branes matter localizes and at points where three branes intersect Yukawa couplings emerge. These properties are captured by the geometry of the elliptic fibration of $ Y_4 $. For further details we refer to the great review \cite{Weigand:2010wm}.

These F-theory vacua allow in general only for local weakly coupled descriptions due to the local $ SL(2,\mathbb{Z}) $ symmetry of our theory. Vacua for which such a global weak coupling limit exists are scarce and were first discussed by Sen \cite{Sen:1996vd,Sen:1997gv}. This scarcity renders the F-theory framework much more general than the regular weakly coupled Type IIB description. In the dual M-theory picture on the Calabi-Yau fourfold $ Y_4 $ the degrees of freedom of the seven-branes are geometric moduli of $ Y_4 $, geometrizing the seven-branes and accounting for their full dynamics. This geometrization blurrs the distinction of closed (bulk) and open (brane) string degrees of freedom in the dual type IIB orientifold compactifications and allows a description of the effective four-dimensional physics of Type IIB string theory at finite coupling (general $ \tau $). 

From another point of view, there is no need any more for the Type IIB interpretation that distinguishes between open and closed strings as both are captured by the geometric deformations of the Calabi-Yau fourfold geometry. In general this distinction does not even exist! This motivates the study of the geometry of Calabi-Yau fourfolds and its deformations as it is essential to the computation of the effective field theory dynamics as for example described in \cite{Bizet:2014uua,Cota:2017aal}. In particular, it is possible to construct explicit examples of these Calabi-Yau fourfolds and to derive the couplings of the theory in terms of geometrical quantities. A simple class of these in general very complicated constructions, toric hypersurfaces, will be central to the second part of this work. 

\section{Novel features of Calabi-Yau fourfolds with non-trivial odd cohomology}

In this short subsection we want to highlight the novel results that were obtained in course of our studies. As we conveyed so far, F-theory on elliptically fibered Calabi-Yau fourfolds provide an interesting arena for string phenomenology, the study of the low-energy effective theories of string theory. The goal of this active research field is to derive observable consequences of string theory. As emphasized earlier, F-theory provides enough flexibility to derive the details of its effective theory, but also contains brane dynamics and is inherently non-perturbative. To achieve these features it uses the machinery of algebraic geometry that provides us with powerful tools that allow calculations of spectra and couplings.

As F-theory can be interpreted as a generalization of the weakly coupled Type IIB orientifold vacua, it also contains novel features and unifies complicated physics of the perturbative theory. The particular feature we study in this work are non-trivial harmonic three-forms on the underlying Calabi-Yau fourfold. The corresponding massless modes have no analogue in Calabi-Yau threefolds that are characterized by their geometric moduli, the K\"ahler moduli determining volumina and the complex structure moduli that describe the shape of the underlying threefold. The three-form moduli of the F-theory description, however, defy a simple geometric interpretation and behave differently as they arise not from the metric of the theory, but from the three-form gauge-potential of M-theory or Type IIA string theory. 

Consequently, the inclusion in the topological string theory framework like in \cite{Witten:1991zz,Mayr:1996sh,Alim:2012gq} is elusive, even for type II compactifications on Calabi-Yau fourfolds. This can be related to the fact that in the conformal field theory approach, for example using orbifolds of Landau-Ginzburg vacua \cite{Vafa:1989xc,Intriligator:1990ua}, the states corresponding to the three-forms arise from twisted sectors and are not marginal deformations of the $ \cN=(2,2) $ SCFT as they have $ R $-charges $(2,1)$. Consequently, the topological twists of the SCFT that lead to the familiar A-model (K\"ahler deformations) and to the B-model (complex structure deformations) do not apply. Therefore, their kinetic terms, usually given by the metric on the moduli space of marginal deformations of the SCFT can not be described by the Zamolodchikov metric \cite{Zamolodchikov:1986gt,Candelas:1989qn}. Due to this complications and the fact that the simplest examples of Calabi-Yau fourfolds have a trivial three-form cohomology, the inclusion of three-forms is hardly discussed in the literature. In course of this thesis we will find that in the models when non-trivial three-forms are present, we can no longer discuss K\"ahler and complex structure deformations independently as the metric of the three-form moduli depends on both.

Another approach to include the three-form moduli into the low-energy effective theory is to start with a supergravity theory and perform a dimensional reduction on a general Calabi-Yau fourfold. This was done first by Haack and Louis \cite{Haack:1999zv} that compactified Type IIA supergravity on a fourfold with non-trivial non-trivial three-forms. For technical reasons it became evident that in the expansion of the three-form gauge-potential a three-form basis depending on the complex structure moduli was necessary. This was artificially solved by introducing tensors mapping a topological (constant) harmonic three-form basis to a non-constant three-form basis. Consistency required these tensors to depend in a complicated way on the complex structure moduli described by a set of differential equations. A geometric interpretation of these tensors and a solution of the differential equations was, however, not found. 

First progress in this direction was made ten years later in \cite{Grimm:2010ks} as a refined ansatz for the harmonic three-forms was put forward. This enabled a clearer derivation of the effective theory and showed that a certain subset of the three-form moduli are dual to $ U(1) $ gauge-bosons in the weak coupling description of F-theory in four dimensions. The complex structure dependence of this novel ansatz for the harmonic three-forms was captured by a matrix valued function $ f_{\cA \cB}(z) $ that depends on the complex structure moduli $ z $ and determines the coupling of the resulting $ U(1) $ gauge-theory. In a later publication \cite{Grimm:2014vva}, it was observed that a different kind of three-form moduli gives rise to axions in the effective theory and again $ f_{\cA \cB} $ determined the coupling and hence the axio-decay constant relevant for inflationary models. Due to the crucial role of the matrix $ f_{\cA \cB} $ as coupling in the effective theory, the question we strive to answer in this thesis is
\begin{center}
	\textit{What is $ f_{\cA \cB} $ ?}
\end{center} 

In course of this work we will not only give a beautiful geometric interpretation of this function $ f_{\cA \cB} $, but also find the right field basis to include the three-form moduli in the general supergravity framework and calculate their couplings. Furthermore will we extend the usual toric constructions of Calabi-Yau fourfolds to account for this novel three-form sector and discuss the geometry of elliptic fibrations with non-trivial three-form cohomology. As we go through our discussion we will encounter a number of interesting and often puzzling features that deserve further research, as for instance the shift-symmetries of the effective theories investigated in \cite{Corvilain:2016kwe}. In the upcoming section we will give a brief outline of this thesis.

\section{Outline}

We start this thesis by introducing the geometric properties of Calabi-Yau fourfolds and their harmonic forms in \autoref{FourfoldBasics} where we focus especially on the harmonic three-forms whose physical properties are determined by the complex structure dependence of their so called normalized period matrix. The first part of the thesis focuses on effective theories of Calabi-Yau fourfolds. 

In \autoref{IIAreduction} we perform the dimensional reduction of Type IIA supergravity on a general Calabi-Yau fourfold and find a $ \cN =(2,2) $ dilaton-supergravity in two dimensions. Here we determine the massless spectrum and the kinetic potential of the resulting supergravity. We conjecture an extension of the usual K\"ahlerpotential of supergravity theories by a term depending on the dilaton of the theory to account for the kinetic coupling of the novel three-form scalars. Subsequently we apply mirror symmetry in \autoref{mirror_section}, a duality of the same supergravity theories for two different Calabi-Yau fourfolds, to determine the structure of the normalized period matrix at the large complex structure point in moduli space. Following this discussion we extend our analysis to the dimensional reduction of eleven-dimensional supergravity in \autoref{F-theoryapp}. Afterwards we lift the resulting three-dimensional $ \cN=2 $ supergravity to the effective theory of F-theory on elliptically fibered Calabi-Yau fourfolds leading to an effective $ \cN=1 $ supergravity description in four dimensions. We close by discussing the implications of non-trivial three-form cohomology and defer some technical details of the first part into two appendices in \autoref{3d-2dreduction},\autoref{detailed_dual}.

In the second part of this thesis we construct explicit Calabi-Yau fourfold examples via hypersurfaces in toric varieties. We begin in \autoref{ToricSection} with the general construction of toric spaces and their hypersurfaces. We focus on Calabi-Yau fourfolds realized as so called semi-ample hypersurfaces avoiding the Lefschetz-hyperplane theorem that forbids non-trivial three-form cohomology. This is followed by a discussion of the origin of non-trivial harmonic forms on the hypersurface and we state well known formulas for the number of the various harmonic forms in terms of toric data. We added several clarifications and extensions of the published work to include also non-toric and non-algebraic deformations of toric hypersurfaces. In \autoref{JacobianSection} we discuss the moduli space of the three-form scalars which is called intermediate Jacobian of a Calabi-Yau fourfold whose complex structure dependence reduces to the complex structure dependence of Riemann surfaces. The K\"ahler dependence is captured by a so called generalized sphere-tree that can be computed in terms of the ambient space. We determine in this situation the geometrical quantities necessary to derive the effective theories discussed before. This part is completed by a lengthy discussion of two simple examples of Calabi-Yau fourfold geometries with non-trival three-form cohomology in \autoref{ExampleSection}. We conclude the thesis in \autoref{ConclusionSection} giving an outlook of possible future research directions.

This thesis is based on two papers \cite{Greiner:2017ery,Greiner:2015mdm} in collaboration with Thomas W. Grimm and one proceedings article \cite{Greiner:2017thb}.
\begin{itemize}
 \item S.~Greiner and T.~W.~Grimm, 
 ``Three-form periods on Calabi-Yau fourfolds: Toric hypersurfaces and F-theory applications'', \\
  JHEP \textbf{1705} (2017) 151, arXiv:1702.03217 [hep-th].
 \item S.~Greiner and T.~W.~Grimm,
  ``On Mirror Symmetry for Calabi-Yau Fourfolds with Three-Form Cohomology'', \\
  JHEP \textbf {1609} (2016) 073, arXiv:1512.04859 [hep-th].
  \item S.~Greiner,
  ``On Mirror Symmetry for Calabi-Yau Fourfolds with Three-Form Cohomology'', \\
  PoS CORFU \textbf{2016} (2017) 102, arXiv:1704.07658 [hep-th].
\end{itemize}

\newpage
 
 \chapter{Calabi-Yau fourfolds with non-trivial three-form cohomology} \label{FourfoldBasics}
 
 In this section we lay the groundwork for our dimensional reductions int the coming sections by introducing the central points of Calabi-Yau fourfold geometries. Here we begin with a precise definition of Calabi-Yau fourfolds and move then on to the geometric properties like the K\"ahler- and complex structure. For the mathematical details we refer to \cite{Greene:1996cy,Voisin03hodgetheory,Moroianu:2004}. In this first part we will also account for the non-trivial harmonic forms of this class of manifolds that will later determine the spectrum of our effective theories. In the second part of this section we will discuss in detail the harmonic three-forms for which we will choose a special and novel ansatz that will be central to the whole body of this thesis.
 
 \section{Basic properties of Calabi-Yau fourfolds}
 
 We define in this thesis a Calabi-Yau fourfold $ Y_4 $ as a eight-dimensional Riemannian manifold with metric $ g $ and holonomy group the full special unitary group $ SU(4) $. As a consequence, $ Y_4 $ will be K\"ahler and hence admit a closed K\"ahler two-form $ J $ compatible with both the complex structure and the metric of $ Y_4 $. In addition, as was conjectured by Calabi and shown by Yau, there is a unique Ricci-flat K\"ahler metric within the class of $ J \in H^{1,1}(Y_4) $. On such Calabi-Yau fourfolds exists a unique, up to rescaling by a holomorphic function, nowhere-vanishing holomorphic $(4,0)$-form $ \Omega \in H^{4,0}(Y_4)$. This holomorphic $ (4,0) $-form $ \Omega $ only depends on the complex structure, whereas the K\"ahler form $ J $ depends only on the K\"ahler moduli of the underlying space. The complex structure moduli can be interpreted as defining the shape of $ Y_4 $ and the K\"ahler moduli specify the volumes of its cycles.
 
 From $ \Omega $ and $ J $ we can construct top-forms of degree eight, which have to be related. This relation is given by
 \begin{equation} \label{omega_squared}
  \frac{1}{4!} J \wedge J \wedge J \wedge J = \frac{1}{|\Omega|^2} \Omega \wedge \bar \Omega \, , \quad | \Omega |^2 = \frac{1}{\cV} \int_{Y_4} \Omega \wedge \bar \Omega
 \end{equation}
 where we defined indirectly the total volume of $ Y_4 $ that only depends on $ J $ and hence the K\"ahler moduli as
 \begin{equation} \label{cy_volume}
  \cV = \frac{1}{4!} \int_{Y_4} J \wedge J \wedge J \wedge J \, .
 \end{equation}
 
 Due to the $ SU(4) $ holonomy of the underlying space the existence of a complex covariantly constant spinor with definite chirality. This allows to obtain $ J $ and $ \Omega $ as bilinear contractions of this spinor and hence to show their existence. The invariant spinor is obtained from the splitting of the general chiral spinor representation $ \textbf{8}_s $ of an eight-dimensional Riemannian manifold that admits spinors as
 \begin{align}
  Spin(8) \, &\rightarrow \, SU(4) \nonumber \\
  \textbf{8}_s \, & \rightarrow \, \textbf{6} \oplus \textbf{1} \oplus \textbf{1} \, ,
 \end{align}
 where the two singlets combine to the complex covariantly constant spinor. Therefore compactifications of $ Y_4 $ without fluxes to a flat space-time preserve $ 1/8 $ of the supercharges in the effective theory.
 
 Another important feature of Calabi-Yau fourfolds is the very restrictive Hodge diamond displaying the dimensions of the complex cohomology groups
\beq
 h^{p,q} = \text{dim} \, H^{p,q}(Y_4) \, , \qquad \bigoplus_{p=0} ^n H^{n-p,p}(Y_4) = H^n (Y_4, \mathbb{C}) \, ,
\eeq
in the following way: \\

\vspace*{.4cm}
\arraycolsep=2,0pt\def\arraystretch{1.4}
\setlength{\unitlength}{0.6cm}
\begin{picture}(21,8) (1,0)\label{HodgeDiamond}
 \put(5,8){$ h^{0,0} $}
 \put(4,7){$ h^{1,0} $} \put(6,7){$ h^{0,1} $}
 \put(3,6){$ h^{2,0} $} \put(5,6){$ h^{1,1} $} \put(7,6){$ h^{0,2} $}
 \put(2,5){$ h^{3,0} $} \put(4,5){$ h^{2,1} $} \put(6,5){$ h^{1,2} $} \put(8,5){$ h^{0,3} $}
 \put(1,4){$ h^{4,0} $} \put(3,4){$ h^{3,1} $} \put(5,4){$ h^{2,2} $} \put(7,4){$ h^{1,3} $} \put(9,4){$ h^{0,4} $}
 \put(2,3){$ h^{4,1} $} \put(4,3){$ h^{3,2} $} \put(6,3){$ h^{2,3} $} \put(8,3){$ h^{1,4} $}
 \put(3,2){$ h^{4,2} $} \put(5,2){$ h^{3,3} $} \put(7,2){$ h^{2,4} $}
 \put(4,1){$ h^{4,3} $} \put(6,1){$ h^{3,4} $}
 \put(5,0){$ h^{4,4} $}
 
 \put(11,4){$ = $}
 
 \put(17,8){$ 1 $}
 \put(16,7){$ 0 $} \put(18,7){$ 0 $}
 \put(15,6){$ 0 $} \put(17,6){$ h^{1,1} $} \put(19,6){$ 0 $}
 \put(14,5){$ 0 $} \put(16,5){$ h^{2,1} $} \put(18,5){$ h^{2,1} $} \put(20,5){$ 0 $}
 \put(13,4){$ 1 $} \put(15,4){$ h^{3,1} $} \put(17,4){$ h^{2,2} $} \put(19,4){$ h^{3,1} $} \put(21,4){$ 1 $}
 \put(14,3){$ 0 $} \put(16,3){$ h^{2,1} $} \put(18,3){$ h^{2,1} $} \put(20,3){$ 0 $}
 \put(15,2){$ 0 $} \put(17,2){$ h^{1,1} $} \put(19,2){$ 0 $}
 \put(16,1){$ 0 $} \put(18,1){$ 0 $}
 \put(17,0){$ 1 $}
 \setlength{\unitlength}{0.05cm}
 \multiput(20,73)(10,-10){7}{\line(1,-1){6}}       
\end{picture}
\vspace*{.03cm}

\noindent

Due to the severe restrictions arising from the specific holonomy group there can be no non-trivial holomorphic forms beside the trivial zero-form and $ \Omega $. Hodge-duality and Poincar\'e duality lead to symmeties of Hodge numbers between the horizontal and the vertical axis. The axis indicated on the left hand side of the graphic is the symmetry axis of mirror symmetry, a relation between two different manifolds with Hodge numbers symmetric with respect to this axis. This will be discussed in great detail later in this work.

Applying index theorems, one can show that also $ h^{2,2} $ can be calculated from $ h^{1,1}, h^{1,2} $ and $ h^{3,1} $ as
\beq \label{h22fromIndex}
 h^{2,2}(Y_4) = 2(22 + 2h^{1,1} + 2h^{3,1} - h^{2,1}) \, ,  \quad \chi(Y_4)
 = 6(8 + h^{1,1} + h^{3,1} - h^{2,1})
\eeq
where  $ \chi = \sum_{p,q} (-1)^{p+q} h^{p,q}$ is the Euler characteristic of $Y_4$. This makes $ h^{1,1}, h^{1,2} $ and $ h^{3,1} $ the independent quantities of $ Y_4 $ that will specify the spectrum of the resulting effective theory. The Hodge number $ h^{1,1} $ counts the number of independent K\"ahler deformations denoted by $ \delta v^\Sigma $. We have the expansion for the K\"ahlerform
\begin{equation} \label{KaehlerFormExpand}
 J = v^\Sigma \omega_\Sigma \, , \quad \Sigma = 1, \ldots, h^{1,1} \, ,
\end{equation}
where $ \omega_\Sigma $ are harmonic $ (1,1) $-forms.
 In contrast $ h^{3,1} $ counts the number of complex structure deformations denoted by $ \delta z^\cK $ and we use the complex structure coordinates
 \begin{equation}
  z^\cK \, , \quad \cK = 1, \ldots, h^{3,1} \, .
 \end{equation} 
This is very similar to the threefold case. In contrast to the threefold, however, also $ h^{2,1} $ counts the number of three-form moduli 
\begin{equation}
 N_\cA \, , \quad \cA = 1, \ldots, h^{2,1} \, ,
\end{equation}
 which have no threefold analogue. These new deformations arise from the expansion of a gauge three-form into a special ansatz for the harmonic three-forms on $ Y_4 $ which will be discussed in the next section.
 
 \section{Special ansatz for three-forms} \label{threeFormAnsatz}
 
 In this section we want to discuss the ansatz for the harmonic three-forms that will keep us busy throughout this work. In course of the upcoming dimensional reductions, it will become necessary to calculate
 \begin{equation}
  \ast \phi^{(3)} \in H^{5}(Y_4, \mathbb{C}) \, , \quad \phi^{(3)} \in H^{3}(Y_4, \mathbb{C}) \, .
 \end{equation}
 The problem with this ansatz is, that the Hodge-star operator $ \ast $ of $ Y_4 $ will depend on the metric $ g $ and therefore also on the K\"ahler- and complex structure moduli in our theory. Therefore, even if we choose $ \phi^{(3)} $ topological, $ \ast \phi^{(3)} $ will still depend on the other moduli. It can be shown that for a Calabi-Yau fourfold $ Y_4 $ we can calculate the Hodge star of a three-form if it has certain Hodge type. For a Calabi-Yau fourfold the cohomology group of three-forms splits as
 \begin{equation}
  H^3(Y_4, \mathbb{C}) = H^{2,1}(Y_4) \oplus H^{1,2} (Y_4) \, .
 \end{equation}
 The two subspaces correspond to $ \pm i J \wedge $ subspaces of the Hodge-star operator $ \ast $ on $ Y_4 $ as
 \begin{equation} \label{starPsi}
  \ast \psi = i J \wedge \phi \, \quad \psi \in H^{2,1}(Y_4) \, .
 \end{equation}
 Therefore it is desirable to choose a basis of three-forms with a definite Hodge-type
 \begin{equation}
  \psi_\cA \in H^{2,1}(Y_4) \, , \quad \bar \psi_\cA \in H^{1,2}(Y_4) \, , \quad \cA = 1, \ldots, h^{2,1} \, .
 \end{equation}
 The split of $ H^{3}(Y_4, \mathbb{C}) $ into these two subspaces is done via the induced complex structure $ \cJ $ that varies with the complex structure moduli $ z^\cK $. This is a real endomorphism on $ H^3(Y_4,\mathbb{C}) $ that squares to the identity. $ H^{2,1} (Y_4) $ and $ H^{1,2} (Y_4) $ are the $ \pm i $ eigenspaces of this endomorphism. Therefore we can choose a real basis  $ \alpha_\cA, \beta^\cA \in H^3(Y_4, \mathbb{R}) $ for a fixed complex structure that satisfies
 \begin{equation} \label{ReferenceComplexStructure}
  \cJ(\alpha_\cB) = \delta_{\cB \cA} \beta^\cA \, , \quad \cJ(\beta^\cB) = - \delta^{\cB \cA} \alpha_\cA \, .
 \end{equation}
 From this we can construct the desired basis of $ (2,1) $-forms as
 \begin{equation}
  \psi_\cA = \alpha_\cA + i \delta_{\cA \cB} \beta^\cB \, , \quad \Rightarrow \quad \cJ(\psi_\cA) = \delta_{\cA \cB} \beta^\cB - i \alpha_\cA = - i \psi_\cA \, .
 \end{equation}
 From the general theory of complex structure variations it is known that $ H^{2,1}(Y_{4,z}) $ varies holomorphically with the complex structure moduli $ z^\cK $. Here and in the following we denote by $ Y_{4,z} $ the Calabi-Yau fourfold with a specific complex structure $ z^\cK $ that is a small deformation around a reference point where we defined the topological basis \eqref{ReferenceComplexStructure}. Therefore, we can make the ansatz
 \begin{equation} \label{psi_def}
  \psi_\cA (z) = {\Pi_\cA} ^\cB (z) \alpha_\cB + \tilde \Pi_{\cA \cB}(z) \beta^\cB \, \in H^{2,1}(Y_{4,z}) \, ,
 \end{equation}
 and since both matrices are invertible as seen before (they are proportional to the identity matrix at the reference point in complex structure space) we can make the ansatz
 \begin{equation}
  \psi_\cA (z) =  \alpha_\cA + i f_{\cA \cB} (z) \beta^\cB \, \in H^{2,1}(Y_{4,z}) \, ,
 \end{equation}
 where the real part $ \text{Re} \, f_{\cA \cB} $ of the holomorphic endomorphism-valued function $ f(z)_{\cA \cB} $ is invertible for small deformations around the reference complex structure. The matrices $  {\Pi_\cA} ^\cB$ and $ \Pi_{\cA \cB} $ are called period matrices and we will refer to them as three-form period matrices. The three-form periods are the column-vectors of these matrices. The quotient of these two matrices 
 \begin{equation} \label{fThreeFormDef}
  f_{\cA \cB} = {(\Pi^{-1})_\cA} ^\cC \tilde \Pi_{\cC \cB}
 \end{equation}
 is called the three-form normalized period matrix. This matrix is one of the central topics of this thesis and we will explore both its physical applications as well as calculate it for explicit examples.
 
 In course of this work, it will also be important to calculate the Hodge star of
 \begin{equation}
   \frac{\partial}{\partial z^\cK} \, \psi_{\cA} \in H^{2,1}(Y_4) \oplus H^{1,2} (Y_4) \, .
 \end{equation}
 So we also want the first derivatives of our three-form basis to have a definite Hodge-type. This can be achieved by a rescaling with the inverse of the real part of $ f_{\cA \cB} $ we denote by $ \text{Re} \, f^{\cA \cB} = {(\text{Re} \, f)^{-1}} ^{\cA \cB}$
 \begin{equation} \label{Psi_def}
  \Psi ^\cA (z, \bar z) = \frac{1}{2} \text{Re} \, f ^{\cA \cB}\big(  \alpha_\cB - i \bar f_{\cB \cC} \beta^\cC \big) \,  \in H^{1,2}(Y_4) \, .
 \end{equation}
 This new basis is, however, not anymore (anti-)holomorphic in the complex structure moduli. The $ 1/2 $ factor in front has only cosmetically reasons. The advantage is that we can calculate
 \begin{align}
  \frac{\partial}{\partial z^\cK} \, \Psi ^\cA &= - \text{Re} \, f^{\cA \cB} \partial_{z^\cK} \text{Re} \,f _{\cB \cC} \Psi^\cC \, , \\
  \frac{\partial}{\partial \bar z^\cK}\,  \Psi ^\cA &= - \text{Re} \, f^{\cA \cB} \partial_{\bar z^\cK} \text{Re} \, f _{\cB \cC} \bar \Psi^\cC \, .
 \end{align}
 and hence also the derivatives of $ \Psi^\cA $ have a definite Hodge type. In course of this work, we will find that the complex scalars $ N_\cA $ appear in the effective actions of $ M $-theory and Type IIA-string theory. These have kinetic terms determined by a positive bilinear form
\begin{equation}
 H(\Psi^\cA, \Psi^\cB) = \int_{Y_4} \Psi^\cA \wedge \ast \, \Psi^\cB = i v^\Sigma \int_{Y_4} \omega_\Sigma \wedge \Psi^\cA \wedge \bar \Psi^\cB \, .
\end{equation}
up to a constant factor. Here we used the Hodge-star operator $ \ast $ on $ Y_4 $ which simplifies on a $ (1,2) $-form to $ \ast \Psi^\cA = -i J \wedge \Psi^\cA $ and we expanded the K\"ahler form $ J $ as in \eqref{KaehlerFormExpand}. We can use the expansion of $ \Psi_\cA $ in \eqref{Psi_def} into topological three-forms to find
\begin{equation} \label{HPsiDef}
 H(\Psi^\cA, \Psi^\cB) = - \frac{1}{2} \text{Re} \, f^{\cB \cC} v^\Sigma \big({M_{\Sigma \cC}}^\cA + i f_{\cC \cD} {M_\Sigma} ^{\cD \cA} \big)
\end{equation}
where we introduced the topological intersection numbers
\begin{equation} \label{Mdef}
 {M_{\Sigma \cA}}^\cB = \int_{Y_4} \omega_\Sigma \wedge \alpha_\cA \wedge \beta^\cB \, , \quad {M_\Sigma} ^{\cA \cB} = \int_{Y_4} \omega_\Sigma \wedge \beta^\cA \wedge \beta^\cB \, ,
\end{equation}
that will play an important role throughout this thesis. Note that we can choose without loss of generality a basis of topological three-forms with $ \alpha_\cA \wedge \alpha_\cB = 0 $. In contrast the intersection numbers $  {M_\Sigma} ^{\cA \cB} $ do not vanish in general. The form \eqref{HPsiDef} will be important for the derivation of effective theories, for the geometric calculations on Calabi-Yau hypersurfaces in toric ambient spaces, however, evaluating $ H $ on the holomorphic $(2,1)$-forms \eqref{psi_def} is more convenient. This is related to \eqref{HPsiDef} via multiplication with appropriate multiples of $ \text{Re} \, f_{\cA \cB} $ and reads
\begin{equation} \label{HpsiDef}
 H(\psi_\cA, \psi_\cB) = 2 \, \text{Re} \, f_{\cA \cC} v^\Sigma \big({M_{\Sigma \cB}}^\cC + i f_{\cB \cD} {M_\Sigma} ^{\cD \cC} \big) \, .
\end{equation}
From the previous analysis we see that not only the normalized period matrix $ f_{\cA \cB} $, but also the intersection numbers $  {M_{\Sigma \cA}}^\cB,  {M_\Sigma} ^{\cA \cB} $ need to be calculated to fully understand the metric $ H $. As we will see in the upcoming sections these two quantities are related by mirror symmetry at certain points in moduli space. To understand this further, we will first perform dimensional reductions of Type IIA supergravity on a Calabi-Yau fourfold $ Y_4 $ in the next section.
 
\chapter{Dimensional reduction of Type IIA supergravity}  \label{IIAreduction}

In this section we perform the dimensional reduction of Type IIA supergravity on 
a Calabi-Yau fourfold $Y_4$. Such reductions have  already been
performed in \cite{Haack:1999zv,Haack:2001jz,Haack:2000di,Gates:2000fj}. Our analysis follows 
\cite{Haack:1999zv,Haack:2001jz,Haack:2000di}, but we will apply in addition the improved understanding about 
the three-form cohomology of \autoref{threeFormAnsatz}.

\section{The effective action from a Calabi-Yau fourfold reduction} \label{IIAreduction_details}

The Kaluza-Klein reduction of Type IIA supergravity can be trusted in the
limit in which the typical length scale of the physical volumes of submanifolds of $Y_4$ are sufficiently 
large compared to the string scale. This limit is referred to 
as the large volume limit. Furthermore, these typical length scales 
set the Kaluza-Klein scale which we assume to be sufficiently above the 
energy scale of the effective action. We therefore keep only the massless Kaluza-Klein modes 
in the following reduction. 

Our starting point will be the bosonic part of the ten-dimensional Type IIA action in string-frame given by \footnote{
Note that for convenience we have set $\kappa^2 = 1$.}
\begin{align} \label{10Daction}
S^{(10)}_{\textrm{IIA}} &= \int e^{-2\check \phi_{\textrm{IIA}}} \left( \frac{1}{2}\check R \, \check \ast 1 + 2 d\check \phi_{\textrm{IIA}} \wedge \check \ast d\check \phi_{\textrm{IIA}} - \frac{1}{4} \check H_3 \wedge \check \ast \check  H_3 \right) \nonumber \\
									&\quad - \frac{1}{4} \int \Big( \check F_2 \wedge \check \ast \check F_2 + \check {\mathbf F}_4 \wedge \check \ast \check {\mathbf{F}}_4+ \check B_2 \wedge \check F_4 \wedge \check F_4\Big) \, ,
\end{align}
where $\check \phi_{\textrm{IIA}}$ is the ten-dimensional dilaton, $ \check H_3 = d \check B_2 $ is the field strength of the NS-NS two-form 
$ \check B_2 $, and $\check F_p = d\check C_p$ are the field strengths of the R-R $p$-forms $\check C_1$ and $\check C_3$. 
We also have used the modified field strength $ \check {\mathbf{F}}_4 = \check F_4 - \check C_1 \wedge \check H_3 $. 
Here and in the following we will use a check to indicate ten-dimensional fields. 

The background solution around which we want to consider the effective theory is 
taken to be of the form $\mathbb{M}_{1,1} \times Y_4$, where $\mathbb{M}_{1,1}$
is the two-dimensional Minkowski space-time, and $Y_4$ is a Calabi-Yau fourfold 
with properties introduced in \autoref{FourfoldBasics}. As pointed out there 
such a manifold admits one complex covariantly constant spinor of definite chirality. 
This spinor can be used to dimensionally reduce the $\cN=(1,1)$ supersymmetry 
of Type IIA supergravity to obtain a two-dimensional $\cN=(2,2)$ supergravity theory.
In particular, the two ten-dimensional gravitinos of opposite chirality reduce to two 
pairs of two-dimensional Majorana-Weyl gravitinos with opposite chirality. We will have more to 
say about the supersymmetry properties of the two-dimensional action in \autoref{22dilatonSugra}.
Furthermore, recall that $Y_4$ admits a 
Ricci-flat metric $g^{(8)}_{mn}$ and one can thus check that a metric of the form 
\beq  \label{metric-ansatz}
   d\check s^2 = \eta_{\mu \nu} dx^\mu dx^\nu + g^{(8)}_{mn} dy^m dy^n\ , 
\eeq
solves the ten-dimensional equations of motion in the absence of background fluxes.\footnote{The 
inclusion of background fluxes complicates the reduction further. In particular, it 
requires to introduce a warp-factor. The M-theory reduction with warp-factor was recently 
performed in \cite{Grimm:2014xva,Grimm:2014efa,Grimm:2015mua}.}
Note that in \eqref{metric-ansatz} we denote 
by $ x^\mu $ the two-dimensional coordinates of the space-time $\mathbb{M}_{1,1} $, 
whereas the eight-dimensional real coordinates on the Calabi-Yau fourfold $ Y_4 $ are denoted 
by $ y^m $. 

The massless perturbations around this background both consist of fluctuations 
of the internal metric $g^{(8)}_{mn}$ that preserve the Calabi-Yau condition as
well as the fluctuations of the form fields $\check B_2$, $\check C_1,\check C_3$ and the dilaton $\check \phi_{\textrm{IIA}}$.  
The metric fluctuations give rise to the real K\"ahler structure moduli $v^\Sigma$, $\Sigma=1,\ldots, h^{1,1}(Y_4)$ that preserve the 
complex structure and are given by 
\begin{equation} \label{Kaehlermoduli}
	g_{i\bar \jmath } + \delta g_{i \bar \jmath} = -i J_{i \bar \jmath} = -i v^\Sigma \, (\omega_{\Sigma})_{ i \bar \jmath} \, ,
\end{equation}
where $J$ is the K\"ahler form on $Y_4$ and $ \omega_\Sigma$ comprises a real basis of 
harmonic $(1,1)$-forms spanning $H^{1,1}(Y_4) $. The K\"ahler structure moduli 
appear also in the expression of the total string-frame volume $\cV$ of $Y_4$ given by 
\beq
    \cV \equiv \int_{Y_4} \ast 1  = \frac{1}{4!} \int_{Y_4} J \wedge J \wedge J \wedge J \, .
\eeq
In addition to the K\"ahler structure moduli one finds a set of complex structure moduli $z^\cK$, $\cK = 1,\ldots, h^{3,1}(Y_4)$.
These fields parameterize the change in the complex structure of $Y_4$ preserving the class 
of its K\"ahler form $J$. Infinitesimally they are given by the fluctuations $\delta z^\cK$ as 
\begin{equation} \label{CSmoduli}
	\delta g_{\bar \imath \bar \jmath} = - \frac{1}{3 | \Omega| ^2 } \overline{\Omega}_{\bar \imath}^{\ lmn} (\chi_{\cK})_{lmn \bar \jmath}\, \delta z^{\cK} \, ,	
\end{equation}
where $\Omega$ is the $(4,0)$-form, the $\chi_{\cK}$ form a 
basis of harmonic $(3,1)$-forms spanning $H^{3,1}(Y_4)$, and $|\Omega|^2$
was already given  in \eqref{omega_squared}. 

The Kaluza-Klein ansatz for the remaining fields takes the form 
\begin{align} \label{BC-expand}
   &\check B_2 = b^\Sigma \omega_\Sigma \, , \qquad \check C_1 = A \, ,\\
   &\check C_3 = V^\Sigma \wedge \omega_\Sigma + N_\cA \Psi^\cA + \bar N_\cA \bar \Psi^\cA \, , \nn 
\end{align}
where $\Psi^\cA$ is a basis of harmonic 
$(1,2)$-forms spanning $H^{1,2}(Y_4)$ as introduced in \eqref{Psi_def}. 
A discussion of the properties of $\Psi^\cA$ was already given in \autoref{threeFormAnsatz}.
Finally, we dimensionally reduce the Type IIA dilaton by dropping its 
dependence on the internal manifold $Y_4$. It turns out to be convenient 
to define a two-dimensional dilaton $\phi_{\textrm{IIA}}$ in 
terms of the ten-dimensional dilaton $\check \phi_{\textrm{IIA}}$ as
\beq \label{def-2dilaton}
   e^{2 \phi_{\textrm  {IIA}}}  \equiv  \frac{e^{2 \check \phi_{\textrm  {IIA}}}}{\cV}\ .
\eeq   
In summary, we find in the two-dimensional $\cN=(2,2)$ supergravity theory 
the $2 h^{1,1}(Y_4) + 1$ real scalar fields $v^\Sigma (x)$, $b^\Sigma(x)$, $\phi_{\textrm {IIA}}(x)$ as 
well as the $h^{3,1}(Y_4) + h^{2,1}(Y_4)$ complex scalar fields $z^\cK$, $ N_\cA $. 
In addition there are $h^{1,1}(Y_4) + 1$ vectors $A$, $V^\Sigma$, which carry, however, 
no physical degrees of freedom in a two-dimensional theory if they are not involved 
in any gauging. Since the effective action considered here contains 
no gaugings, we will drop these in the following analysis. 

To perform the dimensional reduction one inserts the expansions \eqref{Kaehlermoduli}, 
\eqref{CSmoduli}, \eqref{BC-expand}, and \eqref{def-2dilaton} into
 the Type IIA action \eqref{10Daction}.
 It reduces 
to the two-dimensional action 
\begin{align} \label{2Daction}
  S^{(2)} &= \int  e^{-2 \phi_{\textrm  {IIA}}} \Big( \frac{1}{2} R \ast 1 + 2 d \phi_{\textrm  {IIA}} \wedge \ast \phi_{\textrm{  IIA}} \Big) \\
  &\quad - \int  e^{-2 \phi_{\textrm  {IIA}}}  \Big( G_{\cK \bar \cL}\, dz^\cK \wedge * d\bar z^\cL + G_{\Sigma \Lambda} \, d t^\Sigma \wedge * d\bar t^\Lambda \Big) \nn \\
  &\quad  -  \int \frac{1}{2} v^\Sigma d_\Sigma{}^{\cA \cB} D N_\cA \wedge * D \bar N_\cB 
                                                + \frac{i}{4} d_\Sigma{}^{\cA \cB}  db^\Sigma \wedge \big(N_\cA D \bar N_\cB - DN_{\cB} \bar N_\cA \big)  \, . \nn
\end{align}
We note that the NS-NS part, which 
is summarized in the first line of \eqref{10Daction}, reduces to the first line of \eqref{2Daction}, 
while the R-R part, i.e.~the second line of  \eqref{10Daction}, reduces to the second line of \eqref{2Daction}.

Let us introduce the various objects appearing in the action \eqref{2Daction}. 
First, we have defined the complex coordinates 
\beq
    t^\Sigma \equiv b^\Sigma + i v^\Sigma\, ,
\eeq
which combine the K\"ahler structure moduli with the B-field moduli. 
Furthermore, we have introduced the metric \footnote{The second 
equality follows from the cohomological identity $ \ast \, \omega_\Sigma = -\frac{1}{2} \omega_\Sigma \wedge J \wedge J + \frac{1}{36} \cV^{-1}\cK_\Sigma J \wedge J \wedge J$.}
\beq \label{GABmetric}
    G_{\Sigma \Lambda} = \frac{1}{4 \cV } \int_{Y_4} \omega_\Sigma \wedge \ast \, \omega_\Lambda 
    = - \frac{1}{8 \cV} \Big( \cK_{\Sigma \Lambda} - \frac{1}{18 \cV} \cK_\Sigma \cK_\Lambda \Big)\, ,
\eeq
where $\cV$, $\cK_\Sigma$ and $\cK_{\Sigma \Lambda}$ are given in terms of the 
 quadruple intersection numbers $\cK_{\Sigma \Lambda }$ as
 \begin{align}
   &\cK_{\Sigma \Lambda \Gamma \Delta} = \int_{Y_4} \omega_\Sigma \wedge \omega_\Lambda \wedge \omega_\Gamma \wedge \omega_\Delta \, , \\
    &\cV =\frac{1}{4!} \cK_{\Sigma \Lambda \Gamma \Delta} v^\Sigma v^\Lambda v^\Gamma v^\Delta \, ,\quad \cK_\Sigma =  \cK_{\Sigma \Lambda \Gamma\Delta}   v^\Lambda v^\Gamma v^\Delta \, , \quad 
    \cK_{\Sigma \Lambda} =  \cK_{\Sigma \Lambda \Gamma \Delta}   v^\Gamma v^\Delta \, . \nn 
 \end{align}
With these definitions at hand, we can further evaluate the metric 
$G_{\Sigma  \Lambda}$ and show that it can be obtained from a K\"ahler potential as
\beq
    G_{\Sigma  \Lambda} =- \partial_{t^\Sigma} \partial_{\bar t^\Lambda} \log \, \cV  \, .
\eeq
Also the metric $G_{\cK \bar \cL}$ is actually a K\"ahler metric. It only depends on the complex structure 
moduli $z^\cK$ and takes the form 
\beq
    G_{\cK \bar L} = - \frac{\int_{Y_4} \chi_{cK} \wedge \bar \chi_\cL}{\int_{Y_4} \Omega \wedge \bar \Omega} 
    = - \partial_{z^\cK} \partial_{\bar z^\cL} \log \int_{Y_4} \Omega \wedge \bar \Omega\, .
\eeq
Note that both $G_{\Sigma \Lambda}$ and $G_{\cK \bar \cL}$ are actually positive definite and therefore 
define physical kinetic terms in \eqref{2Daction}. Both terms scale
with the dilaton $\phi_{\textrm{IIA}}$ and it is easy to check that this dependence cannot be 
removed using a Weyl-rescaling of the two-dimensional metric. We will show in 
\autoref{22dilatonSugra} that this is consistent with the form of the $\cN=(2,2)$ dilaton 
supergravity. 

Let us now turn to the R-R part of the action \eqref{10Daction} and 
discuss the couplings appearing in the second line of \eqref{2Daction}.
First, we introduce the coupling function
\beq \label{evaluate_d}
    {d_\Sigma}^{\cA \cB} \equiv  i \int_{Y_4} \omega_\Sigma \wedge \Psi^\cA \wedge \bar \Psi^\cB 
    = - \frac{1}{2} (\R f)^{\cA \cC} {M_{\Sigma \cC}}^\cB  \, .
\eeq
Here and in the following we assume the second intersection number $ {M_{\Sigma } }^{\cB \cC} $ in \eqref{Mdef} to vanish for simplicity. This is in particular 
satisfied for the examples we will consider in \autoref{ToricSection}.
We have used the fact that we can choose the three-form basis $ (\alpha_\cA, \beta^\cA) $ with $ \alpha_\cA \wedge \alpha_\cB = 0 $ to evaluate the second equality, and $ {M_{\Sigma } }^{\cB \cC} = 0 $ to 
show the third equality. 
One also checks the relation
\beq \label{def-H}
   H^{\cA \cB} \equiv \int_{Y_4} \Psi^\cA \wedge * \bar \Psi^\cB  
   =  i \int_{Y_4} J \wedge \Psi^\cA \wedge \bar \Psi^\cB =  v^\Sigma {d_\Sigma}^{\cA \cB} \, ,
\eeq
where we have used \eqref{starPsi} for the (1,2)-forms $\Psi^\cA$. This contraction gives precisely the 
positive definite metric of the complex scalars $N_\cA$ in \eqref{2Daction}.
It turns out to be convenient to write
\beq \label{splitH}
   H^{\cA \cB}  = v^\Sigma {d_\Sigma}^{\cA \cB} =   - \frac{1}{2} (\R \, f)^{\cA \cC} v^\Sigma {M_{\Sigma \cC}}^\cB  \equiv - \frac{1}{2} (\R \, f)^{\cA \cC} \, \R\, {h_\cC}^\cB\, ,
\eeq 
where $ {h_\cC}^\cB = - i t^\Sigma {M_{\Sigma \cC}}^\cB $ with the intersection number $ {M_{\Sigma \cC}}^\cB $ of \eqref{Mdef}.
Note that $H^{\cA \cB}$ thus depends non-trivially on the complex structure moduli $z^\cK$
through the holomorphic functions $f_{\cA \cB}$ and on the K\"ahler structure moduli $t^\Sigma$
through the holomorphic function ${h_{\cC}}^\cB $. $ H^{\cA \cB} $ is the metric on the three-form moduli space 
as defined in \eqref{HPsiDef}. 
Second, we note that the modified derivative $DN_\cA $ appearing in \eqref{2Daction} is a shorthand for
\beq
  D N_\cA = d N_\cA - 2 \, \text{Re} \, N_\cC (\text{Re} \, f)^{\cC \cB} \partial_{z^\cK} (\text{Re} \, f_{\cB \cA} ) \, dz^\cK\, .
\eeq
We use the notation $ D \bar N_\cA = \overline{DN}_\cA $ in the action of \eqref{2Daction}.
Using this expression one easily reads off the coefficient function in front of $dN_\cA \wedge \ast dz^\cK$
and checks that it can be obtained by taking derivatives of a real function. In the next subsection 
we show that this is true for all terms in \eqref{2Daction} and discuss the connection with two-dimensional 
supersymmetry. 

\section{Comments on two-dimensional $\cN=(2,2)$ supergravity} \label{22dilatonSugra}

Having performed the dimensional reduction we next want to comment 
on the supersymmetry properties of  the 
action \eqref{2Daction}. As pointed out already 
in the previous subsection the  counting of covariantly constant spinors on 
the Calabi-Yau fourfold suggests that the two-dimensional effective theory admits 
$\cN=(2,2)$ supersymmetry. It was pointed out in \cite{Gates:2000fj} 
that, at least in the case of $h^{2,1}(Y_4)=0$ one expects to be 
able to bring the action \eqref{2Daction} into the standard form of 
an two-dimensional $\cN=(2,2)$ dilaton supergravity. In 
this work the dilaton supergravity action was constructed using 
superspace  techniques. Earlier works in this direction include \cite{Gates:1994gx,deWit:1992xr,Grisaru:1995dr}.
In the following we comment on this matching for $h^{2,1}(Y_4)=0$ 
and then discuss the general case in which $h^{2,1}(Y_4)>0$.

In order to display the supergravity actions we first have to introduce two sets 
of multiplets containing scalars in two-dimensions: (1) a set of chiral multiplets
with complex scalars $\phi^\kappa$, 
(2) a set of twisted-chiral multiplets with complex scalar $\sigma^\Sigma$.
In a superspace description these multiplets obey the two
inequivalent linear spinor derivative constraints leading to 
irreducible representations.   

To discuss the actions we first focus on the 
case $h^{2,1}(Y_4)=0$ and follow the constructions of \cite{Gates:2000fj}. 
For simplicity we will not include gaugings or a scalar potential. 
The superspace action used in \cite{Gates:2000fj} is 
given by 
\beq \label{suspace1}
  S^{(2)}_{\textrm  {dil}}  = \int d^2 x d^4 \theta E^{-1} e^{-2V - \cK}\, . 
\eeq
Here $E^{-1}$ is the superspace measure, $V$ is a real superfield with $V| = \varphi$ as lowest component,
and $\cK$ is a function of the chiral and twisted-chiral multiplets with lowest components $\phi^\kappa$ 
and $\sigma^\Sigma$, respectively. To display the bosonic part of the action \eqref{suspace1} 
we first set 
\beq
    e^{-2 \tilde \varphi} = e^{-2 \varphi - \cK}\, ,
\eeq
where $\cK (\phi^\kappa,\bar \phi^\kappa,\sigma^\Sigma,\bar \sigma^\Sigma)$ is evaluated as a function of 
the bosonic scalars. With this definition at hand one finds the bosonic action
\begin{align} \label{bosonic_dilatonsugra}
 S^{(2)}_{\textrm  {dil}} &= \int e^{-2\tilde{\varphi}}  \left( \frac{1}{2}R *1 + 2d \tilde{\varphi} \wedge * d\tilde{\varphi} 
 -  \cK_{\phi^\kappa \bar \phi^\lambda} \, d\phi^\kappa \wedge * d \bar {\phi}^\lambda \right. \\
 &\quad  +  \cK_{ \sigma^\Sigma \bar  \sigma^\Lambda }\, d \sigma^\Sigma \wedge * d \bar {\sigma}^\Lambda
   - \left. \cK_{\phi^\kappa \bar  \sigma^\Lambda} \, d \phi^\kappa \wedge d \bar \sigma^\Lambda
	  - \cK_{\sigma^\Sigma \bar \phi^\lambda} \, d \bar {\phi}^\lambda \wedge d\sigma^\Sigma \right) , \nonumber 
\end{align}
where $\cK_{\phi^\kappa \bar \phi^\lambda } = \partial_{\phi^\kappa} \partial_{\bar \phi^\lambda} \cK $, 
$\cK_{ \phi^\kappa \bar \sigma^\Sigma } = \partial_{\phi^\kappa} \partial_{\bar \sigma^\Sigma} \cK $ 
with a similar notation for the other coefficients.
It is now straightforward to compare \eqref{bosonic_dilatonsugra} with the action 
\eqref{2Daction} for the case $h^{2,1}(Y_4)=0$, i.e.~in the absence of any complex scalars $N_\cA$. 
One first identifies 
\beq
  \tilde \varphi = \phi_{\textrm  {IIA}} \, ,\qquad \phi^\kappa = z^\cK\, , \qquad \sigma^\Sigma = t^\Sigma\, ,
\eeq
and then infers that 
\beq \label{hatKsimple}
 \cK = - \log \int_{Y_4} \Omega \wedge \bar \Omega + \log \cV \, .
\eeq
Note that we find here a positive sign in front of the logarithm of $\cV$. This is related 
to the fact that there is an extra minus sign in the kinetic terms of the twisted-chiral 
fields $\sigma^\Sigma$ in \eqref{bosonic_dilatonsugra}. Clearly, the kinetic terms of the 
complex structure deformations $z^\cK$ and complexified 
K\"ahler structure deformations $t^\Sigma$ in the action \eqref{2Daction} have both positive definite kinetic 
terms.\footnote{Our discussion differs here from the one in \cite{Gates:2000fj}, where the 
sign in front of $\log \cV$ was claimed to be negative.}

Let us now include the complex scalars $N_\cA$. It is important to note that 
the action \eqref{2Daction} cannot be brought into the form \eqref{bosonic_dilatonsugra}.
In fact, we see in \eqref{2Daction} that the terms independent of the two-dimensional 
metric do not contain an $\phi_{\textrm{IIA}}$-dependent pre-factor, while the terms of this type 
in \eqref{bosonic_dilatonsugra} do admit an $\tilde \varphi$-dependence. Any field
redefinition in \eqref{2Daction} involving the dilaton seems to introduce new 
undesired mixed terms that cannot be matched with \eqref{bosonic_dilatonsugra} either.
However, we note that the action \eqref{2Daction} actually can be 
brought to the form
\begin{align} \label{bosonic_dilatonsugra_extended}
 S^{(2)} &= \int e^{-2\tilde{\varphi}} \left( \frac{1}{2}R *1 + 2d \tilde{\varphi} \wedge * d\tilde{\varphi} 
 -  \tilde K_{\phi^\kappa \bar \phi^\lambda} \, d\phi^\kappa \wedge * d \bar {\phi}^\lambda \right. \\
	&\quad  +  \tilde  K_{ \sigma^\Sigma \bar \sigma^\Lambda}\, d \sigma^\Sigma \wedge * d \bar {\sigma}^\Lambda 
    - \left. \tilde K_{ \phi^\kappa \bar  \sigma^\Lambda} \, d \phi^\kappa \wedge d \bar \sigma^\Lambda
	  - \tilde K_{ \sigma^\Sigma \bar \phi^\lambda} \, d \bar {\phi}^\kappa \wedge d\sigma^\Sigma \right) , \nonumber 
\end{align}
where $\tilde K$ is now allowed to be dependent on $\tilde \varphi$ and given by 
\beq \label{tildeK}
    \tilde K = \cK + e^{2 \tilde \varphi} \cS\, , 
\eeq
Similar to $\cK$, the new function $\cS$ is allowed to depend on the chiral and twisted-chiral scalars $\phi^\kappa, \sigma^\Sigma$,
but is taken to be independent of $\tilde \varphi$. The action \eqref{bosonic_dilatonsugra_extended}
trivially reduces to \eqref{bosonic_dilatonsugra} for $\cS=0$. Note that the new terms 
induced by $\cS$ do not scale with $e^{-2\tilde \varphi}$.
Comparison with \eqref{2Daction} reveals that one can identify
\beq \label{id_fields}
  \tilde \varphi = \phi_{\textrm  {IIA}} \, ,\qquad \phi^\kappa = (z^\cK,N_\cA)\, , \qquad \sigma^\Sigma = t^\Sigma\, ,
\eeq
and introduce the generating functions 
\begin{align} \label{idhatKS}
   &\cK = - \log \int_{Y_4} \Omega \wedge \bar \Omega + \log \cV \, , \\
   & \cS =     H^{\cA \cB} \, \R \, N_\cA \, \R \, N_\cB \,  ,\qquad  H^{\cA \cB} \equiv v^\Sigma {d_\Sigma}^{\cA \cB} \, . \nn
\end{align}
To show this, it is useful to note that ${d_\Sigma}^{\cA \cB}$ can be evaluated 
as in \eqref{evaluate_d} and depends on the complex structure moduli through 
the holomorphic function $f_{\cA \cB}(z)$ only. 

Let us close this subsection with two remarks. First, note 
that \eqref{bosonic_dilatonsugra_extended} is
expected to be compatible with $\cN=(2,2)$ supersymmetry and gives an extension 
of the two-dimensional dilaton supergravity action \eqref{suspace1}. 
A suggestive form of the extended superspace action is  
\beq \label{suspace2}
  S^{(2)} = \int d^2 x d^4 \theta E^{-1} \big(e^{-2V - \cK} + \cS\big) \, ,
\eeq
where $\cS$ is now evaluated as a function of the chiral and twisted-chiral superfields.
It would be interesting to check that \eqref{suspace2} indeed correctly reproduces the 
bosonic action \eqref{bosonic_dilatonsugra_extended} with $\tilde K$ as in \eqref{tildeK}. 

Second, the action \eqref{bosonic_dilatonsugra_extended} with the identification \eqref{id_fields}
can also be straightforwardly obtained by dimensionally reducing M-theory, or rather 
eleven-dimensional supergravity, first on $Y_4$ and then on an extra circle of radius $r$. 
The reduction of M-theory on $Y_4$ was carried out in \cite{Haack:1999zv,Haack:2001jz}. We give the resulting 
three-dimensional action in \eqref{3daction} and briefly recall 
this reduction in \autoref{Mreduction_details} when considering applications to F-theory.
Using the standard relation of eleven-dimensional supergravity on a circle and Type IIA supergravity,
one straightforwardly identifies 
\beq \label{id_vA}
   r = e^{-2\phi_{\textrm  {IIA}}} \, , \qquad     e^{2 \phi_{\textrm {IIA}}} v^\Sigma = \frac{v^\Sigma_{\textrm  M}}{\mathcal{V}_{\textrm  M}} \equiv L^\Sigma \, , 
\eeq
where $v^\Sigma_{\textrm  M}$ and $\cV_{\textrm  M}$ are the analogs of $v^\Sigma$ and $\cV$
used in the M-theory reduction. Note that the scalars $L^\Sigma$ are the appropriate fields to appear in three-dimensional 
vector multiplets. Inserting the identification \eqref{id_vA} 
into \eqref{tildeK} together with \eqref{id_fields}, \eqref{idhatKS} one finds
\begin{align}  \label{def-tildeKM}
   \tilde K^{\textrm  M} =  &- \log \int_{Y_4} \Omega \wedge \bar \Omega + \log \Big(\frac{1}{4!} \cK_{\Sigma \Lambda \Gamma \Delta}L^\Sigma L^\Lambda L^\Gamma L^\Delta \Big) \nonumber \\
   &+    L^\Sigma {d_\Sigma}^{\cA \cB} \,  \R \, N_\cA \,  \R \, N_\cB \, ,
\end{align}
where we have dropped the logarithm containing the circle radius. Indeed $\tilde K^{\textrm  M}$ agrees 
precisely with the result found in \cite{Haack:1999zv,Haack:2001jz,Grimm:2010ks} from the M-theory reduction. 
The general discussion of the circle reduction of a three-dimensional un-gauged $\cN=2$ supergravity theory 
to a two-dimensional $\cN=(2,2)$ supergravity theory can be found in \autoref{3d-2dreduction}.

\section{Legendre transforms from chiral and twisted-chiral scalars} \label{sec_Legendre}

In this subsection we want to introduce an operation that allows to 
translate the dynamics of certain chiral multiplets to twisted-chiral multiplets and 
vice versa. More precisely, we will assume that some of the scalars, say the scalars 
$\lambda_\cA$, in the $\cN=(2,2)$ supergravity action have continuous shift 
symmetries, i.e.~$\lambda_\cA \rightarrow \lambda_\cA + c_\cA $ for constant $c_\cA$. 
These scalars therefore only appear with derivatives $d\lambda^\cA $ in the action. 
By the standard duality of massless $p$-forms to $(D-p-2)$-forms in $D$ dimensions, 
one can then replace the scalars $\lambda_\cA$ by 
dual scalars $ \lambda'^{\, \cA}$. Accordingly, one has to adjust 
the complex structure on the scalar field space by performing a 
Legendre transform. 
In the following we will give representative  examples
of how this works in detail. We will see that this duality, in particular 
as described in the first example, becomes crucial 
in the discussion of mirror symmetry of \autoref{mirror_section}.

As a first example, let us consider the above theory with complex scalars 
$z^\cK,N_\cA$ in chiral multiplets and complex scalars $t^\Sigma$ in twisted-chiral 
multiplets. The kinetic potential for these fields $\tilde K$
was given in \eqref{tildeK} with \eqref{idhatKS}.
Two facts about this example are crucial for the following discussion. 
First, the fields $N_\cA$ admit a shift symmetry $N_\cA \rightarrow N_\cA + i c_\cA $ in the 
action, i.e.~the kinetic potential $\tilde K$ given in \eqref{idhatKS} is independent of $N_\cA - \bar N_\cA$.
Second, the $N_\cA$ only appear in the term $\cS$ of 
the kinetic potential and thus carry no dilaton pre-factor in the 
action. One can thus straightforwardly 
dualize $N_\cA - \bar N_\cA$ into real scalars $\lambda'^{\, \cA}$. 
The new complex scalars $N'^{\, \cA}$ are then given by  
\beq \label{N'}
   N'^{\, \cA} =  \frac{1}{2}\frac{\partial \cS}{\partial \R \, N_\cA}+ i \lambda'^{\, \cA}\ ,
   \eeq
where we have included a factor of $1/2$ for later convenience. 
Furthermore, the new kinetic potential $\tilde K'$ is now a function 
of $z^\cK,\ t^\Sigma,\ N'^{\, \cA}$ and given by the Legendre transform
\beq \label{tildeK'}
   \tilde K'  = \tilde K - 2\, e^{2\tilde \varphi} \R \, N'^{\, \cA}  \R \, N_\cA \ ,
\eeq
where $ \R \, N_\cA $ has to be evaluated as a function of $\R \, N'^{\,\cA}$
and the other complex fields by solving \eqref{N'} for $\R \, N_\cA$.
One now checks that the scalars $N'^{\, \cA}$ actually reside in 
twisted-chiral multiplets. Using the  transformation \eqref{N'} and 
\eqref{tildeK'} in the action \eqref{bosonic_dilatonsugra_extended} simply yields a dual description in which 
certain chiral multiplets are consistently replaced by twisted-chiral multiplets.  
It is simple to  evaluate \eqref{N'}, \eqref{tildeK'} for $\cS$ given in
\eqref{idhatKS} to find
\bea
  &N'^{\, \cA} &=  H^{\cA \cB}\, \R\, N_\cB  + i \lambda'^{\, \cA}\ , \label{N'ex} \\
  &\tilde K'  &= \cK -   e^{2 \phi_{\textrm{IIA}}} H_{\cA \cB} \, \R \, N'^{\, \cA} \, \R \, N'^{\, \cB}\ ,\label{tildeK'ex}
\eea
where $H^{\cA \cB}$ is the inverse of the matrix $H_{\cA \cB}$ introduced in \eqref{def-H}, \eqref{splitH}.
It is interesting to realize that upon inserting \eqref{N'ex} into \eqref{tildeK'ex} 
one finds that $\tilde K'$ evaluated as a function 
of $N_\cA $ only differs by a minus sign in front of the term linear in $e^{2 \phi_{\textrm{IIA}}}$
from the original $\tilde K$. This simple transformation arises from the 
fact that $\tilde K$ is only quadratic in the $N_\cA$. This observation 
will be crucial again in the discussion of mirror symmetry in \autoref{mirror_section}.

As a second example, we briefly want to discuss a dualization that 
transforms all multiplets containing scalars to become chiral. 
The detailed computation for a general $\cN=(2,2)$ setting can be 
found in \autoref{detailed_dual}. 
For the example of \autoref{22dilatonSugra} we focus on the twisted-chiral 
multiplets with complex scalars 
$t^\Sigma$. These admit a shift symmetry $t^\Sigma \rightarrow t^\Sigma + c^\Sigma$ for 
constant $c^\Sigma$, such that $\R\, t^\Sigma$ only appears with derivatives in 
the action. Accordingly, the kinetic potential $\tilde K$
is independent of $t^\Sigma + \bar t^\Sigma$ as seen in  \eqref{tildeK} with \eqref{idhatKS}. 
Due to the shift symmetry we can dualize the scalars $t^\Sigma + \bar t^\Sigma$
to scalars $\rho_\Sigma$. However, note that by using the kinetic potential  \eqref{tildeK}, \eqref{idhatKS}
there are couplings of $t^\Sigma$ 
in \eqref{bosonic_dilatonsugra_extended} that have a dilaton 
factor $e^{\tilde \varphi}$, and others that are independent 
of $e^{\tilde \varphi}$. 
This seemingly prevents us from performing a straightforward Legendre transform 
to bring the resulting action to the form \eqref{bosonic_dilatonsugra_extended}
with only chiral multiplets. Remarkably, the special properties of the 
kinetic potential \eqref{tildeK}, \eqref{idhatKS}, however, allow us to nevertheless achieve this goal
as we will see in the following. 

The action \eqref{bosonic_dilatonsugra_extended} for a setting with only chiral multiplets with 
complex scalars $M^I$ takes the form 
\beq \label{S2Kaehler}
   S^{(2)} = \int e^{-2\tilde{\varphi}} \left( \frac{1}{2}R *1 + 2d \tilde{\varphi} \wedge * d\tilde{\varphi} 
 -  \mathbf{K}_{M^I \bar M^J} \, dM^I\wedge * d \bar M^J \right)\ ,
\eeq
where $  \mathbf{K}_{M^I \bar M^J} = \partial_{M^I} \partial_{\bar M^J} \mathbf{K}$.
In other words, the potential $\mathbf{K}$ is in this case actually a K\"ahler potential 
on the field space spanned by the complex coordinates $M^I$. 
For our example  \eqref{tildeK}, \eqref{idhatKS} the scalars $M^I$ 
consist of $z^\cK$, $N_\cA$, and $T_\Sigma$, where $T_\Sigma$ are the duals of the 
complex fields $t^\Sigma$.
We make the following ansatz for the dual coordinates 
$T_\Sigma$
\beq \label{TA_Ansatz1}
   T_\Sigma =  e^{-2\tilde \varphi} \frac{\partial \tilde K}{\partial \I\, t^\Sigma} + i \rho_\Sigma =
    e^{-2\tilde \varphi} \frac{\partial \cK }{\partial \I\, t^\Sigma}+ \frac{\partial \cS}{\partial \I\, t^\Sigma} + i \rho_\Sigma \ , 
\eeq
  and the dual potential $\mathbf{K}$ 
\beq \label{TA_Ansatz2}
    \mathbf{K} = \tilde K - e^{2\tilde \varphi} \R \, T_\Sigma \, \I \, t^\Sigma\ .  
\eeq
These expressions describe the standard Legendre transform for $\I \,t^\Sigma$, but 
crucially contain dilaton factors $e^{2\tilde \varphi}$. This latter fact 
allows to factor out $e^{-2\tilde \varphi}$ as required in \eqref{S2Kaehler},
but requires to perform a two-dimensional Weyl rescaling as we will discuss 
below. Using  \eqref{tildeK} with \eqref{idhatKS} one 
straightforwardly evaluates
\bea \label{TAbfK}
   &T_\Sigma &= e^{-2\phi_{\textrm  {IIA}}} \frac{1}{3!} \frac{\cK_{\Sigma}}{\cV} + {d_\Sigma}^{\cA \cB}\, \R \, N_\cA \, \R \, N_\cB  +  i \rho_\Sigma \, , \\
   & \mathbf{K} &= - \log \int_{Y_4} \Omega \wedge \bar \Omega + \log \cV \, .
\eea
Clearly, upon using the map \eqref{id_vA} 
this result is familiar from the study of M-theory compactifications 
on Calabi-Yau fourfolds \cite{Haack:1999zv,Haack:2001jz,Grimm:2010ks}. Also note that the contribution 
$\cS$ present in the kinetic potential \eqref{tildeK} is removed by the Legendre transform
in $\mathbf{K} $  and reappears in a more involved definition of the coordinates $T_\Sigma$. 

At first it appears that \eqref{TA_Ansatz1} induces new mixed terms involving 
one $d\tilde \varphi$ due to the dilaton dependence in front of the derivatives of $\cK$. 
Interestingly, these can be removed by a two-dimensional Weyl rescaling 
if $\cK$ satisfies the conditions 
\beq \label{hatKcond}
  \cK_{t^\Sigma} \, \cK^{t^\Sigma \bar t^\Lambda}\, \cK_{\bar t^\Lambda} = k \, ,
  \qquad  \cK_{v^\Sigma} \ d \, \I \, t^\Sigma = d f \, , 
\eeq
for some constant $k$ and some real field dependent function $f$. 
In this expression $\cK^{t^\Sigma \bar t^\Lambda}$ is the inverse of $\cK_{t^\Sigma \bar t^\Lambda}$ 
and $\cK_{v^\Sigma} \equiv \partial_{\I\, t^\Sigma} \cK$.
In fact, one can perform the rescaling $\tilde g_{\mu \nu} = e^{2 \omega} g_{\mu \nu}$,
which transforms the Einstein-Hilbert action as 
\beq \label{Weyl-EH}
   \int  e^{-2 \tilde \varphi} \, \frac{1}{2} \, \tilde R \, \tilde \ast \, 1  = 
    \int e^{-2 \tilde \varphi} \left( \frac{1}{2} R \, \ast \, 1 - 2 d\omega \wedge \ast d \tilde \varphi  \right) \, ,
\eeq
while leaving all other terms invariant. 
Using \eqref{Weyl-EH} to absorb the mixed terms one needs to chose
\beq
 \omega = - \frac{k}{2} \tilde \varphi - \frac{f}{2} \, .
\eeq
The details of this computation can be found in \autoref{detailed_dual}. 
Indeed, for the example \eqref{idhatKS} one finds $f= \text{log}\ \cV$
and $k=-4$. Remarkably, the condition \eqref{hatKcond}
essentially states that $\cK$ has to satisfy a no-scale like condition. 
A recent discussion and further references on the subject of 
studying four-dimensional supergravities satisfying 
such conditions can be found in \cite{Ciupke:2015ora}.

\chapter{Mirror symmetry at large volume and large complex structure} \label{mirror_section}

In \autoref{IIAreduction} we have determined the two-dimensional action obtained from 
Type IIA supergravity compactified on a Calabi-Yau fourfold. We commented 
on its $\cN=(2,2)$ supersymmetry structure which relies on the proper identification 
of chiral and twisted-chiral multiplets in two dimensions. In this section 
we are exploring the action of mirror symmetry. More precisely, we consider 
pairs of geometries $Y_4$ and $\hat Y_4$ that are mirror manifolds \cite{Greene:1993vm,Mayr:1996sh,Klemm:1996ts}.
From a string theory world-sheet perspective one expects the two theories 
obtained from string theory on $Y_4$ and $\hat Y_4$ to be dual. This implies 
that after finding the appropriate identification of coordinates the two-dimensional effective 
theories should be identical when considered at dual points in moduli space. 
We will make this more precise for the large volume and large complex structure 
point in this section. Note that in contrast to mirror symmetry for Calabi-Yau threefolds the mirror 
theories encountered here are both arising in Type IIA string theory.\footnote{This can be seen 
immediately when employing the SYZ-understanding of mirror symmetry as T-duality \cite{Strominger:1996it}. 
Mirror symmetry is thereby understood as T-duality along half of the compactified 
dimensions, i.e.~$Y_4$ is argued to contain real four-dimensional tori along which T-duality can be 
performed. Clearly, this inverts an even number of dimensions for Calabi-Yau fourfolds.}

\section{Mirror symmetry for complex and K\"ahler structure deformations}

Mirror symmetry arises from the observation that the conformal field theories associated with
$Y_4$ and $\hat Y_4$ are equivalent. 
It describes the identification of  
Calabi-Yau fourfolds $Y_4$, $\hat Y_4$
with Hodge numbers
\beq
     h^{p,q}(Y_4) = h^{4-p,q}(\hat {Y}_4)\, .
\eeq
Note that this particularly includes the non-trivial conditions 
\bea 
   &h^{1,1}(Y_4) = h^{3,1} (\hat Y_4) \, , \qquad h^{3,1}(Y_4) = h^{1,1} (\hat Y_4) \, ,&  \label{h11=h31} \\
    &h^{2,1}(Y_4) = h^{2,1}(\hat Y_4) \, ,  \label{h21=h21} &
\eea
The first identification \eqref{h11=h31} together with the observations made in 
\autoref{IIAreduction} implies that mirror symmetry exchanges K\"ahler structure deformations 
of $Y_4$ ($\hat Y_4$) with complex structure deformations of $\hat Y_4$ ($Y_4$). 
 Accordingly one needs to 
exchange chiral multiplets and twisted-chiral multiplets in the effective $\cN=(2,2)$ supergravity 
theory. The second identification \eqref{h21=h21} seems to suggest that for the fields $N_\cA$
the mirror map is trivial. However, as we will see in \autoref{3form_mirror} this is not the case 
and one has to equally change from a chiral to a twisted-chiral description.  

To present a more in-depth discussion of mirror symmetry we first 
need to introduce some notation. All fields and couplings obtained  
by compactification on $Y_4$ are denoted as in \autoref{IIAreduction}. 
To destinguish them from the quantities obtained in the  $\hat Y_4$ 
reduction we will dress the latter with a hat. In particular for the fields we write
\bea
   Y_4\, :& \quad & \phi_{\textrm{IIA}}\, ,\ t^\Sigma\, ,\ z^\cK \,  ,\ N_\cA \, , \\
   \hat Y_4 \,:& \quad & \hat \phi_{\textrm{IIA}}\, ,\ \hat t^\cK\, ,\ \hat z^\Sigma \,  ,\ \hat N_\cA \, . \nn
\eea 
Note that we have exchanged the indices on $\hat t^K$ and $\hat z^\Sigma$
in accordance with the fact that complex structure and K\"ahler structure 
deformations are interchanged by mirror symmetry.  In other words,
$K= 1, \ldots, h^{1,1}(\hat Y_4)$ and $A = 1, \ldots, h^{3,1}(\hat Y_4)$
is compatible with the previous notation due to \eqref{h11=h31}.
Similarly we will adjust the notation for the couplings. For example, 
the functions introduced in \eqref{idhatKS} and \eqref{fThreeFormDef}, \eqref{def-H} are 
\bea
   Y_4 \, :& \quad &  f_{\cA \cB}(z) \, ,\ H^{\cA \cB}(v,z) \, , \\
   \hat Y_4 \,:& \quad &\hat f_{\cA \cB}(\hat z) \, , \ \hat H^{\cA \cB}(\hat v,\hat z)\,  .
\eea
The functional form of the various couplings will in general differ for 
 $Y_4$ and $\hat Y_4$. A match of the two mirror-symmetric 
effective theories should, however, be possible when identifying 
the mirror map, which we denote formally by $\cM[\cdot]$.

We want to focus on the sector of the theory independent of the three-forms. 
Recall that in the two-dimensional effective theory obtained from 
$Y_4$ the kinetic terms of the complex structure moduli $z^K$ and 
K\"ahler structure moduli $t^\Sigma$ are obtained from the kinetic potential \eqref{hatKsimple}, \eqref{idhatKS} as
\begin{align} \label{hatKY4}
 \cK(Y_4) =  \log \Big( \frac1{4!} \cK_{\Sigma \Lambda \Gamma \Delta}\, \I \,t^\Sigma \, \I\, t^\Lambda \, \I \, t^\Gamma\, \I \, t^\Delta \Big)- \log \int_{Y_4} \Omega \wedge \bar \Omega \, ,
\end{align}
when used in the action \eqref{bosonic_dilatonsugra}. 
Mirror symmetry exchanges the K\"ahler moduli $t^\Sigma$ of $Y_4$ with the 
complex structure moduli $\hat z^\Sigma$ of $\hat Y_4$. The expression
\eqref{hatKY4} was computed at the large volume 
point in K\"ahler moduli space, i.e.~with the assumption that $\I \, t^\Sigma \gg 1$ in string units. 
Accordingly one has to evaluate  $\cK(\hat Y_4)$ at the large complex structure 
point as 
\beq
   \int_{\hat Y_4} \hat \Omega \wedge  \hat { \bar \Omega} = 
   \frac1{4!} \cK_{\Sigma \Lambda \Gamma \Delta}\, \I \,\hat z^\Sigma \, \I \,\hat z^\Lambda \, \I \,\hat z^\Gamma\, \I \,\hat z^\Delta \, ,
\eeq
where now $\I \, \hat z^\Sigma \gg 1$.
Similarly, one has to proceed for the K\"ahler moduli part of the kinetic potential $\cK(\hat Y_4)$
and evaluate $\cK(Y_4)$ at the large complex structure point 
\beq
   \int_{Y_4}   \Omega \wedge \bar{   \Omega} = \frac1{4!} \hat \cK_{\cK \cL \cM \cN}\, \I\,   z^\cK \, \I  \, z^\cL \, \I  \, z^\cM\, \I  \, z^\cN \, ,
\eeq
where $\hat \cK_{KLMN}$ are now the quadruple intersection numbers on the geometry $\hat Y_4$. 
Therefore, at the large volume and large complex structure point the two effective theories obtained 
from $Y_4$ and $\hat Y_4$ are identified under the mirror map
\beq \label{mirror_tz}
 \mathcal{M} \big[t^\Sigma \big]= \hat{z}^\Sigma\ , \quad \mathcal{M}\big[z^\cK\big]= \hat{t}^\cK\ ,
\eeq
and  
\beq \label{mirror_phiIIA}
 \mathcal{M} \big[ \cK(Y_4) \big] = - \cK(\hat Y_4)\ , \qquad \cM \big[ \phi_{\textrm{IIA}}\big] = \hat \phi_{\textrm{IIA}} \ .
\eeq
It is important to stress that a sign change occurs when applying the mirror map to $\cK$. 
This can be traced back to the fact that scalars in chiral and twisted-chiral multiplets have different 
sign kinetic terms in the actions \eqref{bosonic_dilatonsugra}, \eqref{bosonic_dilatonsugra_extended}. 
The quantum corrections to $\cK$ were discussed using mirror symmetry in \cite{Greene:1993vm,Mayr:1996sh,Klemm:1996ts,Grimm:2009ef} 
and localization in \cite{Honma:2013hma,Halverson:2013qca} (using and extending
the results of \cite{Benini:2012ui,Doroud:2012xw,Jockers:2012dk}). 

\section{Mirror symmetry for non-trivial three-forms} \label{3form_mirror}

Let us next include the moduli $N_\cA$ arising for Calabi-Yau fourfolds $Y_4$ with non-vanishing 
$h^{2,1}(Y_4)$. In \autoref{IIAreduction} we have seen that these complex scalars  
are part of chiral multiplets. Their dynamics was described by 
the real function $\cS$ in the kinetic potential $\tilde K$ given in \eqref{tildeK} and \eqref{idhatKS}. 
For completeness we recall that
\beq \label{SY4}
  \cS(Y_4) =   H^{\cA \cB}\, \R\, N_\cA \, \R\,N_\cB \, ,\qquad  H^{\cA \cB} \equiv v^\Sigma {d_\Sigma}^{\cA \cB}\, , 
\eeq
where $ {d_\Sigma}^{\cA \cB} $ is a function of the complex structure moduli of $Y_4$. 
Mirror symmetry should map the fields $N_\cA$ to scalars $\hat N_\cA$ arising 
in the reduction on the mirror Calabi-Yau fourfold $\hat Y_4$, i.e.~one 
should have 
\beq \label{cMN}
   \cM \big[ N_\cA \big] = Q_\cA( \hat N, \hat z, \hat t)\ ,
\eeq
where we have allowed the image of $N_\cA$ to be a non-trivial 
function that will be determined in the following. 
In fact, note that the map cannot be as simple as $\cM(N_\cA) = \hat N_\cA$.
As already pointed out in  \cite{Gates:2000fj} 
the mirror duals $\cM(N_\cA)$ need to be, in contrast to the $N_\cA$, parts 
of \textit{twisted}-chiral multiplets. To 
achieve this we need to  use the 
results of \autoref{sec_Legendre}.

Let us therefore consider the reduction on $\hat Y_4$ using the same notation as 
in \autoref{IIAreduction} but with hatted symbols. The two-dimensional theory 
will contain a set of complex scalars $\hat N_\cA$ that reside in chiral multiplets. 
We can transform them to scalars in twisted-chiral multiplets using \eqref{N'ex} 
and \eqref{tildeK'ex}. In other words, we find a dual description with 
scalars $\hat N'^\cA$ defined as 
\beq \label{hatN'}
   \hat N'^{\, \cA} =  \hat H^{\cA \cB}\, \R\, \hat N_\cB  + i \hat \lambda'^{\, \cA}\, , 
\eeq
where $ \hat H^{\cA \cB}$ is a function of the mirror complex structure moduli $\hat z^\Sigma$
and K\"ahler moduli $\hat v^\cK$. The dual kinetic potential takes the form
\beq \label{tildeK'hatY4}
  \tilde K'(\hat Y_4) =  \cK(\hat Y_4) -  e^{2 \hat \phi_{\textrm{IIA}}} \hat H_{\cA \cB}\ \R\, \hat N'^{\, \cA} \ \R\, \hat N'^{\, \cB} \, . 
\eeq
The mirror map \eqref{mirror_tz}, \eqref{mirror_phiIIA} and  \eqref{cMN} 
exchanges chiral and twisted-chiral states and therefore has to take the form 
\bea
 \cM \big[ N_\cA \big] &= \hat N'^{\, \cA}(\hat N,\hat z,\hat t) \, , \quad \mathcal{M} \big[ t^\Sigma \big] &= \hat{z}^\Sigma \, , 
  \quad \mathcal{M}\big[ z^\cK \big] = \hat{t}^\cK \, ,  \label{full_mirror1} \\
  \mathcal{M} \big[ \tilde K(Y_4)\big] &= - \tilde {K}'(\hat Y_4) \, ,  \  \ \quad \cM\big[ \phi_{\textrm{IIA}}\big] &= \hat \phi_{\textrm{IIA}} \, . 
  \label{full_mirror2}
\eea
and is evaluated as a function of $\hat N_\cA$, $\hat z^\Sigma$ and $\hat t^\cK$ by using \eqref{hatN'}.

Using these insights we are now able to infer the mirror image of the function $H_{\cA \cB}$ appearing 
in $\tilde K(Y_4)$. 
To do that, we apply the mirror map to the kinetic potential $\tilde K$. Note 
that 
\beq
   \cM \big[ \tilde K(Y_4) \big] = - \cK(\hat Y_4) + e^{2 \hat \phi_{\textrm{IIA}}} \cM \big[ \cS(Y_4) \big]\ , 
\eeq
where we have used \eqref{mirror_phiIIA}. Furthermore, we insert \eqref{full_mirror2} into \eqref{SY4}
to find 
\beq
 \cM \big[ \cS(Y_4) \big] = \sum_{\cA , \cB} \cM \big[ H^{\cA \cB} \big] \R\, \hat N'^\cA  \, \R\, \hat N'^\cB \,  .
\eeq
We next apply \eqref{full_mirror2} together with \eqref{tildeK'hatY4} which requires 
\beq
     \sum_{\cA , \cB} \cM \big[ H^{\cA \cB}\big] \R\, \hat N'^\cA  \, \R\, \hat N'^\cB \overset{!}{=} \hat H_{\cA \cB}\ \R\, \hat N'^{\, \cA} \ \R\, \hat N'^{\, \cB}\, ,
\eeq 
and thus enforces
\beq \label{cMH}
    \cM \big[ H^{\cA \cB} \big]  \overset{!}{=} \hat H_{\cA \cB}\ .
\eeq
We therefore find that the mirror map actually identifies $H^{\cA \cB}$ with the \textit{inverse} $\hat H_{\cA \cB}$ of $\hat H^{\cA \cB}$. 
This inversion is crucial and stems from the exchange of chiral an twisted-chiral multiplets under mirror symmetry. 
In the final part of this section we evaluate the condition \eqref{cMH} at the large complex structure 
point, since $H^{\cA \cB}$ given in \eqref{SY4} was computed  at large volume.

Using the mirror map we are now able to determine the holomorphic function $f_{\cA \cB}$ appearing 
in the definition of $H_{\cA \cB}$ at the large complex structure point.
Note that \eqref{splitH} translates on $Y_4$ and $\hat Y_4$ to 
\bea \label{recall_HhatH}
   &H^{\cA \cB} &=  - \frac{1}{2}  \R \, f ^{\cA \cC} \, \R\, {h_\cC}^\cB \, , \qquad   {h_\cC}^\cB = - i t^\Sigma {M_{\Sigma \cC}}^\cB \, , \qquad \\
   &\hat H^{\cA \cB} &=  - \frac{1}{2}  \R \, \hat f ^{\cA \cC} \, \R\, {\hat h_\cC} {}^\cB \, ,\qquad   \hat h_\cC {}^\cB =-i \hat t^\cK \hat M_{\cK \cC}{}^\cB \, , \nn
\eea
where on the mirror geometry we introduced the intersection numbers
\begin{equation}
  \hat M_{\cK \cA}{}^\cB = \int_{\hat Y_4} \hat{\omega}_{\cK} \wedge \hat{\alpha}_\cA \wedge \hat{\beta}^\cB \, .
\end{equation}
Using \eqref{full_mirror1}, \eqref{full_mirror2}, \eqref{cMH}, and \eqref{recall_HhatH} in the
mirror map one infers that a possible identification is \footnote{Note that 
in general the basis $(\alpha_\cA,\beta^\cA)$ might not directly map to $(\hat \alpha_\cA,\hat \beta^\cA)$
on the mirror geometry $\hat Y_4$. In this expression we have assumed that there is no non-trivial 
base change under mirror symmetry. }
\beq
   \R \, f_{\cA \cB}= \I \, z^\cK {\hat M_{\cK \cA}}{}^\cB\ .
\eeq
By holomorphicity of $f_{\cA \cB}$ we finally conclude
\beq   \label{lin_fresults}
f_{\cA \cB}= -i z^\cK {\hat M_{\cK \cA}}{}^\cB \, .
\eeq
Having determined the function $f_{\cA \cB}$ at the large complex 
structure point we have established a complete match of the 
two two-dimensional effective theories obtained from $Y_4$ and
$\hat Y_4$ under the mirror map $\cM[\cdot]$. The result \eqref{lin_fresults}
is not unexpected. In fact, from the variation of Hodge-structures one 
could have expected a leading linear dependence on $z^K$. Furthermore, 
we will find agreement with a dual Calabi-Yau threefold result when using 
the geometry $Y_4$ as F-theory background and performing the orientifold limit. 
This will be the task of the next section.

\chapter{Applications for F-theory and Type IIB orientifolds} \label{F-theoryapp}

In this section we want to apply the result obtained by using 
mirror symmetry to compactifications of F-theory and their orientifold limit.  
The F-theory effective action is studied via the M-theory to 
F-theory limit. Therefore, we will briefly review in \autoref{Mreduction_details} 
the dimensional reduction of M-theory on a smooth Calabi-Yau fourfold including three-form moduli. 
This  reduction was already performed in \cite{Haack:2001jz}, but we will use the insights we have gained in the previous sections to include the three-form moduli more conveniently. 
In \autoref{MFlimit} we will then restrict to a certain class 
of elliptically fibered Calabi-Yau fourfolds and perform the M-theory to F-theory limit. 
This allows us to identify the characteristic data determining the
 four-dimensional $ \cN = 1 $ F-theory effective action in terms of the geometric 
 quantities of the internal space \cite{Grimm:2010ks}. We note that for certain fourfolds the holomorphic 
 function $f_{\cA \cB}$ lifts to a four-dimensional gauge coupling function. 
 Starting from these F-theroy settings we will then perform the weak string coupling limit in \autoref{orientifold_limit}.
 In this limit $f_{\cA \cB}$ can be partially computed by using mirror symmetry for Calabi-Yau threefolds
 and we show compatibility with the fourfold result of \autoref{mirror_section}.

\section{M-theory on Calabi-Yau fourfolds} \label{Mreduction_details}

In this subsection we review the dimensional reduction of M-theory on a Calabi-Yau fourfold $ Y_4 $ in the large volume limit without fluxes. The ansatz here is similar to the one used for Type IIA supergravity in \autoref{IIAreduction_details}.

We start with eleven-dimensional supergravity as the low-energy limit of M-theory. 
Its bosonic two-derivative action is given by
\begin{equation} \label{11Daction}
S^{(11)} = \int \frac{1}{2} \check R \, \check\ast \, 1 - \frac{1}{4} \check F_4 \wedge\check \ast \, \check F_4 - \frac{1}{12}\check C_3 \wedge \check F_4 \wedge \check F_4\ ,
\end{equation}
with $\check F_4 = d \check C_3 $ the eleven-dimensional three-form field strength. This will be dimensionally reduced on the background
\beq
   d\check s ^2 = \eta^{(3)}_{\mu \nu} dx^\mu dx^\nu + g^{(8)}_{mn} dy^m dy^n\ , 
\eeq
where $ \eta^{(3)} $ is the metric of three-dimensional Minkowski space-time $ \mathbb{M}_{2,1} $ and $ g^{(8)} $ the metric of the Calabi-Yau fourfold $ Y_4 $. This is the analog to \eqref{metric-ansatz}
and, as we briefly discussed at the end of \autoref{22dilatonSugra}, the Type IIA supergravity vacuum can be obtained by 
a circle-reduction of this Ansatz. 

To perform the dimensional reduction one inserts similar expansions of \eqref{Kaehlermoduli}, 
\eqref{CSmoduli} and \eqref{BC-expand} into the eleven-dimensional action \eqref{11Daction}. For the metric deformations consisting of K\"ahler and complex structure deformations, this is exactly the same as \eqref{Kaehlermoduli} and
\eqref{CSmoduli}, hence we obtain $ h^{1,1}(Y_4) $ real scalars $ v^\Sigma_{\textrm  M}$ by expanding the M-theory K\"ahler form $J_{\textrm  M}$ as
\beq \label{JM}
   J_{\textrm  M} = v^\Sigma _{\textrm  M}\omega_\Sigma
\eeq
and $ h^{3,1}(Y_4) $ complex scalars $ z^\cK $ in three dimensions. Since the eleven-dimensional three-form $\check C_3 $ is the common origin of the Type IIA fields 
$\check {B}_2,\ \check{C}_3$, we expand
\beq \label{11Dthree-form}
\check C_3 = V^\Sigma \wedge \omega_\Sigma + N_\cA \Psi^\cA + \bar{N}_\cA \bar{\Psi}^\cA \, .
\eeq
This yields  $ h^{2,1}(Y_4) $ three-dimensional complex scalars $ N_\cA $ and $ h^{1,1}(Y_4) $ vectors $ V^\Sigma $.
The latter combine with the real scalars $ v^\Sigma_{\textrm  M} $ into three-dimensional vector multiplets, whereas $ z^\cK, N_\cA $ give 
rise to three-dimensional  chiral multiplets. Combining the expansions \eqref{Kaehlermoduli}, 
\eqref{CSmoduli} and \eqref{11Dthree-form} with the action \eqref{11Daction} by using the notation of \autoref{IIAreduction_details} and \autoref{22dilatonSugra} we thus obtain the three-dimensional effective action 
\footnote{The action has been Weyl-rescaled to the three-dimensional Einstein 
frame by introducing ${g}^{\textrm  new}_{\mu \nu} = \mathcal{V}^{-2} g^{\textrm  old}_{\mu \nu} $}
\begin{align} \label{3daction}
S^{(3)} &= \int \frac{1}{2} R \ast 1 - G_{\cK \bar \cL} dz^\cK \wedge \ast d \bar z^\cL 
					- \frac{1}{2} d \log \mathcal{V}_{\textrm  M} \wedge \ast d \log \mathcal{V}_{\textrm  M}   \\
					 &\quad -  G_{\Sigma \Lambda}^{\textrm  M} dv^\Sigma_{\textrm  M} \wedge \ast dv^\Lambda_{\textrm  M} 
					 -  \mathcal{V}_{\textrm  M}^2  G^{\textrm  M}_{\Sigma \Lambda} dV^\Sigma \wedge \ast dV^\Lambda \nonumber \\
					 &\quad - \frac{1}{2}v^\Sigma_{\textrm  M}\, d_\Sigma{}^{\cA \cB} DN_\cA \wedge \ast D \bar N_\cB + \frac{i}{4}  d_\Sigma{} ^{\cA \cB} dV^\Sigma \wedge \big(N_\cA D \bar N_\cB  - \bar N_\cB DN_\cA \big) \, . \nonumber 
\end{align}
Note the $G_{\Sigma \Lambda}^{\textrm  M}$ takes the same functional form as \eqref{GABmetric}, but uses the 
M-theory K\"ahler structure deformations $v^\Sigma_{\textrm  M}$.

The three-dimensional action given in \eqref{3daction} is an $ \cN = 2 $ supergravity theory.
The proper scalars in the vector multiplets are denoted by $L^\Sigma$ and 
are expressed in terms of the $v^\Sigma_{\textrm  M}$ as $L^\Sigma = \frac{v^\Sigma_{\textrm  M}}{\cV_{\textrm  M}}$, 
as  already given in \eqref{id_vA}. The complex scalars in the chiral multiplets 
are collectively denoted by $\phi^\kappa = (z^\cK, N_\cA)$.
The action \eqref{3daction} can then be written using a kinetic potential $\tilde K^{\textrm  M}$ as
\begin{align} \label{chirallinear3D}
S^{(3)} = \int & \frac{1}{2} R^{(3)} \ast 1   + \frac{1}{4} \tilde{K}_{L^\Sigma L^\Lambda}^{\textrm  M} dL^\Sigma \wedge \ast dL^\Lambda 
     + \frac{1}{4} \tilde{K}^{\textrm  M}_{L^\Sigma L^\Lambda} dV^\Sigma \wedge \ast dV^\Lambda \\
					 &  
					- \tilde{K}^{\textrm  M}_{\phi^\kappa \bar \phi^\lambda}\, d\phi^\kappa \wedge \ast d\bar  \phi^\lambda +  dV^\Sigma \wedge \I (\tilde K^{\textrm  M}_{L^\Sigma \phi^\kappa} d \phi^\kappa )\, , \nonumber 
\end{align}   
where $\tilde{K}^{\textrm  M}_{L^\Sigma L^\Lambda}  = \partial_{L^\Sigma} \partial_{L^\Lambda} \tilde K$, 
$\tilde{K}^{\textrm  M}_{\phi^\kappa \bar \phi^\lambda} = \partial_{\phi^\kappa} \partial_{\bar \phi^\lambda} \tilde K^{\textrm  M}$, 
and $\tilde K^{\textrm  M}_{L^\Sigma \phi^\kappa} =  \partial_{L^\Sigma} \partial_{\phi^\kappa} \tilde K^{\textrm  M} $.
Comparing \eqref{3daction} with \eqref{chirallinear3D} the 
kinetic potential obtained for this M-theory reduction therefore reads 
\begin{align} \label{def-tildeKM_Recall}
     \tilde K^{\textrm  M} =  &- \log \int_{Y_4} \Omega \wedge \bar \Omega + \log \Big(\frac{1}{4!} \cK_{\Sigma \Lambda \Gamma \Delta}L^\Sigma L^\Lambda L^\Gamma L^\Delta \Big) \\
     				        &+    \, L^\Sigma d_\Sigma{}^{\cA \cB}  \ \R \, N_\cA  \, \R \, N_\cB \, , \nonumber 
\end{align}
and was already given in \eqref{def-tildeKM}. 
Recalling the discussion at the end of \autoref{22dilatonSugra} it is not hard to check 
that \eqref{3daction} reduces to the Type IIA result found 
in \autoref{IIAreduction_details} upon a circle compactification. 
The detailed circle reduction is performed for a general three-dimensional 
un-gauged $\cN=2$ theory in \autoref{3d-2dreduction}.

\section{M-theory to F-theory lift} \label{MFlimit}

Let us now lift the result \eqref{chirallinear3D} of the M-theory reduction on a general smooth Calabi-Yau fourfold $ Y_4 $ to a four-dimensional effective F-theory compactification. To do so, we need to restrict $ Y_4 $ to be an elliptic fibration $ \pi: Y_4 \rightarrow B_3 $ over a base manifold $ B_3 $ which is a three-dimensional complex K\"ahler manifold. This four-dimensional theory exhibits $ \cN=1 $ supersymmetry. 
In the following we will not need to consider the full four-dimensional theory, but 
will rather  focus on the kinetic terms of the complex scalars and vectors without 
including gaugings or a scalar potential. 
Supersymmetry ensures that these can be written in the form \cite{Wess:1992cp}
\begin{align} \label{S4gen}
S^{(4)} &=  \int \frac{1}{2} R \ast 1 
     - K^{\textrm  F}_{M^I \bar M^J} \, d M^I \wedge \ast \, d\bar{M}^J \\
     &\quad - \int \frac{1}{2} \R \, \mathbf{f}_{\Lambda \Sigma}
      F^\Lambda \wedge \ast \, F^\Sigma + \frac{1}{2} \I \,\mathbf{f}_{\Lambda \Sigma} F^\Lambda \wedge F^\Sigma \, .  \nonumber
\end{align}
In this expression we denoted by $M^I$ the bosonic degrees of freedom in chiral multiplets,
and by $F^\Lambda$ the field strengths of vectors $A^\Lambda$. 
The metric $K^{\textrm  F}_{M^I \bar M^J} $ is K\"ahler and thus can be obtained from 
a K\"ahler potential $K^{\textrm  F}$ via $K^{\textrm  F}_{M^I \bar M^J} =  \partial_{M^I} \partial_{\bar M^J} K^{\textrm  F}$.
The gauge-kinetic coupling function $\mathbf{f}_{\Lambda \Sigma} $ is holomorphic 
in the complex scalars $M^I$. 

In order to determine the K\"ahler potential $K^{\textrm  F}$ and the gauge coupling 
function  $\mathbf{f}_{\Lambda \Sigma} $  via M-theory one next would have 
to compactify \eqref{S4gen} on a circle. The resulting 
three-dimensional theory then has to be pushed to the Coulomb 
branch and all massive modes, including the excited 
Kaluza-Klein modes of all four-dimensional fields, have to 
be integrated out. The resulting three-dimensional effective theory 
can then, after a number of dualizations, be compared with 
the M-theory effective action \eqref{3daction}.  
Performing all these steps is in general complicated. 
However, a relevant special case
has been considered in \cite{Grimm:2010ks} and 
will be the focus in the following discussion.\footnote{The geometries of the other two cases will be considered later in \autoref{ExampleSection}. They require more involved geometries as we will see.}
Despite the fact that we could refer 
to \cite{Grimm:2010ks} we will try to keep the derivation of 
$K^{\textrm  F}$ and $\mathbf{f}_{\Lambda \Sigma} $ 
in this subsection self-contained.

Let us therefore assume that $Y_4$ is an elliptically fibered 
Calabi-Yau fourfold that satisfies the conditions
\beq  \label{special_geom}
 h^{2,1}(Y_4)= h^{2,1}(B_3)\ , \qquad \quad h^{1,1}(Y_4) = h^{1,1}(B_3) + 1\ .
\eeq
It is not hard to use toric geometry to construct examples
that satisfy these conditions (see, for example, refs.~\cite{Grimm:2009yu}). 
From the point of view of F-theory, or Type IIB string theory, the first 
condition in \eqref{special_geom} implies 
that all scalars $N_\cA$ in \eqref{3daction} lift to R-R vectors $A^\cA $ in 
four dimensions. In other words, one can compactify 
Type IIB on the base $B_3$ and obtain vectors $A^l$
by expanding the R-R four-form as 
\beq \label{C4expand}
     C_4 = A^\cA \wedge \alpha_\cA - \tilde A_\cA \wedge \beta^\cA + \ldots\ .
\eeq 
The vectors $\tilde A_\cA$ are the magnetic duals of the $A^\cA $
and can be eliminated by using the self-duality of the field-strength 
of $C_4$. 

The second condition in \eqref{special_geom}
implies that there are no further vectors in the four-dimensional 
theory, i.e.~there are no massless vector degrees of 
freedom arising from seven-branes. 
The two-forms used in \eqref{JM} and \eqref{11Dthree-form} split simply as
\beq
   \omega_\Sigma = (\omega_0, \omega_\sigma)\ ,
\eeq
where $ \omega_0$ is the Poincar\'e-dual of the base divisor $B_3$
and $\omega_\sigma$ is the Poincar\'e-dual of 
the vertical divisors $ D^\sigma = \pi^{-1} (D^\sigma_{\textrm  b}) $ stemming from 
divisors $D^{\sigma}_{\textrm  b}$ of  $B_3$. 
Accordingly one 
splits the three-dimensional vector multiplets in \eqref{chirallinear3D} 
as
\beq
    L^\Sigma = (R, L^\sigma )\ , \qquad V^\Sigma = (A^0, A^\sigma) \ . 
\eeq
One can now evaluate the kinetic potential \eqref{def-tildeKM_Recall}
for the special case \eqref{special_geom}. 
The only relevant non-vanishing quadruple 
intersection numbers are given by 
\beq \label{base-triple}
   \cK_{0\sigma \lambda \gamma} = \int_{Y_4} \omega_0 \wedge \omega_\sigma \wedge \omega_\lambda \wedge \omega_\gamma \equiv \cK_{\sigma \lambda \gamma}\ , 
\eeq
which are simply the triple intersections $\cK_{\sigma \lambda \gamma}$ of the base $B_3$. Crucially, 
for an elliptic fibration one has $\cK_{\sigma \lambda \gamma \delta} = 0$. 
Furthermore, note that due to \eqref{special_geom} all non-trivial 
three-forms come from the base $B_3$ and we can chose the 
basis $(\alpha_\cA , \beta^\cA)$ such that 
\beq \label{special_C}
   M_{0\cA}{}^\cB = \int_{Y_4} \omega_0 \wedge \alpha_\cA \wedge \beta^\cB = \delta_\cA^\cB \, , \qquad M_{\sigma \cA}{}^{\cB} = 0\, , 
\eeq
with $M_{\Sigma \cA}{}^\cB$ introduced in \eqref{Mdef}.
Inserting \eqref{base-triple} and \eqref{special_C} into \eqref{def-tildeKM_Recall} one finds
\begin{align} \label{tildeKM_Special}
     \tilde K^{\textrm  M} =  &- \log \int_{Y_4} \Omega \wedge \bar \Omega 
         + \log \Big(\frac{1}{3!} \cK_{\sigma \beta \gamma} L^\sigma L^\beta L^\gamma \Big) + \log(R)  \\
         & - \frac{1}{2} R\, \R \, f^{\cA \cB}  \, \R\, N_\cA  \, \R \, N_\cB \, ,  \nonumber
\end{align}
where we have used that $L^\Sigma d_\Sigma{}^{\cA \cB} =-\frac{1}{2} L^\Sigma M_{\Sigma \cA}^\cB \R \, f^{\cA \cB} =-\frac{1}{2} R \, \R \, f^{\cA \cB}$,
and we have dropped terms in the logarithm that are higher order in $R$.

In order to compare this kinetic potential with the result of the 
circle reduction of \eqref{S4gen} we next have to dualize $(L^\sigma, A^\sigma)$
into three-dimensional complex scalars $T_{\sigma}$, and $N_\cA$ into three-dimensional 
vectors $(\xi^\cA, A^\cA)$. Due to our assumption \eqref{special_geom}
leading to \eqref{special_C} we can perform these dualizations independently.
The change from $(L^\sigma,A^\sigma)$ to $\R \, T_\sigma = \partial_{L^\sigma} \tilde K^{\textrm  M}$
is similar to \eqref{TAbfK}. It is  conveniently 
parameterized by the base K\"ahler deformations $v^\sigma_{\textrm  b}$
and the base volume $\cV_{\textrm  b}$ defined as \cite{Grimm:2004uq,Grimm:2010ks}
\beq \label{def-vb}
     L^\sigma = \frac{v^\sigma_{\textrm  b}}{\cV_{\textrm  b}} \ , \qquad 
     \cV_{\textrm  b} = \frac{1}{3!} \cK_{\sigma \beta \gamma} v^{\sigma}_{\textrm  b} v^{\beta}_{\textrm  b} v^{\gamma}_{\textrm  b}\, .
\eeq 
The dualization of the complex scalars $N_k$ into three-dimensional vectors 
is similar to the dualization yielding \eqref{N'}, \eqref{tildeK'} and \eqref{N'ex}, \eqref{tildeK'ex}. First, one introduces 
\beq
\xi^\cA = \partial_{\R \, N_\cA} \tilde K^{\textrm  M} \ , \qquad  \tilde K^{\textrm  M \rightarrow F}  = \tilde K^{\textrm  M} -   \xi^\cA \R \, N_\cA \, ,
\eeq 
and then dualizes the 
field $\I \, N_\cA$ with a shift symmetry  into the vector $A^\cA$.   
Together both Legendre transforms yield
\beq \label{tildeKFM}
   \tilde K^{\textrm  M \rightarrow \textrm F} =  - \log \int_{Y_4} \Omega \wedge \bar \Omega 
         - 2 \log \, \cV_{\textrm  b}\, + \log \, R   + \frac{1}{2 R}\, \R \, f_{\cA  \cB}  \, \xi^\cA \xi^\cB \, ,
\eeq
which has to be evaluated as a function of $z^\cK$, $\xi^\cA $ and 
\beq   \label{Talphabase}
 T_{\sigma} 
      = \partial_{L^\sigma} \tilde K^{\textrm  M}  + i \rho_\sigma
         =  \frac{1}{2!} \cK_{\sigma \beta \gamma}  v^\beta_{\textrm  b} v^\gamma_{\textrm  b}  + i \rho_\sigma\ .
\eeq
The kinetic potential \eqref{tildeKFM} is now in the correct frame to be lifted to 
four space-time dimensions. 

To derive  $K^{\textrm  F}$, $\mathbf{f}_{\cA \cB} $ 
one reduces \eqref{S4gen} on a circle of radius $r$ 
with the usual Kaluza-Klein ansatz 
the four-dimensional metric and vectors as
\beq
g^{(4)}_{\mu \nu} =
\begin{pmatrix}
 g^{(3)} _{pq} + r^2 A^0 _p A^0 _q & r^2 A^0 _q \\
 r^2 A^0 _p & r^2
\end{pmatrix},
\qquad A^\cA _\mu = (A^\cA_p + A^0 _p \zeta^\cA, \zeta^\cA)\ ,
\eeq
where we introduced the three-dimensional indices $ p,q = 0,1,2 $ and the Kaluza-Klein vector $ A^0 $. Note that we use for three-dimensional vectors the same symbol $ A^\cA $ as in four dimensions. 
Furthermore, we introduced the new three-dimensional real scalars $ r, \zeta^\cA $ into the theory. 
We next define 
\beq
R = r^{-2}\ , \quad \xi^{\hat \cA} = ( R, R \zeta^\cA)\ , \quad A^{\hat \cA} = (A^0, A^\cA) \ .
\eeq
The three-dimensional theory obtained by reducing \eqref{S4gen} has thus 
the field content: chiral multiplets with complex scalars $M^I$ and vector multiplets 
$(\xi^{\hat \cA},A^{\hat \cA} )$.
Its action can be written in the form \eqref{chirallinear3D} 
with a kinetic potential 
\beq \label{redK}
 \tilde K(M,\bar{M}, \xi) = K^F (M,\bar{M}) + \log (R) - \frac{1}{R} \R\, \mathbf{f}_{\cA \cB }(M) \xi^\cA \xi^\cB \  ,
\eeq
when replacing $L^\Sigma \rightarrow \xi^{\hat \cA}$, $V^\Sigma \rightarrow A^{\hat  \cA}$, and $\phi^\kappa \rightarrow M^I$.
Finally, comparing \eqref{redK} with \eqref{tildeKFM} implies that one finds $M^I = \{T_\sigma, z^\cK \} $
\bea
K^{\textrm  F} &=& -\log (\int_{Y_4} \Omega \wedge \bar \Omega) - 2 \log{\cV_\Lambda}\ , \label{F-theoryKpot}\\
 \mathbf{f}_{\cA \cB}&=& \frac{1}{2} f_{\cA \cB}\, .
\eea
In the next section, we want to derive the orientifold limit of this result relating the data of F-theory on $ Y_4 $ to Type IIB supergravity with $ O7/O3 $-planes on the closely related Calabi-Yau three-fold $ Y_3 $, a double cover of $ B_3 $. 

\section{Orientifold limit of F-theory and mirror symmetry} \label{orientifold_limit}
  
In this final subsection we investigate the orientifold limit of the F-theory
effective action introduced above. 
More precisely, we assume that the F-theory compactification on the 
elliptically fibered geometry $Y_4$ admits a weak string coupling limit as 
introduced by Sen \cite{Sen:1996vd,Sen:1997gv}. This limit takes one to a 
special region in the complex structure moduli space of $Y_4$
in which the axio-dilaton $\tau = C_0 + ie^{-\phi_{\textrm  IIB}}$, given by the complex 
structure of the two-torus fiber of $Y_4$, is almost everywhere 
constant along the base $B_3$. The locations where $\tau$
is not constant are precisely the orientifold seven-planes (O7-planes).  
In the weak string coupling limit the geometry $Y_4$ can 
be approximated by 
\begin{equation} \label{orientifold_projection}
Y_4 \cong (Y_3 \times T^2)/\tilde {\sigma}
\end{equation}
where we introduced the involution $ \tilde {\sigma} = (\sigma, -1,-1) $ with $ \sigma $ being a holomorphic and isometric 
orientifold involution such that $ Y_3 / \sigma = B_3 $. The two one-cycles of the torus are both odd under the involution, but its volume form is even. 
It was shown in \cite{Sen:1996vd,Sen:1997gv} that the double cover $Y_3$ of $B_3$ is actually a Calabi-Yau threefold. The location of the 
O7-planes in $Y_3$ is simply the fixed-point set of $\sigma$. More refined picture of the weak coupling limit will be developed in \autoref{ExampleSection} 
that holds for the examples considered there.

In the limit \eqref{orientifold_projection} we can check compatibility of the mirror symmetry results of 
\autoref{mirror_section} with the mirror symmetry of the Calabi-Yau threefold $Y_3$.
By using the mirror fourfold $\hat Y_4$ of $Y_4$ we have found that the normalized period matrix $f_{\cA \cB}$ is linear in the large 
complex structure limit of $Y_4$. Here we recall that the weak string coupling expression gives a compatible result. Using 
the mirror $\hat Y_3$ of $Y_3$ one shows that the period matrix $f_{\cA \cB}$ is linear in the large 
complex structure limit of $Y_3$. This can be depicted as follows:
\beq
\begin{array}{ccc}
  \text{F-theory on}\ Y_4  &\quad  \xrightarrow{\quad \text{weak coupling} \quad } & \, \text{Type IIB orientifolds}\ Y_3 / \sigma \qquad \\
  && \\
  && \quad \updownarrow \quad \text{physical mirror duality}\\
  && \\
  && \, \text{Type IIA orientifolds}\ \hat Y_3 / \hat \sigma \qquad
\end{array}
\eeq
Note that mirror symmetry of $Y_3$ and $\hat Y_3$ gives a physical map between Type IIB and Type IIA orientifolds.
The mirror map between $Y_4$ and $\hat Y_4$ has no apparent physical meaning in F-theory. Nevertheless, 
using the geometry $Y_4$ in Type IIA compactifications it can be used to calculate 
$f_{\cA \cB}$ as we explained in \autoref{mirror_section}. 

Let us now introduce the function $f_{\cA \cB}$ for the geometry \eqref{orientifold_projection}. 
In the orientifold setting one splits the cohomologies of $Y_3$ as $H^{p,q} (Y_3)= H^{p,q}_+ (Y_3) \oplus H^{p,q}_- (Y_3)$,
which are the two eigenspaces of $\sigma^*$. We denote their dimensions as $h^{p,q}_\pm (Y_3)$.
As reviewed, for example, in \cite{Denef:2008wq} 
the complex structure moduli $z^\cK$ of $Y_4$ split into three sets of fields at weak string coupling.
First, there is the axio-dilaton $\tau$, which is now a modulus of the effective theory as it is constant over the internal space. 
Second, there are $h^{2,1}_-$ complex structure moduli $z^{\alpha}$  of the quotient $Y_3/\sigma$. 
Third, the remaining number of complex structure deformations of $Y_4$  correspond to 
D7-brane position moduli. The last set are open string degrees of freedom
and are not captured by the geometry of $Y_3$. For simplicity, we will not include 
them in the following discussion. 
With this simplifying assumption  one finds that the pure complex structure part of the F-theory 
K\"ahler potential \eqref{F-theoryKpot} splits as
\beq \label{Omegasplit}
-\log (\int_{Y_4} \Omega \wedge \overline{\Omega} ) = - \log \big[ -i(\tau - \bar{\tau})\big] - \log \Big[ i \int_{Y_3} \Omega_3 \wedge \bar{\Omega}_3 \Big] +\ldots \, ,
\eeq
where $\Omega_3$ is the $(3,0)$-form on $Y_3$ that varies holomorphically 
in the complex structure moduli $z^{\alpha}$.  
The dots indicate that further corrections arise that are suppressed at weak string coupling $-i(\tau - \bar \tau) \gg 1$. 
Taking the weak coupling limit for the K\"ahler potential \eqref{F-theoryKpot} of the 
K\"ahler structure deformations is more straightforward. The deformations are counted by 
$h^{1,1}_+(Y_3)$ and identified with the K\"ahler structure deformations $v_{\textrm  b}^\sigma$ of the 
base $B_3$ introduced in \eqref{def-vb}. The orientifold K\"ahler potential for this 
set of deformations is then simply the second term in \eqref{F-theoryKpot} and 
the K\"ahler coordinates are given by \eqref{Talphabase}.

Turning to the gauge theory sector, we note that the number of R-R vectors $A^\cA $ arising 
from $C_4$ as in \eqref{C4expand} are counted by $h^{2,1}_+(Y_3)$ in the orientifold setting. 
The gauge coupling function for these vectors is determined as function of 
the complex structure moduli $z^{\alpha}$ of $Y_3$ in \cite{Grimm:2004uq}.\footnote{Note that we 
have slightly changed the index conventions with respect to \cite{Grimm:2004uq} in order to 
match the F-theory discussion. } 
It is given by 
\beq \label{f-ori}
  f_{\cA \cB}(z^\alpha) = -i \cF_{\cA \cB}|(z^{\alpha}) \equiv \partial_{z^\cA} \partial_{z^\cB} \cF | (z^{\kappa})\ ,
\eeq
where $\cF$ is the pre-potential determining the moduli-dependence of the 
$\Omega_3$  of the geometry $Y_3$. To evaluate \eqref{f-ori} one first 
splits the complex structure moduli of $Y_3$ into $h^{2,1}_-(Y_3)$ fields 
 $z^{\alpha}$ and $h^{2,1}_+(Y_3)$ fields $z^\cA$. The pre-potential $\cF(z^{\kappa},z^\cA)$
 of $Y_3$ at first depends on both sets of fields. 
 Then one has to take derivatives of $\cF$ with respect to $z^\cA$ and 
 afterwards set these fields to constant background values compatible 
 with the orientifold involution $\sigma$. This freezing of the $z^\cA$ is indicated 
 by the symbol $|$ in \eqref{f-ori}. 
 Using mirror symmetry for Calabi-Yau threefolds it is well-known 
 that the pre-potential at the large complex structure point of $Y_3$ 
is a cubic function of the complex structure moduli $z^{\alpha}$ and $z^\cA$.
Taking derivatives and evaluating the expression on the orientifold moduli 
space one thus finds
\beq
f_{\cA \cB}(z^\alpha) =  -i z^{\alpha} \hat \cK_{\alpha \cA \cB} \ ,
\eeq
where $ \hat \cK_{\alpha \cA \cB } = \int_{\hat{Y}_3} \hat{\omega}_{\alpha} \wedge \hat{\omega}_\cA \wedge \hat{\omega}_\cB $
are the triple intersection numbers of the mirror threefold $\hat Y_3$.
This result agrees with the one for Type IIA orientifolds, which have 
been studied at large volume in \cite{Grimm:2004ua}.
Hence, we find consistency with the F-theory result \eqref{lin_fresults} obtained by using 
mirror symmetry for $Y_4$ at the large complex structure point. To obtain a
complete match of the results the 
intersection matrix $\hat M_{\alpha \cA}{}^\cB$ of $\hat Y_4$ is identified with the triple intersection $\hat \cK_{\alpha \cA \cB } $
of $\hat Y_3$. 

To close this section we stress again that we have only discussed the matching with 
the orientifold limit for special geometries satisfying \eqref{special_geom}. 
Furthermore, we have not included the open string degrees of freedom on the 
orientifold side. Clearly, our result for $f_{\cA \cB}$ obtained in \autoref{mirror_section} 
can be more generally applied. For example, a simple generalization 
is the inclusion of $h^{1,1}_-(Y_3)$ moduli $G^\cA$ into the orientifold setting, 
which arise in the expansion of the complex two-form $C_2-\tau B_2$. 
In F-theory the same degrees of freedom appear
from the expansion \eqref{11Dthree-form} into non-trivial three-forms $\Psi_\cA$ that have two legs 
in the base $B_3$ and one leg in the torus fiber, i.e.~are not present in 
the geometries satisfying \eqref{special_geom}.
In the orientifold setting one finds that the fields $G^\cA$ correct the complex 
coordinates \eqref{Talphabase}. We read off the result from \cite{Grimm:2004uq} to find \footnote{Note that 
compared with \cite{Grimm:2004uq} we have redefined $\rho_\sigma$ to make the terms
in $T_\sigma$ involving the $G^\cA$ real.}
\beq
T_\sigma =\frac{1}{2!} \cK_{\sigma \beta \gamma}  v^\beta_{\textrm  b} v^\gamma_{\textrm  b}  
    +  \frac{1}{2 \, \I \tau}\ \cK_{\sigma \cA \cB}\ \I \, G^\cA \I \, G^\cB  + i \rho_\sigma \, .
 \eeq
Comparing this expression with \eqref{TAbfK} we read off that 
\beq \label{TwoFormScalarsEff}
   N^\cA = i G^\cA\ , \qquad d_{\sigma \cA \cB} = \frac{1}{2} \frac{1}{ \I \tau} \cK_{\sigma \cA \cB}\ , \qquad f_{\cA \cB} (\tau)= i \tau \delta_{\cA \cB}\ ,
\eeq
in order to match the F-theory result as already done in \cite{Grimm:2005fa}. 
Again we find that the result is linear in one of the complex structure moduli, namely the field $\tau$, 
of the Calabi-Yau fourfold $Y_4$ in the orientifold limit \eqref{orientifold_projection}. 

In \autoref{ExampleSection} we will generalize these results further and also include Wilson line moduli in the discussion. 
These are proper open string degrees of freedom that can also be derived from the three-forms of the fourfold geometry.
In order to gain intuition about the general situation we will study explicit examples using toric techniques that we will introduce in the next section.

\chapter{Geometry of toric Calabi-Yau fourfold hypersurfaces} \label{ToricSection}

In this section we will introduce the basic notions of Calabi-Yau fourfold hypersurfaces in toric varieties as already studied in \cite{Mayr:1996sh,Klemm:1996ts}. A variety is the algebraic equivalent of a manifold that allows for singularities and is hence a more flexible geometric object. We start out with some basic definitions in \autoref{BasicToricConstructionSection} introducing concepts like toric varieties, their homogeneous coordinate rings and the Newton polyhedron of a divisor. In \autoref{GysinSection} we explain how to use the Gysin-sequence to deduce the non-trivial cohomology groups of a semi-ample Calabi-Yau fourfold hypersurface from its toric subvarieties. The geometry of these toric subvarieties is then explained in \autoref{divisorSection} and reduced to the study of toric divisors that have certain fibration structures. Introducing the Poincar\'e residue in \autoref{PoincareSection} to represent non-trivial forms as rational functions of the homogeneous coordinates allows us to study the complex structure dependence of these forms in \autoref{HodgeVariationSection}. The fibration structures combined with the Gysin-sequence arguments allow us to compute the Hodge numbers of the fourfold in \autoref{HodgeNumberSection} which determines the massless spectrum of the effective theories derived in previous sections. The field of toric geometry is a broad subject and we give the condensed treatment necessary to construct Calabi-Yau fourfolds. For a more thorough introduction and conventions we refer to \cite{Fulton:1993,0036-0279-33-2-R03} for the basics of toric geometry and for general algebraic geometry to \cite{griffiths2011principles}. Great introductions are also given in \cite{Hosono:1993qy,Greene:1996cy}

\section{Basic construction of toric Calabi-Yau hypersurfaces} \label{BasicToricConstructionSection}

Let us start with the basic construction of hypersurfaces with trivial anti-canonical class in five dimensional toric varieties, i.e~ toric Calabi-Yau fourfold hypersurfaces. We begin with the $ d $-dimensional complex ambient space $ \cA_d $ which will be a toric variety. The special cases we will consider are $ d \leq 5 $ as we will later also discuss the toric subvarieties $ \cA_d $ of $ \cA_5 $.

This toric ambient space is defined by a convex polyhedron $ \Delta^\ast \subset N_{\mathbb{Q}} $ in the rational extension of a lattice $ N \simeq \mathbb{Z}^d $. The integral points of the polyhedron will be denoted $ \nu_i ^\ast \in \Delta^\ast \cap N $ and the rays from the origin through $ \nu^\ast _i $ we denote by $ \tau_i $. A toric variety is constructed from a so called fan $ \Sigma $, as for example explained in detail in \cite{Fulton:1993}. A fan $ \Sigma $ is in our definiton a set of convex rational polyhedral cones $ \sigma $ that are spanned by one-dimensional rays $ \tau = \mathbb{Q}^+ \nu^\ast $ where $ \tau $ is a ray from the origin of the lattice $ N $ through a lattice point $ \nu^\ast \in N \cap \Delta^\ast $ of the polyhedron $ \Delta^\ast $ as
\begin{equation}
 \sigma = \{ r_1 \nu^\ast _1 + \ldots + r_s \nu^\ast _s \,  | \, \nu^\ast \in N \cap \Delta^\ast \, , \, r_i \in \mathbb{Q}^+\} \, .
\end{equation} 
A cone is called simplicial if it is generated by linearly independent vectors. The simplest example is the cone spanned by the unit vectors $ e_i $ in $ N $. The dual lattice $ M $ of $ N $ is defined as
\begin{equation}
 M = \text{Hom} (N,\mathbb{Z})
\end{equation}
with $ M_\mathbb{Q} $ its rational extension. Identifying $ M \simeq \mathbb{Z}^d $ we have $ m(n) = \langle m, n \rangle = \sum_i m_i n_i $ using the standard inner product on $ \mathbb{Z}^d $. The dual cone $ \sigma^\vee $ of $ \sigma $ is
\begin{equation}
 \sigma^\vee = \{ u \in M_\mathbb{Z} \, | \, \langle u, v \rangle \geq 0 \, , \, \forall v \in \sigma \} \, .
\end{equation}
To the integral points in $ S_\sigma = \sigma^\vee \cap M $ we can associate abstract characters $ \chi^\nu $ satisfying $ \chi^{\nu_1} \cdot \chi^{\nu_2} $ that form a commutative algebra $ \mathbb{C}[S_\sigma] $ with unit $ 1 = \chi^0 $ and therefore gives rise to a complex affine variety
\begin{equation}
 \cA_\sigma = Spec(\mathbb{C}[S_\sigma]) \, , \quad S_\sigma = \sigma^\vee \cap M \, .
\end{equation}
The characters $ \chi^\nu $ can be interpreted as monomials of a polynomial ring. The cone $ \sigma $ generated by the unit vectors $ e^\ast _i $ of $ \mathbb{Z}^d $, the dual cone is also generated by the unit vectors $ e^\ast _i $ of the dual and we can associate to them the coordinates $ X_i = \chi^{e_i} $ and hence we have
\begin{equation}
 \cA_\sigma = \mathbb{C}^d = Spec(\mathbb{C}[X_i]) \, .
\end{equation}
Simplicial cones will in general result in affine varieties with orbifold singularities along a toric subvariety. A simple example is given by the cone $ \sigma $ spanned by $ 2 e_1 + e_2, e_2 $ in $ \mathbb{Q}^2 $ leading to
\begin{equation}
 \mathbb{C}[S_\sigma] = \mathbb{C}[X, Y^2 X^{-1}] \simeq \frac{\mathbb{C}[X,Y,Z]}{XY - Z^2} \, , \quad \cA_{\sigma} = \mathbb{C} ^2/\mathbb{Z}_2 \, ,
\end{equation}
with the singularity at $ (0,0) \in \mathbb{C}^2 /\mathbb{Z}_2 $. which is a cone.
These affine toric varieties $ \cA_\sigma $ can be glued together to form a toric variety $ \cA_d $ and the corresponding information is contained in the fan $ \Sigma $. To ensure a well-defined gluing of the cones and hence the affine varieites $ \cA_\sigma $, $\sigma \in \Sigma $ need to intersect each other only in faces that are cones that are also part of $ \Sigma $. If we want to obtain a compact or rather complete toric variety $ \cA_d $, the cones of $ \Sigma $ need to cover the full space $ N_\mathbb{Q} $. We will denote the fan $ \Sigma(\Delta^\ast) $ a fan that is defined by the polyhedron $ \Delta^\ast $ and the lattice $ N $. Another interpretation is to view $ \cA_d $ as a compactification of $ (\mathbb{C}^\ast)^d $ with $ \mathbb{C}^\ast = \mathbb{C} - \{ 0 \} $ an algebraic torus giving the toric variety its name.

For simplicity we will assume in the following that all cones are simplicial and hence each correspond to a simplicial subpolyhedron of a face of $ \Delta^\ast $. In higher dimensions $ d>3 $ this can not be guaranteed, but will simplify our discussion. The polyhedron $ \Delta^\ast $ is triangulated, i.e.~ it is divided into subpolyhedra with integral vertices that generate together with the origin the convex cones for the fan $ \Sigma(\Delta^\ast) $. This is called a star-triangulation of $ \Delta^\ast $, since all cones have their tip at the same point in the lattice, the origin. We will work with a maximal star-triangulation of $ \Delta^\ast $ with all rays $ \tau $ through integral points $ \nu^\ast \in N \cap \Delta^\ast $. Such a triangulation is in general not unique and we will assume henceforth that all cones arise from such a maximal star-triangulation and are simplicial and hence $ \cA_d $ will only have $ \mathbb{Z}_n $-orbifold singularities along subspaces of codimension greater or equal to one.

The Calabi-Yau fourfold hypersurface $ Y_4 $ in $ \cA_5 $ will be described via the language of divisors and line-bundles over a toric variety $ \cA_d $. A Weil-divisor $ D_{Weil} $ is a formal sum of codimension one irreducible subvarieties $ V_i $
\begin{equation}
 D_{Weil} = \sum b_i V_i \, , \quad \text{codim}(V_i) = 1 \, , \quad b_i \in \mathbb{Z} \, .
\end{equation}
A Cartier divisor $ D $ is a set of non-zero rational functions over each affine coordinate patch that can be glued together to form a line-bundle $ \cO(D) $ over $ \cA_d $. If this line-bundle is trivial, i.e. defined by a global non-zero rational function on $ \cA_d $, the divisor is principal. A Cartier-Divisor defines a Weil-Divisor as
\begin{equation}
 D_{Cartier} = \sum_{ \text{codim}(V)} \text{ord}_V(D) \cdot V \, ,
\end{equation}
where $ \text{ord}_V(D) $ is the order of vanishing of the defining function of $ D $ along the subvariety $ D $. Two Weil-divisors are linear equivalent if they differ by a principal divisor and their group of equivalence classes is called Chow-group $ A_{d-1}(\cA_d) $. The T-invariant principal divisors are given by the global rational functions $ \chi^u $, $ u \in M $ which is therefore $ d $-dimensional. A T-invariant Cartier divisor is given by a set $ (\cA_\sigma, \chi^{u(\sigma)}) $ where $ u(\sigma) \in S_\sigma $ and $ \cA_\sigma $ cover $ \cA_d $. As the toric divisors $ D_i $ of codimension one are defined by the rays $ \tau_i = \nu^\ast _i \cdot \mathbb{Q} $ in $ \nu^\ast _i \in \Delta^\ast \cap N $, we have
\begin{equation}
 [div(\chi^{u(\sigma)})] = \sum_i \langle u(\sigma), \nu^\ast _i) \rangle D_i \, ,
\end{equation}
we will discuss the toric subvarieties of $ \cA_d $ later in more detail. On the toric varieties we consider we have the isomorphisms
\begin{equation}
 Pic(\cA_d) \otimes \mathbb{C} \simeq A_{d-1}(\cA_d) \otimes \mathbb{C}  \simeq H^2(\cA_d, \mathbb{C}) \, .
\end{equation}
Therefore, each toric divisor $ D_i $, $ \nu^\ast _i \in \Delta^\ast \cap N $ modulo rational equivalence defines an element of $ H^2(\cA_d, \mathbb{C}) $ and a line bundle $ \cO(D) \in Pic(\cA_d) $. The principal divisors defined by one global rational function are give rise to trivial line-bundles.

The defining rational function of a divisor $ [D] $ can hence be viewed as a global section of a line-bundle $ L_\Delta \in Pic(\cA_d) $ over $ \cA_d $. The global sections $ H^0(\cA_d, L_\Delta) $ of $ L_\Delta $ are defined via a so called Newton-polyhedron $ \Delta \subset M_\mathbb{Q} $. A basis of global sections of $ L_\Delta $ is given abstractly by
\begin{equation}
 \chi^\nu \in H^0(\cA_d, L_\Delta) \, , \quad \nu \in \Delta \cap M \, .
\end{equation}
The relation to a divisor $ D_\Delta = \sum_i b_i D_i $ with $ D_i $ the toric divisors is given by
\begin{equation}
  \Delta = \{ u \in M_\mathbb{Q} \, | \,  \langle u, v \rangle \geq -b_i \, , \, \forall  v \in \Delta^\ast \} \, .
\end{equation}
A specific global section of the line bundle $ L_\Delta $ is hence defined by the linear combination
\begin{equation}
 p_\Delta (a_j) = \sum_{\nu_j \in \Delta \cap M} a_j \chi^\nu \, \in H^0(\cA_d, L_\Delta)  \, , \quad a_j \in \mathbb{C} \, .
\end{equation}
Varying the $ a_j $ preserves hence the class $[D_\Delta] \in A_{d_1}(\cA_d) $ of the divisor, but changes in general its representative $ D $. This enables us to describe a family of hypersurfaces by one divisor-class.

To each of the toric divisors $ D_i $ we can associate a coordinate $ X_i $ such that $ D_i = \{ X_i = 0 \} $. These coordinates form the ring of homogeneous coordinates of $ \cA_d $ as defined in \cite{Cox:1993fz} we denote it by
\begin{equation}
 S_d = \mathbb{C} [X_i \, , \, \nu^\ast _i \in \Delta^\ast \cap N] \, . 
\end{equation}
This ring has a natural grading by divisor classes $ \alpha \in A_{d-1}(\cA_d) $. For a given monomial 
\begin{equation}
 f = \prod_i X_i ^{b_i} \ , \quad \text{deg} (f) = \alpha \, , \quad \alpha = [\sum_i b_i D_i] \, .
\end{equation}
We can interpret the elements of $ S_d $ of degree $ \beta $ as the global holomorphic section vanishing over $ \beta $ as
\begin{equation}
 H^0(\cA_d, \cO_{\cA_d}(D)) = S_\beta \, , \quad \beta \in [D]
\end{equation}
as for example shown in \cite{1993alg.geom..6011B,Cox:1993fz}. The coordinate ring $ S_d $ and the toric variety $ \cA_d $ are related by the proj-construction
\begin{equation}
 \cA_d = \text{Proj} (S_d) = \text{Proj} (\mathbb{C}[X_i \, , \, \nu^\ast _i \in \Delta^\ast \cap N])\, ,
\end{equation}
relating a graded ring to a projective variety. The projective toric varieties we consider are therefore in particular K\"ahler.
This ring is called the homogenous coordinate ring of $ \cA_d $, since we can choose global homogeneous coordinates, denoted by
\begin{equation}
 [X_1 : X_2 : \ldots ] =[\underline X_i] = [\underline{\lambda ^{\ell_i} X_i}] \, , \quad \sum_i \nu^\ast _i \ell_i = 0 \, .
\end{equation}
These are equivalence classes under rescalings by powers of a non-zero complex number $ \lambda \in \mathbb{C}^\ast = \mathbb{C} - \{ 0 \} $ which is an algebraic torus from which the name toric variety is derived. The $ \ell_i \in \mathbb{Z}^{\ell(\Delta^\ast)-1-d}$ induce the grading of $ S_d $ and generate a cone in $ \mathbb{Q}^{\ell(\Delta^\ast)-1-d} $ where $ \ell(\Delta^\ast) -1 $ denotes the number integral points in the boundary of $ \Delta^\ast $.

It can be shown \cite{Fulton:1993,Mavlyutov:2000hw} that $ A_{d-1} (\cA_d) $ the Chow group of $ \hat \cA_d $ is given by
\begin{equation}
  A_{d-1} (\cA_d) = \frac{\mathbb{C}[D_i]}{\langle P[\Sigma(\Delta^\ast)] \rangle}
\end{equation}
with the projective equivalence
\begin{equation}
 P[\Sigma(\Delta^\ast)] = \langle \sum_{\nu^\ast _i \in \Delta^\ast \cap N} \langle m, \nu^\ast _i \rangle D_i \, : \, m \in M \rangle
\end{equation}
which is a $ d $-dimensional sub-module of $ A_{d-1} (\cA_d) $. The equivalence classes are denoted by $ [D_i] $. We note here that it can be shown that the toric divisors generate the full cohomology of a complete simplicial toric variety $ \cA_d $ which reads
\begin{equation} \label{ambient_cohomology}
 H^\ast (\cA_d, \mathbb{C}) = \frac{\mathbb{C}[D_i]}{\langle P[\Sigma(\Delta_d ^\ast)] \oplus \text{SR} \rangle}
\end{equation}
where the $ [D_i] $ have grading $ (1,1) $ viewed as their dual two-forms $ \omega_i \in H^{1,1}(\cA_d)$. In particular, all non-trivial cohomology classes of toric varieties have Hodge-type and can be represented by the intersection of $ p $ toric divisors. The Stanley-Reissner ideal $ \text{SR} $ provides the information for the product in the cohomology corresponding to the intersection of divisor-classes.
\begin{equation}
 \text{SR} =  \{ D_{i^1} \cdots D_{i^k} \, | \, \{ \tau_{i^1}, \ldots \tau_{i^k} \}  \not \subset \sigma \, , \, \forall \sigma \in \Sigma(\Delta ^\ast) \} \, .
\end{equation}
In addition, we note that from description we can calculate $ h^{1,1}(\cA_d) $ as
\begin{equation}
 h^{1,1} (\cA_d) = \ell(\Delta^\ast ) - (d + 1) \, .
\end{equation}
Here $ \ell(\Delta^\ast) $ counts the number of integral points in $ \Delta^\ast \subset N $ and subtracts the dimension of $ P[\Sigma(\Delta ^\ast)] $ which is $ d $ and the $ -1 $ corresponds to the origin, the integral point of $ \Delta^\ast $ to which no divisor is associated.

Let us illustrate this with a simple example, the two dimensional complex projective space $ \mathbb{P}^2 $. The polyhedron is spanned by $ e_1, e_2, -e_1 -e_2 \in N \simeq \mathbb{Q}^2 $ and each two of these give rise to a simplicial cone, isomorphic to $ \mathbb{C}^2 $. Another way to see this is to define
\begin{equation}
 S_2 = \mathbb{C}[X_1,X_2,X_3] \, , \quad \cA_2 = Proj(S_2) = \mathbb{P}^2 \, ,
\end{equation}
where $ X_i $ all have the same degree. All the $ [D_i] $ are linearly equivalent as
\begin{equation}
 P[\Sigma(\Delta^\ast)] = \langle (D_1 - D_3), (D_2 - D_3) \rangle
\end{equation}
and can be represented by the hyperplane class $ [H] = [D_i] \in A_1(\mathbb{P}^2) $. The SR-ideal is simply given by $ H^3 \simeq D_1 D_2 D_3 $ and the full cohomology is therefore
\begin{equation}
 H^\ast (\mathbb{P}^2, \mathbb{C}) = \mathbb{C}[H] / H^3 \, . 
\end{equation}
This is a first example of a weighted projective space that we will consider in more detail in \autoref{WeightedProjectiveSpaceSection}.

The special divisor class that will give rise to a Calabi-Yau hypersurface is the so called anti-canonical divisor class of $ \cA_d $ given by
\begin{equation}
 D_\Delta = - K_{\cA_d} = \sum_{\nu^\ast _i \in \Delta^\ast \cap N} D_i \, .
\end{equation}
The details of this construction were established in the seminal paper of Batyrev \cite{Batyrev:1994hm}. The corresponding anti-canonical hypersurface $ D_\Delta $ and all members of the same class are Calabi-Yau if the associated anti-canonical line-bundle $ L_\Delta $ is reflexive or equivalently, if both polyhedra are convex and contain only one interior point. This interior point can then always be shifted to the origin of $ M $. In the second case, we can describe $ \Delta $ as
\begin{equation}
 \Delta = (\Delta^\ast)^\ast = \{ u \in M_\mathbb{Q} \, \big| \, \langle u, v \rangle \geq -1 \, , \, \forall v \in \Delta^\ast \} \, .
\end{equation}
The resulting hypersurface will in general be singular and we denote it by $ Y_{d-1} ^{\text{sing}} $. This is the vanishing set of a global section of $ -K_{\cA_d} $ given by
\begin{equation}
 p_\Delta (a_j, X_i) = \sum_{\nu \in \Delta \cap M} a_j \prod_{\nu^\ast _i \in \Delta^\ast \cap N} X_i ^{\langle \nu_j, \nu^\ast _i \rangle + 1} \quad \in S_d (-K_{\cA_d}) \, .
\end{equation}
We have chosen implicitly a maximal star-triangulation of $ \Delta^\ast $ by associating to every integral point $ \nu^\ast _i \in \Delta^\ast \cap N $ a homogeneous coordinate $ X_i $. Due to the Bertini-theorem, that basically states that we can always change the $ a_j $ infinitesimally varying the hypersurface in its divisor-class, we can assume that all singularities of $ Y_{d-1} ^{\text{sing}} $ arise from the ambient space $ \cA_d $. The $ n $-dimensional toric subvarieties $ \cA_n $ of $ \cA_d $ along which we have singularities are then intersected by $ Y_{d-1} ^{\text{sing}} $ also in hypersurfaces of codimension one. 

We can resolve these singularities by adding new homogeneous coordinates corresponding to integral points of $ N $ that will also change $ \Delta^\ast $, its triangulation and therefore also the fan of $ \cA_d $. Adding a new ray corresponding to such an additional integral point in $ N $ that is not contained in the boundary of $ \Delta^\ast $ will also affect the number of integral points in $ \Delta $. Hence this will change the number of possible deformations of the hypersurface. If the chosen integral point is in the boundary of $ \Delta^\ast $ the resolution is called crepant, preserving the anti-canonical class $ -K_{\cA_d} $. We will henceforth assume that we can resolve the singularities of $ \cA_d $ that will be inherited by the hypersurface via crepant resolutions. \footnote{For $ d >4 $ this is in general not possible, but it simplifies our discussion.} In addition, we are going to assume that we can always choose a transverse and quasi-smooth hypersurface in the anti-canonical hypersurface class, intersecting all toric subvarieties of the ambient space in smooth varieties of codimension one. The resolved smooth Calabi-Yau hypersurface will be denoted by $ Y_{d-1} $ and the fully resolved ambient space $ \hat \cA_d $.

With the most important basics and notations introduced, we will discuss the precise origin and representations of the non-trivial cohomology classes of the Calabi-Yau hypersurface $ Y_4 $ in $ \cA_5 $ in the upcoming section.

\section{Cohomology of $ Y_4 $ via the Gysin-sequence} \label{GysinSection}

In this subsection, we want to describe the origin of the non-trivial cohomology classes that will give rise to the massless fields in the spectrum of our effective field theories. The key-point here is that we need to circumvent the Lefschetz-hyperplane theorem for quasi-smooth Fano-hypersurfaces to obtain non-trivial three-form cohomology on $ Y_4 $. This is done via quasi-Fano or semiample hypersurfaces as discussed by Mavlyutov in \cite{Mavlyutov:2000dj}.

The standard way to calculate the cohomology groups of the hypersurface $ Y_4 $ is to apply the Lefschetz-hyperplane theorem as for example stated in \cite{1993alg.geom..6011B}. This theorem states that for a quasi-smooth hypersurface $ D $ of a $ (d+1) $-dimensional complete simplicial toric variety $ \cA_{d+1} $ defined by an ample divisor we have the isomorphism
\begin{equation}
 \iota^\ast \, : \, H^j(\cA_{d+1}, \mathbb{C}) \,  \xrightarrow{\quad \simeq \quad } \, H^j(D, \mathbb{C}) \,  , \quad j \leq d - 1
\end{equation}
This is the induced map of the inclusion $ \iota: \, D \hookrightarrow \cA_{d+1} $. Furthermore, we have the inclusion
\begin{equation}
 \iota^\ast \, : \, H^d (\cA_{d+1}, \mathbb{C}) \, \hookrightarrow \, H^d (D, \mathbb{C}) \,  .
\end{equation}
This implies basically that all non-trivial cohomology of degree less than $ d $ is induced from the ambient toric variety $ \cA_{d+1} $. Combing this with \eqref{ambient_cohomology} which implies
\begin{equation}
 H^{i,j} (\cA_{d+1}) \neq 0 \quad \Rightarrow \quad i = j \, ,
\end{equation}
we see that this imposes strong restrictions on the cohomology of the hypersurface. In particular, it follows that a quasi-smooth four-dimensional Calabi-Yau hypersurface in a toric ambient space with ample anti-canonical line-bundle can not have non-trivial three-forms. A variety with ample anti-canonical line-bundle is called a Fano variety. A line-bundle over a variety is ample, iff for every point we can find a global section that doesn't vanish over that point. An example for this is the sextic, the anti-canonical hypersurface in $ \mathbb{P}^5 $ which does not support odd cohomology, as studied in \cite{Braun:2014xka}. Therefore, we want to consider ambient spaces more general than regular projective spaces.

As was shown in \cite{1998math.....12163M} the non-trivial cohomology classes of degree less than $ d $ in a semiample hypersurface $ D_\Delta $ in a complete simplicial toric variety $ \cA_{d+1} $ arise from the toric divisors $ D_i $ of $ \cA_{d+1} $
\begin{equation}
 D_i = \{ X_i = 0 \} \, , \quad D^\prime _i = D_i \cap D_\Delta \, , \quad \nu^\ast _i \in \Delta^\ast \cap N \, .
\end{equation}
In toric geometry, see \cite{Fulton:1993} section 3.4. it can be shown that for an ample Cartier divisor over a complete toric variety with polyhedron $ \Delta^\ast $ we have a one-to-one correspondence between vertices of $ \Delta $ and maximal-dimensional cones in $ \Delta^\ast $. Due to the fact that crepant resolutions of a Calabi-Yau hypersurface subdivide the maximal cones of $ \Delta^\ast $, but leave the dual polyhedron $ \Delta $ invariant, a crepant resolution renders the resulting ambient space non-Fano. The exceptional divisors of the toric resolutions are the $ D_i $ that will induce non-trivial cohomology on the hypersurfaces $ D_\Delta $. These divisors carry themselves non-trivial cohomology that lifts to the full Calabi-Yau hypersurface. Therefore, we call a hypersurface semiample if it is obtained from crepant resolutions of an ample hypersurface in a possibly singular Fano toric ambient space. This can be viewed as starting from a polyhedron $ \Delta^\ast $ with a triangulation that only contains rays through the vertices defining the singular space $ \cA_{d+1} $. Then we add subsequently the rays through all integral points of $ \Delta^\ast $ such that we resolve all singularities on the hypersurface which leads to the new toric ambient space $ \cA_{d+1} $ that is simplicial and complete. In terms of $ n $-dimensional polyhedra $ \Delta^\ast _n $, we hence call a hypersurface $ k $-semiample, $ k \leq n $, if the Newton-polyhedron $ \Delta_k $ of the hypersurface class has dimension $ k $. A semiample hypersurface of $ \cA_{n} $ is therefore $ n $-semiample.

The precise composition of the cohomology groups of a Calabi-Yau fourfold $ Y_4 $ realized as a semiample divisor in the toric simplicial complete ambient space $ \cA_5 $ is given by a number of exact sequences as found in equation $ (7) $ of \cite{1998math.....12163M}. The first one, inducing non-trivial two-forms is given by
\begin{equation}
 \bigoplus_{\nu^\ast _i} H^{0,0}(D_i ^\prime) \,  \xrightarrow{\quad \oplus_i \iota_{i,\ast} \quad } \, H^{1,1}(Y_4) \,  , \quad \iota_i \, : \, D^\prime _i \, \hookrightarrow \, Y_4 \, .
\end{equation}
We note that $ \bigoplus_{\nu^\ast _i} H^{0,0}(D_i ^\prime)   $ basically provides a two-form for every toric divisor class of $ Y_4 $. It is, however, possible to obtain $ h^{0,0} (D_i ^\prime) > 1 $ after intersecting with the hypersurface, but  we always have $ h^{0,0} (D_i) = 1 $ for the purely toric setting. We will see this in more detail in the next section, when we discuss the cohomology of toric divisors of a Calabi-Yau hypersurface. The map $ \iota_{i,\ast} $ is the Gysin-map induced by the inclusion of the toric divisor $ D^\prime _i $ into $ Y_4 $, which we will discuss shortly. It can be viewed as dual to taking the non-trivial cycles of $ D^\prime _i $ as cycles of $ Y_4 $.

The next sequence clarifies the origin of non-trivial three-forms in $ Y_4 $ and reads
\begin{equation}
 \bigoplus_{\nu^\ast _i} H^{1,0}(D_i ^\prime) \,  \xrightarrow{\quad \oplus_i \iota_{i,\ast} \quad } \, H^{2,1}(Y_4) \, .
\end{equation}
The requirement that a toric divisor $ D^\prime $ hosts non-trivial one-forms will be a severe restriction, as we will discuss in the next section.

The next sequence we want to mention will allow us to count the $ (3,1) $-forms of $ Y_4 $, which correspond to the complex structure deformations that preserve the Calabi-Yau structure. This is, however, a true exact sequence that does not collapse to one isomorphism
\begin{equation} \label{Gysin_complex_structure}
 0 \longrightarrow \bigoplus_{\nu^\ast _i} H^{2,0}(D_i ^\prime) \,  \xrightarrow{\, \oplus_i \iota_{i,\ast} \, } \, H^{3,1}(Y_4) \longrightarrow \text{Gr}_4 ^W H^{3,1}(Y_4 \cap \mathbb{T}) \longrightarrow 0 \, .
\end{equation}
We will later see, that $ \text{Gr}_4 ^W H^{3,1}(Y_4 \cap \mathbb{T}) $ can be interpreted as the bulk complex structure deformations corresponding to the deformations of the defining section $ p_\Delta $ of the hypersurface. Therefore, these are called algebraic complex structure deformations. The torus $ \mathbb{T} $ is the open torus $ (\mathbb{C}^\ast)^5 \subset \hat \cA_5 $ of which $ \cA_5 $ is a compactification. The union of toric divisors is the complement of $ \mathbb{T} $ in $ \hat \cA_5 $. The other deformations arise from holomorphic $ (2,0) $-forms on toric divisors and are therefore called non-algebraic divisors, since they leave $ p_\Delta $ invariant.

The Gysin map $ \iota_\ast $ of a inclusion $ \iota : N \rightarrow M $ of an $ n $-dimensional submanifold of a compact manifold $ M $ of dimension $ m $ is best described with the following diagram
\begin{equation}
\usetikzlibrary{matrix}
 \begin{tikzpicture}
  \matrix (m) [matrix of math nodes,row sep=3em,column sep=4em,minimum width=2em]
  {
     H_{n-p}(N,\mathbb{C}) & H_{n-p}(M,\mathbb{C}) \\
     H^{p}(N,\mathbb{C}) & H^{m - n + p}(M,\mathbb{C}) \\};
  \path[-stealth]
    (m-1-1)  edge node [above] {$ \iota $} (m-1-2)
    (m-1-2) edge node [right] {$ \text{PD} $} (m-2-2)
    (m-2-1)  edge node [below] {$ \iota_\ast $} (m-2-2)
    (m-1-1) edge node [left] {$ \text{PD} $} (m-2-1)
            ;
\end{tikzpicture}
\end{equation}
Here $ \text{PD} $ is Poincar\'e duality and from this we see that the Gysin-map is the dual map of the inclusion of cycles of submanifolds. This is a purely topological construction of topological manifolds and hence also applies to our setting of varieties. In general there is no reason why this map should be surjective or injective. In our case the toric divisors have the real dimension $ n = 2d - 2 $ and $ m = 2d $. This implies for example that for $ p = 1 $ we have
\begin{equation} \label{Gysin-square}
\usetikzlibrary{matrix}
 \begin{tikzpicture}
  \matrix (m) [matrix of math nodes,row sep=3em,column sep=4em,minimum width=2em]
  {
     H_{5}(D_i ^\prime,\mathbb{C}) & H_{5}(Y_4,\mathbb{C}) \\
     H^{1}(D_i ^\prime,\mathbb{C}) & H^{3}(Y_4,\mathbb{C}) \\};
  \path[-stealth]
    (m-1-1)  edge node [above] {$ \iota_i $} (m-1-2)
    (m-1-2) edge node [right] {$ \text{PD} $} (m-2-2)
    (m-2-1)  edge node [below] {$ \iota_{i,\ast} $} (m-2-2)
    (m-1-1) edge node [left] {$ \text{PD} $} (m-2-1)
            ;
\end{tikzpicture}
\end{equation}
Here we note that due to the fact that all maps are topologically and independent of the metric or the complex structure on $ Y_4 $. Therefore, the Gysin-map is compatible with the splitting into Hodge-type and we obtain the exact sequences we stated before.

In the two upcoming sections, we will discuss the geometry of toric divisors further and also see, how the algebraic deformations of a hypersurface can be desribed in detail via chiral rings and the Poincar\'e residue.

\section{The geometry of toric divisors of a Calabi-Yau hypersurface} \label{divisorSection}

So far we have established that we can obtain the non-trivial forms of various types on a semiample hypersurface $ D_\Delta $ in a simplicial toric ambient space $ \hat \cA_5 $ from toric divisors. These toric divisors correspond to the rays through integral points $ \nu^\ast $ in the boundary of the polyhedron $ \Delta^\ast $ defining $ \hat \cA_5 $. The corresponding rays can be classified by the codimension $ \text{codim}(\theta^\ast) $ of the face $ \theta^\ast \subset \Delta^\ast $ such that $ \nu^\ast \in \text{int}(\theta^\ast) \cap N $. This was already suggested in \cite{Klemm:1996ts} where this idea was used to determine the Hodge-numbers of Calabi-Yau fourfolds. In our case, since we are interested in the precise moduli dependence of the harmonic forms, we will be more explicit.

First, we note that since the toric divisor $ D_i ^\prime = D_i \cap Y_4 $ of the semiample hypersurface $ Y_4 $ in the simplicial toric space $ \hat \cA_5 $ is again a semi-ample hypersurface in a toric variety $ D_i $, we review first the construction of the $ n $-dimensional toric subvarieities $ \cA_n $ of $ \cA_5 $. The subvariety $ \cA_n $ corresponds to a $ (4-n) $-dimensional face $ \theta^\ast $ of $ \Delta^\ast \subset N_\mathbb{Q} $. We construct $ \cA_n $ starting from the $ (5-n) $-dimensional cone $ \sigma \subset N_\mathbb{Q} $ over the face $ \theta^\ast $ with apex at the origin. This cone enables us to define new $ n $-dimensional lattices $ N_n, M_n $ via
\begin{align}
 N_n &= N(\sigma) = N / N_\sigma \, , \quad N_\sigma = N \cap \mathbb{Q} \cdot \sigma \subset N \\
 M_n &= M(\sigma) = M \cap \sigma^\perp \, , \quad \sigma = \theta^\ast \cdot \mathbb{Q}^+ \, . \nonumber 
\end{align}
Here we denoted by $ \mathbb{Q} \cdot \sigma \subset N_\mathbb{Q} $ the $ (5-n) $-dimensional vector space spanned by the elements of $ \sigma $ over $ \mathbb{Q} $. The elements in $ M_\mathbb{Q} $ that pair to zero with $ \sigma \subset N_\mathbb{Q} $ are denoted by $ \sigma^\perp $.
\begin{equation}
 \sigma^\perp = \{ m \in M \, | \, \langle m , n \rangle = 0 \, , \, \forall n \in \sigma \} \, .
\end{equation}
This is a $ n $-dimensional vector space in $ M_\mathbb{Q} $. The fan of the toric subvariety $ \cA_{n,\theta^\ast} $ is given by the set $ Star(\sigma) $ of all cones over faces of $ \Delta^\ast $ that share a face with $ \theta^\ast $ projected to $ N(\sigma) $. The image of these adjacent faces under the projection form again a star subdivison of a polytope $ \Delta^\ast _n $ in $ N(\sigma) $ and all of $ \sigma \cdot \mathbb{Q} $ gets projected to the origin of the quotient lattice $ N(\sigma) $. For details we refer to \cite{Fulton:1993}.

Consequently, we can associate a homogeneous coordinate ring to $ \cA_{n,\theta^\ast} $ as follows
\begin{align} \label{homogeneousCoordinatesSn}
 S_{n, \theta^\ast} &= \mathbb{C}[X_i \, , \, \nu^\ast _i \in \Delta^\ast _n] \, \subset \, \mathbb{C}[X_i \, , \nu^\ast \in \Delta^\ast] / \langle X_i \, , \nu^\ast _i \in \theta^\ast \rangle \\
 &= S_5 / \langle X_i \, , \nu^\ast _i \in \theta^\ast \rangle \, .  \nonumber
\end{align}
These rings inherit the grading structure of the homogeneous coordinate ring $ S_5 $ of $ \cA_5 $, since the other toric divisors either intersect $ \cA_{n, \theta^\ast} $ transversely or not at all. If the divisor $ D_i $ intersects $ \cA_{n, \theta^\ast} $ the coordinate $ X_i $ is also in $ S_{n, \theta^\ast} $, if it does not intersect we can set $ X_i = 1 $ in $ S_{n,\theta^\ast} $. If $ \nu_i ^\ast \in \theta^\ast $, the coordinate $ X_i$ gets projected out, since it is equivalent to zero in $ S_{n,\theta^\ast} $. 

Let us now discuss, how the hypersurface intersects the toric subvariety $ \cA_{n,\theta^\ast} $. We have seen that the polynomial $ p_\Delta $ defining the hypersurface is built from global sections of the anti-canonical bundle of the ambient space $ -K_{\hat \cA_5} $. Therefore, $ p_\Delta \in S(-K_{\hat \cA_5}) $ and its monomials have degree $ [\sum_i D_i] $ where we sum over all toric divisors $ D_i $ of $ \hat \cA_5 $. Restricting now to $ \cA_{n, \theta^\ast} $ where the divisors $ D_i $ vanish for which $ \nu^\ast _i \in \theta^\ast \cap N $, we obtain from the projection
\begin{equation}
 S_5(-K_{\hat \cA_5}) \, \rightarrow S_{n, \theta^\ast} ( -K_{\hat \cA_5}) \, , \quad p_\Delta \mapsto p_\theta = p_\Delta |_{\cA_{n,\theta^\ast}}\, ,
\end{equation}
the hypersurface equation $ p_\theta = 0 $ on $ \cA_{n, \theta^\ast} $. Homogeneous coordinates in $ p_\Delta $ corresponding to divisors not intersecting $ \cA_{n,\theta^\ast} $ will be set to one. The monomials of $ p_\theta $ correspond to the global sections of $ - K_{\hat \cA_5} $ that do not vanish over $ \cA_{n, \theta^\ast} $ determined by the integral points of the dual face $ \theta $ of $ \theta^\ast $ given by
\begin{equation}
 \theta = \{ v \in \Delta \, | \, \langle v, w \rangle\ = -1 \, , \, \forall w \in \theta^\ast \} \, .
\end{equation}

As was discussed in \cite{Mavlyutov:2000hw}, the toric divisors $ D^\prime _i \in D_i \cap Y_4 $ are so called $ \text{dim}(\theta) $-semiample hypersurfaces of the toric varieties $ D_i $ where $ \nu^\ast _i \in \text{int}( \theta^\ast) \cap N $. In the following we will denote pairs of faces as
\begin{equation}
 (\theta^\ast _\alpha, \theta_\alpha) \, , \quad \text{dim}(\theta^\ast _\alpha) = n \, , \quad \alpha = 1, \ldots , k_n \, .
\end{equation}
To each of such pairs we can associate divisors $ D_{l_\alpha} $ with
\begin{equation}
 D_{l_\alpha} \, : \quad \nu^\ast _{l_\alpha} \in \text{int}(\theta^\ast _\alpha) \cap N \, , \quad l_\alpha = 1, \ldots, \ell^\prime (\theta^\ast _\alpha) \, ,
\end{equation}
where $ \ell^\prime (\theta^\ast _\alpha) $ counts the number of interior points of $ \theta^\ast _\alpha $. Let us first consider the toric divisors $ D_{l_\alpha} $ of $ \hat \cA_5 $. The cones corresponding to this subvariety are the rays
\begin{equation}
 \tau_{l_\alpha} = \nu^\ast \cdot \mathbb{Q}^+ \, , \quad D_{l_\alpha} = V(\tau_{l_\alpha}) = \cA_{4, \nu^\ast _{l_\alpha}}\, .
\end{equation}
Here we introduced the alternative expression $ D_{l_\alpha} = V(\tau_{l_\alpha}) $ to make contact with the literature. The divisors $ D_{l_\alpha} $ admit a fibration structure, which can be seen as follows. For $ \nu^\ast _{l_\alpha} $ an inner point of a $ n $-dimensional face $ \theta^\ast _\alpha $, the ray $ \tau_{l_\alpha} $ is contained in the cone $ \sigma_\alpha $ over $ \theta^\ast _\alpha $. This we project to $ N(\tau_{l_\alpha}) $ to obtain the polyhedron $ \Delta^\ast _4 \subset N(\tau_{l_\alpha})  $ of the toric variety $ \cA_{4, \nu^\ast _{l_\alpha}} = D_{l_\alpha} $. Due to the convexity of $ \Delta^\ast $ and $ \nu^\ast _{l_\alpha} \text{int}(\theta^\ast _\alpha) \cap N $, we find that $ \theta^\ast _\alpha $ maps to a subpolyhedron of $ \Delta^\ast _4 $ containing the origin and defining the subvariety $ D_{l_\alpha} $. This subpolyhedron of dimension $ (4-n) $ has an induced triangulation from $ \Delta^\ast $ and defines a simplicial irreducible (connected) and complete (compact) toric variety $ E_{l_\alpha} $. As  described in the literature, for example in \cite{Fulton:1993}, this implies that $ D_{l_\alpha} $ is a fibration over $ \cA_{\theta^\ast _\alpha, n} = V(\sigma_\alpha) = \cA_\alpha $ with fiber given by $ E_{l_\alpha} $. This can be summarized in the following fibration diagram:
\begin{equation} \label{Elalph_fibrD}
 \begin{tikzpicture}
  \matrix (m) [matrix of math nodes,row sep=3em,column sep=4em,minimum width=2em]
  {
     E_{l_\alpha} & D_{l_\alpha} = V(\tau_{l_\alpha}) \\
     & \cA_\alpha = V(\sigma_\alpha) \\};
  \path[-stealth]
    (m-1-1)  edge node [above] {$i_{l_\alpha}$} (m-1-2)
    (m-1-2) edge node [right] {$\pi_{l_\alpha}$} (m-2-2)
            ;
\end{tikzpicture}
\end{equation}
Here $ \pi_{l_\alpha} $ denotes the projection of $ D_{l_\alpha} $ to the base $ \cA_\alpha $ and $ i_{l_\alpha} $ the inclusion of the fiber.

The semiample hypersurface $ D^\prime _{l_\alpha} = D_{l_\alpha} \cap Y_4 $ inherits this fibration structure, since the defining polynomial $ p_{\theta_\alpha} = p_\alpha $ is obtained from $ p_\Delta $ by setting all homogeneous coordinates corresponding to integral points in $ \theta^\ast _\alpha $ to zero. From this we can deduce that the hypersurface equation is independent of the homogeneous coordinates of $ E_{l_\alpha} $ and therefore, we find a similar fibration structure
\begin{equation} \label{Elalph_fibrDprime}
 \begin{tikzpicture}
  \matrix (m) [matrix of math nodes,row sep=3em,column sep=4em,minimum width=2em]
  {
     E_{l_\alpha} & D^\prime _{l_\alpha} = V^\prime (\tau_{l_\alpha}) \\
     & R_\alpha = V ^\prime (\sigma_\alpha)  \\};
  \path[-stealth]
    (m-1-1)  edge node [above] {$i_{l_\alpha}$} (m-1-2)
    (m-1-2) edge node [right] {$\pi_{l_\alpha}$} (m-2-2)
            ;
\end{tikzpicture}
\end{equation}
Here we denoted by $ R_\alpha = V(\sigma_\alpha) $ the $ \text{dim}(\theta) $-semiample hypersurface $ \cA_\alpha \cap Y^{sing}_4 $ defined by the polynomial $ p_\alpha $. We used here $ Y^{sing} _4 $ because without adding the rays through the integral inner points of $ \theta^\ast _\alpha $ the hypersurface is general and contains singularities along $ R_\alpha $. Only after resolving the singularities via the introduction of the exceptional divisors $ E_{l_\alpha} $ is $ \hat \cA_5 $ and therefore $ Y_4 $ resolved. After the resolution, $ R_\alpha $ is not a subvariety of $ Y_4 $, because the fibration $ D^\prime _{l_\alpha} $ may not have a section.

Having established the fibration structure of the toric divisors $ D^\prime _{l_\alpha} $, we can calculate its cohomology. The fibration structure enables us to use the Leray-Hirsch theorem, \cite{hatcher2002algebraic,Voisin03hodgetheory}, since due to the toric fibration structure of $ D_{l_\alpha} $, one can show that the fiber does not degenerate and is locally trivial. The inclusion $ i_{l_\alpha} : E_{l_\alpha} \rightarrow D^\prime _{l_\alpha} $ satisfies furthermore the necessary condition that the elements $ i^\ast _{l_{\alpha}} (c_j) $ for $ c_j \in H^\ast(D^\prime _{l_\alpha}, \mathbb{C}) $ generate $ H^\ast (E_{l_\alpha}, \mathbb{C}) $. Therefore, we find the induced isomorphism of $ \mathbb{C} $-modules
\begin{equation}
 H^\ast (R_\alpha, \mathbb{C}) \otimes_{\mathbb{C}} H^\ast (E_{l_\alpha}, \mathbb{C}) \xrightarrow{\quad \simeq \quad} H^\ast (D_{l_\alpha} ^\prime, \mathbb{C})
\end{equation}
given by the map
\begin{equation}
 b_i \otimes_{\mathbb{C}} i^\ast _{l_\alpha} (c_j) \, \mapsto \, \pi^\ast _{l_\alpha}(b_i) \wedge c_j \, .
\end{equation}

Note that this is an isomorphism of $ \mathbb{C} $-modules and not of rings. Alle morphisms appearing in this construction are independent of the Hodge structure and therefore the Hodge numbers arise from products of Hodge numbers of the base space $ R_\alpha $ and $ E_{l_\alpha} $. Due to the fact that $ E_{l_\alpha} $ is toric and irreducible, i.e.~ connected, its Hodge numbers are restricted to
\begin{equation} 
h^{p,q} (E_{l_\alpha}) = 0 \, , p \neq q \, , \quad h^{0,0}(E_{l_\alpha}) = h^{n,n}(E_{l_\alpha}) = 1 \, , \quad n > 1 \, ,
\end{equation}
For the regular $ n $-semiample hypersurface $ R_\alpha $ of dimension $ n - 1 $, implying that it is ample, we find that
\begin{equation}
 h^{0,0}(R_\alpha) = 1 \, , \quad h^{n-1,0}(R_\alpha) = \ell^\prime(\theta_\alpha)  \, , \quad n > 1 \, .
\end{equation}
This will be discussed in more detail in the upcoming section. We denoted by $ \ell^\prime(\theta_\alpha) $ the number of interior integral points of $ \theta_\alpha $ in $ N_{\mathbb{Q}} $.

Let us now go through the various cases of face dimensions $ (4-n) = \text{dim}(\theta^\ast) $. For $ n = 0 $, we find that $ \cA_0 $ is a set of points in $  \hat \cA_5 $ and hence the corresponding singularities in $ \cA_5 $ will be avoided by a general hypersurface. In the case $ n = 1 $, we find that $ \cA_\alpha \simeq \mathbb{P}^1 $ and the hypersurface will intersect these one-dimensional subvarieties in a number of points counted with multiplicity. The number of points $ pt_\alpha $ is the degree of $ p_\alpha $ and is equal to $ \ell^\prime (\theta_\alpha) + 1 $. Resolving the singularities along $ R_\alpha $ introduces complex three-dimensional toric varieties $ E_{l_\alpha} $ that are irreducible. We have that
\begin{equation} \label{nonToricDivisors}
 \dim(\theta_\alpha) = 1 \, , \quad H^{0,0}(pt_\alpha) = \ell^\prime (\theta_\alpha)  + 1\, .
\end{equation}
The next case is $ n = 2 $, here we find that the toric divisors $ D_{l_\alpha} $ are fibrations over Riemann surfaces $ R_\alpha $ with fibers two-dimensional toric varieties $ E_{l_\alpha} $. The Hodge numbers in this case satisfy
\begin{equation}
 \dim(\theta_\alpha) = 2 \, , \quad H^{1,0}(R_\alpha) = \ell^\prime (\theta_\alpha) \, .
\end{equation}
In the case $ n = 3 $ we have that $ R_\alpha $ are surfaces $ R_\alpha = S_\alpha $ that are ample hypersurfaces of toric varieties and hence satisfy
\begin{equation}
 \dim(\theta_\alpha) = 3 \, , \quad H^{2,0}(S_\alpha) = \ell^\prime (\theta_\alpha) \, , \quad H^{1,0}(S_\alpha) = 0 \, .
\end{equation}
as we will see in the next subsection. The hypersurfaces $ S_\alpha $ are ample divisors of $ \cA_3 $ and hence for them the Lefschetz-hyperplane theorem holds and $ h^{1,0}(S) = 0 $. Contrary, if $ h^{1,0}(S) \neq 0 $, the hypersurface needs to satisfy $ h^{2,0}(S) = 0 $ and we are in the $ n = 2 $ case. We note in particular, that the divisors giving rise to non-algebraic deformations and three-forms can not intersect. 

In this section we have reduced the complex structure dependence of semi-ample divisors to their ample bases, which are ample hypersurfaces in toric varieties. In the next section, we will give an explicit description of the non-trivial holomorphic forms on these hypersurfaces.

\section{Poincar\'e Residue of toric hypersurfaces} \label{PoincareSection}

Let us now discuss the holomorphic $ (n-1) $-forms on semi-ample hypersurfaces in projective and simplicial toric varieties $ \cA_n $, following \cite{1994alg.geom.10017C} for ample hypersurfaces. This was generalized to semiample hypersurfaces in \cite{1998math.....12163M,Mavlyutov:2000hw}. The insight here is that we express the holomorphic $ (n-1) $-forms $ \Omega^{n-1} _R $ of the hypersurface $ R $ as global rational holomorphic forms $ \Omega^n _{\cA_n}(R) $ of the ambient space $ \cA_n $ with poles along $ R $. This will enable us to express the algebraic deformations of the polynomial $ p_\theta $ to explicit forms on the hypersurface, the periods.

A global rational and holomorphic $ n $-form on $ \cA_n $ with poles of first order along the hypersurface $ R $ that is a restriction of the anti-canonical hypersurface in $ \cA_5 $ is given by
\begin{align}
 H^0 (\cA_n, \Omega^n _{\cA_n}(R)) &= \{ \frac{g \, d \omega_{\cA_n}}{p_\theta} \, \big| \, g \in S_n (-K_{\cA_5 |_{\cA_n}} + K_{\cA_n})\} \\
 							&\simeq S_n (-K_{\cA_5 |_{\cA_n}} + K_{\cA_n}) \, .  \nonumber
\end{align}
We introduced the following notation for this: the Cartier divisors class of the restriction of the anti-canonical divisor of $ \cA_5 $ to $ \cA_n $ is denoted by $ -K_{\cA_5}|_{\cA_n} \in A_{n-1}(\cA_n)  $. In $ \cA_n $ we have the hypersurface $ R \subset \cA_n $ defined as the vanishing locus of $ p_\theta \in S_n(-K_{\cA_5} |_{\cA_n}) $. Similarly, we introduced the anti-canonical divisor $ -K_{\cA_n} $ of the subvariety $ \cA_n $ of $ \cA_5 $. Furthermore, we introduced the holomorphic volume-form $ d \omega_{\cA_n} $ of $ \cA_n $ that has degree $ [-K_{\cA_n}] $ as we will show shortly. Finally, we denoted by $ S_n (-K_{\cA_5} |_{\cA_n} + K_{\cA_n}) $ the elements of the homogeneous coordinate ring $ S_n $ of $ \cA_n $ with degree $ [-K_{\cA_5} + K_{\cA_n}] $ as we introduced in \eqref{homogeneousCoordinatesSn}.

For the above expression we introduced the holomorphic volume-form $ d \omega_{\cA_n} $ of $ \cA_n $. To define this, we introduce a fixed integral basis $ \{ m_1, \ldots, m_n\} $ of $ M_n $ and define for each index set $ I = \{ i_1, \ldots, i_n \} $ of $ n $-integral points $ \nu_{i^1} ^\ast , \ldots, \nu_{i^n} ^\ast $ in $ \Delta^\ast _n \cap N_n $ the determinant
\begin{equation}
 \text{det}(\nu^\ast _I) = \text{det} (\langle m_i, \nu^\ast _j \rangle_{1 \leq i,j \leq n}) \, ,
\end{equation}
which can be thought of as the volume of the simplex spanned by the integral points $ \nu^\ast _i $ and the origin. From this we can construct the holomorphic volume-form as
\begin{equation} \label{hol_volume_form}
 d \omega_{\cA_n} = \sum_{|I|=n} \text{det}(\nu^\ast _I) \Big( \prod_{i \notin I} X_i \Big) dX_{i^1} \wedge \ldots \wedge dX_{i^n} \, .
\end{equation}
Here the sum runs over all index sets $ I $ with $ n $ elements $\{ i^1 , \ldots, i^n \}$. If we assign the one-forms $ dX_i $ the same degree as their coordinate counterparts $ X_i $, we see that the degree of $ d \omega_{\cA_n} $ is given by
\begin{equation}
 \text{deg}_{S_n} (d \omega_{\cA_n}) = [ \sum_{\nu^\ast _i \in \Delta^\ast _n \cap N_n} D_i ] = - K_{\cA_n} \, . 
\end{equation}
Note that the holomorphic volume-form is only unique up to a multiple of a constant.
Having introduced the notation, we can now establish the Poincar\'e residue construction to find representations for the holomorphic $ (n-1) $-forms of the Cartier divisor $ R $ with degree $ [-K_{\cA_5} |_{\cA_n}] $ defined by $ p_\theta = 0$ in the toric ambient space $ \cA_n $. We define the Poincar\'e residue as the map
\begin{align}
 H^0(\cA_n, \Omega^n _{\cA_n}(R)) \quad & \rightarrow \quad H^0 (R, \Omega^{n-1}_R) \\
 \frac{g \, d \omega_{\cA_n}}{p_\theta} \quad &\mapsto \quad \int_{\Gamma} \frac{g \, d \omega_{\cA_n}}{p_\theta} \, .  \nonumber
\end{align}
We introduced here a small one-dimensional curve $ \Gamma \in H_1(\cA_n - R, \mathbb{R})$ around $ R $ in the complement of $ R $ in $ \cA_n $. This integral expression defines a holomorphic $ (n - 1) $-form, as applied to a $ (n-1) $-cycles $ \alpha \in H_{n-1}(R) $, we fiber $ \Gamma $ over $ \alpha $ and integrate the rational (meromorphic) form in $ H^0(\cA_n, \Omega^n _{\cA_n}(R)) $ over the resulting $ n $-cycles to obtain a complex number. This map is well-defined, but not injective, since we can partially integrate
\begin{equation}
 \int_{\Gamma} \frac{g^i \partial_{X^i} p_\theta \, d \omega_{\cA_n}}{p_\theta} = 0 \, ,
\end{equation}
which constitutes the kernel of the residue map which can also be denoted by the residue symbol $ \text{Res}_R ( \, \cdot \, ) $.
Therefore, modding out the partial derivatives of $ p_\theta $ of $ S_n $ and define the Jacobian or chiral ring
\begin{equation}
 \cR_\theta = \frac{S_n}{\langle \partial_{X^i} p_\theta \rangle} \, .
\end{equation}
Here we denoted by $ \langle \partial_{X^i} p_\theta \rangle $ the ideal spanned by the partial derivatives of the defining polynomial $ p_\theta $. The ring $ \cR_\theta $ inherits the grading of $ S_n $.
This renders the map
\begin{equation}
 \cR_\theta (-K_{\cA_5}|_{\cA_n} + K_{\cA_n}) \, \quad \hookrightarrow \quad H^0 (R, \Omega^{n-1}_R)
\end{equation}
defined by the Poincar\'e residue injective, as was shown in \cite{Mavlyutov:2000hw}. In the case that the Calabi-Yau hypersurface $ Y_4 ^{sing} $ has only singularities of codimension two arising from the ambient space $ \cA_5 $, as is the case for our semiample situation of the previous section, we see that for $ n = 1 $ the residue just gives a constants over disjoint points, whose location is defined by $ p_\theta $. In this case we find $ \ell(\theta) + 1 $ distinct points over which we have a three-dimensional exceptional divisor. the points are the zeroes of $ p_\theta $. The $ \ell^\prime (\theta) $ count the moduli, the differences between the zeroes in $ \mathbb{P}^1 = \cA_1 $.
\begin{equation}
 n = 1 \, : \, R = \{ pt \, | \,  p_\theta(pt) = 0 \} \, , \quad h^{0,0}(R) = \ell^\prime (\theta) + 1 \, .
\end{equation}
In the case $ n = 2 $, we find the holomorphic one-forms of Riemann-surfaces
\begin{equation}
 n = 2 \, : \quad \cR_\theta (-K_{\cA_5}|_{\cA_2} + K_{\cA_2}) \, \simeq \, H^0 (R, \Omega^{1}_R) = H^{1,0} (R) \, .
\end{equation}
Similarly, we constructed all the holomorphic two-forms of a surface $ S $ as
\begin{equation}
 n = 3 \, : \quad \cR_\theta (-K_{\cA_5}|_{\cA_3} + K_{\cA_3}) \, \simeq \, H^0 (S, \Omega^{2}_S) = H^{2,0} (S) \, .
\end{equation}
The trivial case is now $ n = 5 $ which is the full Calabi-Yau fourfold $ Y_4 $ in the space $ \hat \cA_5 $ and since we are only considering complex structure dependent quantities, it doesn't matter if we resolve via blow-ups of the ambient space. Therefore, we find that
\begin{equation}
 n = 4 \, :  \quad \cR_\Delta (0) \, \simeq \, H^0 (Y_4, \Omega^4 _{Y_4}) = H^{4,0} (Y_4) \, .
\end{equation}
It is easy to see that
\begin{equation}
 \cR_\Delta (0) = S_5 (0) = \langle 1 \rangle \, .
\end{equation}
Note that $ p_\Delta $ can not be linear in any homogeneous coordinate $ X_i $, since the generic monomial in $ p_\Delta $ has degree $ \prod_{\nu^\ast _i \in \Delta \cap N} X_i $.
Therefore, this element has to correspond to the holomorphic four-form $ \Omega $ on $ Y_4 $ and can be represented as
\begin{equation}
 \Omega \sim \int_{\Gamma} \frac{d \, \omega_{\cA_5}}{p_\Delta} \in H^{4,0} (Y_4)\, .
\end{equation}
up to a function holomorphic in the complex structure moduli $ a_j $.

Let us now determine the number of holomorphic $ (n-1) $-forms of the hypersurfaces $ R $, i.e.~ $ h^{n-1,0}(R) $, in terms of the toric data. Given a divisor $ D_\Delta $ with Newton-Polyhedron $ \Delta $ over a toric variety $ \cA $ with polyhedron $ \Delta^\ast $, the degree $ [D_\Delta] $ submodule $ S(D_\Delta) $ of its homogeneous coordinate ring $ S $ is given by
\begin{equation}
 S(D_\Delta) = \bigoplus_{\nu_i \in \Delta \cap M} \mathbb{C} \cdot \prod_{\nu^\ast _i \in \Delta^\ast \cap N} X_i ^{\langle \nu_j, \nu^\ast _i \rangle} \, .
\end{equation}

It can be shown, that the dimension of the quotient $ \cR_\Delta (D_\Delta) $ with $ D_\Delta $ represented by a transverse polynomial $ p_\Delta $ as a $ \mathbb{C} $-module is the given by the number of interior integral points of $ \Delta $
\begin{equation}
 \cR_\Delta (D_\Delta)= \bigoplus_{\nu_j \in \text{int}(\Delta) \cap M} \mathbb{C} \cdot \prod_{\nu_i ^\ast \in \Delta^\ast \cap N} X_i ^{\langle \nu_j,  \nu^\ast _i \rangle} \, .
\end{equation}
Therfore, we find that
\begin{equation}
 h^{n-1,0}(R_{n-1}) = \ell^\prime (\theta) \, , \quad \text{dim}(\theta) = n \, .
\end{equation}

From this we have now seen, how to represent the holomorphic forms on the various toric hypersurfaces we encountered in the Gysin-sequences of the previous sections. In the next section, we will use these representations to gain further insights into the behaviour of these holomorphic forms under complex structure variation.

\section{Hodge variation in semi-ample hypersurfaces}\label{HodgeVariationSection}

In this section, we want to investigate the complex structure variations of the semiample hypersurface $ Y_4 $ in the simplicial toric variety $ \cA_5 $. Due to the fact that these are independent of the K\"ahler moduli and hence the volumina of the blow-up divisors resolving singularities, it does not matter if we blow-up the singularities for the complex structure variations.

The variations around a point $ a \in \cM_c $ in complex structure moduli space $ \cM_c $ of the Calabi-Yau fourfold $ Y_{4,a} $ can be parametrized by
\begin{equation}
 H^1 (Y_{4,a}, \cT Y_4) \simeq H^{3,1} (Y_{4,a}) \, ,
\end{equation}
where the isomorphism is given by contraction with the no-where vanishing holomorphic four-form $ \Omega (a) $. We will drop the $ a $-dependence in the notation for $ Y_{4,a} $ and just write $ Y_4 $ in the following. From the Gysin-sequence \eqref{Gysin_complex_structure} we can deduce the splitting (we are dealing with free groups)
\begin{equation}
H^{3,1}(Y_4) \simeq \text{Gr}_4 ^W H^{3,1}(Y_4 \cap \mathbb{T}) \oplus  \bigoplus_{\nu^\ast _i} H^{2,0}(D_i ^\prime) \, .
\end{equation}
The first part corresponds to the deformations of the polynomial $ p_\Delta $ via monomials $ p_\nu $ with $ \nu \in \Delta \cap M $, therefore we will refer to it as algebraic deformations of the complex structure. It can be shown, \cite{Mavlyutov:2000hw,Mavlyutov:2000dj} that these are given by
\begin{equation}
 H^{3,1}(Y_4) _{alg} \simeq \text{Gr}_4 ^W H^{3,1}(Y_4 \cap \mathbb{T}) \simeq \cR_\Delta (-K_{\cA_5}) \, .
\end{equation}
They are represented by the Poincar\'e residue construction as
\begin{align}
 H^0 (\cA_5, \Omega^5(-2K_{\cA_5})) &\rightarrow H^{3,1}(Y_4) \oplus H^{4,0}(Y_4) \\
 \frac{p_\nu}{p_\Delta ^2} d \omega_{\cA_5} &\mapsto  \int_{\Gamma} \frac{p_\nu}{p_\Delta ^2} d \omega_{\cA_5} \, .  \nonumber
\end{align}
Note that
\begin{equation}
 H^0 (\cA_5, \Omega^5(-2K_{\cA_5})) = S_5 (-K_{\cA_5}) \, ,
\end{equation}
and the kernel is given by $ \langle \partial_i p_\Delta, \, \nu^\ast _i \in \Delta^\ast \rangle $. It was shown by Batyrev in \cite{Batyrev:1994ju} that the dimension of the space of these algebraic complex structure deformations is
\begin{equation}
 h^{3,1} _{alg} = \ell(\Delta) - 6 - \sum_{\text{dim}(\theta) = 4} \ell^\prime (\theta) \, .
\end{equation}
Due to the residue representation we can see that
\begin{equation}
 \frac{\partial}{\partial a_\nu} \, \Omega = \frac{\partial}{\partial a_\nu} \Big( \int_{\Gamma} \frac{1}{p_\Delta} d \omega_{\cA_5} \Big)= -  \int_{\Gamma} \frac{p_\nu}{p_\Delta ^2} d \omega_{\cA_5} \, .
\end{equation}
The remaining complex structure deformations arise from divisors that are blow-ups of singular surfaces $ S_\alpha $
\begin{equation}
 H^{3,1}(Y_4) _{non-alg} \simeq \bigoplus_{\text{dim}(\theta_\alpha) = 3} H^{2,0}(S_\alpha) \simeq \cR_\alpha (-K_{\cA_5} |_{\cA_{3,\alpha}} + \cK_{\cA_{3,\alpha}}) \, .
\end{equation}
The number of these deformations is $ \ell^\prime (\theta_\alpha) $, the number of holomorphic $ (2,0) $-forms on the toric divisors $ D_{l_\alpha} $ times the number of necessary blow-ups $ \ell(\theta^\ast _\alpha) $:
\begin{equation}
 h^{3,1} _{non-alg} = \sum_{\text{dim}(\theta_\alpha) = 3} \ell^\prime (\theta_\alpha) \ell^\prime (\theta_\alpha ^\ast) \, .
\end{equation}
Note that we have in general the map
\begin{align}
 H^0 (\cA_d, \Omega^d(p D) \simeq S_d ( p \beta+ K_{\cA_d}) &\rightarrow \cF^p _{alg} = \bigoplus_{k = 0} ^p H^{d-1-k, k} _{alg} (D)\\
 \frac{q}{p_\Delta ^p} \, d \omega_{\cA_d} &\mapsto  \int_{\Gamma} \frac{q}{p_\Delta ^p} \, d \omega_{\cA_5} \, , \nonumber 
\end{align}
for a quasi-smooth semiample divisor $ D_\Delta $ with Newton-polyhedron $ \Delta $ the zero set of $ p_{\Delta} \in S_d [D_\Delta]$ with degree $ [D_\Delta] \in A_{d-1} (\cA_d)$ and $ \Gamma $ the usual one-dimensional curve in the complement of $ D_\Delta $ in $ \cA_d $. This generates the algebraic part of the $ (d-1) $-dimensional cohomology group of $ D_\Delta $, called horizontal cohomology of $ D_\Delta $. \footnote{Note that in $ \text{Gr}_4 ^W H^{3,1}(Y_4 \cap \mathbb{T}) $ the $ \text{Gr}_4 ^W $ is a weight grading up to order four which corresponds to the pole order of the rational forms necessary to produce all four-forms that arise from the residue construction.}

This divisor $ D_\Delta $ needs not to be Calabi-Yau and hence we can also use this construction for bases $ B_3 $ of elliptically fibered Calabi-Yau fourfolds that are toric hypersurfaces with $ h^{2,1}(B_3) > 0 $ as considered in \autoref{MFlimit}. A simple example is given by the cubic hypersurface $ B_3 = \mathbb{P}^4[3] $ in $ \mathbb{P}^4 $, for which $ p_\Delta $ is just a general degree three homogeneous polynomial in the coordinates $ [X_1, \ldots, X_5] $. In this case $ \Delta $ is not reflexive and its vertices are not integral, but the same construction applies. The $ (2,1) $-forms can be represented as
\begin{equation} \label{cubicExample}
  \gamma_i = \int_\Gamma \frac{X_i}{p_\Delta ^2} \, d \omega_{\mathbb{P}^4} \, \in H^{2,1}(B_3) \, ,
\end{equation}
In particular, we find the non-trivial Hodge numbers $ h^{0,0} = h^{1,1} = h^{2,2} = h^{3,3} = 1 $ (generated by the hyperplane class of $ \mathbb{P}^4 $) and $ h^{2,1} = 5 $. These three-forms do not correspond to the complex structure deformations, as there is no holomorphic no-where vanishing three-form. Note that we have $ h^{3,0} = 0 $ which can be seen from the positive degree of $ d \omega_{\mathbb{P}^4} / p_\Delta $. The geometry of this hypersurface was discussed in detail by Clemens and Griffiths in \cite{10.2307/1970801}.

In the following we will focus on the three-forms of Calabi-Yau hypersurfaces for which a similar construction can be considered. First, however, we will use the obtained representations to count the Hodge numbers of a Calabi-Yau fourfold hypersurface.

\section{Counting the Hodge numbers of a semiample Calabi-Yau fourfold hypersurface} \label{HodgeNumberSection}

The number of components of an $ n $-semiample toric divisors $ D^\prime $ is given by
\begin{align}
 h^{0,0}(D^\prime) = 1 \, , \quad n = 2,3 \, \quad h^{0,0}(D^\prime) =  \ell^\prime(\theta) + 1 \, , \quad n = 1 \,.
\end{align}
For each face $ \theta^\ast $ we have $ \ell^\prime (\theta^\ast) $ toric divisors, where $ \ell^\prime (\theta^\ast) $ counts the number of interior integral points of the $ (4-n) $-dimensional face $ \theta^\ast $ of $ \Delta^\ast $, $ n < 4 $ corresponding to the blow-up divisors necessary to resolve the ambient space $ \cA_5 $. Combining now the insights from the Gysin-sequence of the previous section and the Poincar\'e representation of the holomorphic forms, we can deduce the Hodge numbers of smooth Calabi-Yau fourfold $ Y_4 $. This was already discussed in \cite{Klemm:1996ts}.

We find for the number of $(1,1)$-forms that we have three parts. The first arises from the ambient space $ \cA_{d+1} $ as seen in \eqref{ambient_cohomology} with $ h^{1,1}(\cA_{d+1}) = \ell(\Delta^\ast) - 6 $, to obtain the cohomology of the hypersurface, however, we need to substract the blow-up divisors over points that do not lie on the general hypersurface $ Y_4 $, which are the zero-semiample divisors. To account for non-toric divisors arising from one-semiample divisors we need to add the $ \ell^\ast(\theta) $ components over $ \ell^\ast(\theta^\ast) $ points:
\begin{align}
 h^{1,1}(Y_4) &=  \ell(\Delta^\ast) - 6 - \sum_{\text{dim}(\theta) = 0} \ell^\prime (\theta^\ast) + \sum_{\text{dim}(\theta) = 1}  \ell^\prime(\theta) \ell^\prime (\theta^\ast) \, ,
\end{align}
where $ \ell(\Delta^\ast) $ denotes all integral points of $ \Delta^\ast $. This turns out to be exactly dual to the formula for $ h^{3,1}(Y_4) $ which combines the algebraic and non-algebraic complex structure deformations as 
\begin{align}
 h^{3,1}(Y_4) &= \ell(\Delta) - 6 - \sum_{\text{dim}(\theta) = 4} \ell^\prime (\theta) + \sum_{\text{dim}(\theta) = 3}  \ell^\prime(\theta) \ell^\prime (\theta^\ast) \, ,
\end{align}
counting the number of elements of chiral rings with a certain degree as described in \autoref{HodgeVariationSection}. The duality between the two Hodge numbers of $ h^{1,1} $ and $ h^{3,1} $ is a first manifestation of mirror symmetry on Calabi-Yau fourfolds, as described in \cite{Greene:1993vm,Mayr:1996sh} and we have already seen in \autoref{mirror_section}. The mirror symmetry of toric hypersurfaces can be simply expressed as the exchange of the dual polyhedra $ \Delta $ and $ \Delta^\ast $, as first realized by Batyrev \cite{Batyrev:1994hm}. Due to the fact that the non-trivial three-forms arise solely from toric divisors as discussed around \eqref{Elalph_fibrDprime}, the Hodge number $ h^{2,1} $ per two-dimensional face $ \theta_\alpha $ is the product of one-forms on the the Riemann surface $ R_\alpha $ given by $ \ell^\prime (\theta_\alpha) $ and the number of blow-up divisors necessary to resolve the singularity along $ R_\alpha $ given by $ \ell^\prime(\theta^\ast) $. The result is
\begin{equation} \label{h21count}
 h^{2,1}(Y_4) = \sum_{\text{dim}(\theta) = 2}  \ell^\prime(\theta) \ell^\prime (\theta^\ast) \, ,
\end{equation}
which is invariant under the exchange of $ \Delta $ and $ \Delta^\ast $ as predicted by mirror symmetry.
The remaining non-trivial Hodge number $ h^{2,2} $ can be computed from the other three via index theorems as already seen in \eqref{h22fromIndex}. These split into four orthogonal parts.\footnote{As we do not prove these statements here, this should be interpreted as a conjecture. It is the Calabi-Yau fourfold version of the cohomology splitting shown in \cite{Mavlyutov:2000hw} for general semi-ample hypersurfaces in toric varieties.} The first two parts a horizontal, they arise from the complex structure variations of algebraic and non-algebraic $ (3,1) $-forms and are also primitive, i.e.~ their product with the K\"ahlerform $ J $ vanishes. The second part is the vertical cohomology that arises form wedge products of $ (1,1) $-forms corresponding to intersections of toric and non-toric divisors. \\

%
In the next section we will focus more on the origin and complex structure dependence of the non-trivial three-forms of toric Calabi-Yau fourfold hypersurfaces in toric varieties which was not studied before.

\chapter{The intermediate Jacobian of a Calabi-Yau fourfold hypersurface} \label{JacobianSection}

Using the arguments presented in \autoref{ToricSection}, we deduced that the three-form cohomology of a semiample Calabi-Yau fourfold hypersurface of a toric variety is induced by the holomorphic one-forms of a Riemann surface. The relevant notions of the Riemann surface will be introduced in \autoref{RiemannSection} and then lifted to the full fourfold geometry in \autoref{FourFoldJacobianSection}. In course of this we will identify the quantities of \autoref{threeFormAnsatz} of the toric hypersurface setting and introduce a class of simple example geometries, the weighted projective spaces in \autoref{WeightedProjectiveSpaceSection}.

\section{Three-form periods from Riemann surfaces} \label{RiemannSection}

We want to specify now to the toric divisors of the Calabi-Yau fourfold hypersurface $ \hat Y_4 $ providing non-trivial three-forms. The fibration structure of these divisors with base a Riemann surface will enable us to apply the well-established theory of periods of Riemann surfaces to find a description of the periods of the three-forms on a Calabi-Yau fourfold realized as a toric hypersurface. To do so, we first introduce the general theory of Riemann surfaces necessary to understand the period construction, as is by now textbook material \cite{griffiths2011principles}. Then we will use the representation of holomorphic one-forms on ample toric hypersurfaces as used in \cite{1994alg.geom.10017C,1993alg.geom..6011B,Mavlyutov:2000hw} to find the periods of these Riemann surfaces. These periods satisfy a set of second order differential equations, the Picard-Fuchs equations, as was derived similarly in \cite{Lerche:1991wm}.

In the speical case of two-semiample toric divisors $ D^\prime _{l_\alpha} $ with two-dimensional Newton-polyhedron $ \theta_\alpha $, the fibration structure described in \eqref{Elalph_fibrDprime} reads
\begin{equation}
 \begin{tikzpicture}
  \matrix (m) [matrix of math nodes,row sep=3em,column sep=4em,minimum width=2em]
  {
     E_{l_\alpha} & D^\prime _{l_\alpha} \subset D_{l_\alpha} \\
     & R_\alpha \subset \cA_{2,\alpha} \\};
  \path[-stealth]
    (m-1-1)  edge node [above] {$i_{l_\alpha}$} (m-1-2)
    (m-1-2) edge node [right] {$\pi_{l_\alpha}$} (m-2-2)
            ;
\end{tikzpicture}
\end{equation}
where $ R_\alpha $ is a (compact) Riemann surface embedded in a two-dimensional simplicial complete toric ambient space $ \cA_{2,\alpha} $ by the polynomial $ p_\alpha $. The fiber $ E_{l_\alpha} $ is two-dimensional and toric and the Hodge types of its cohomology is independent of the complex structure of $ \hat Y_4 $. In particular, we see that the full complex structure dependence of the non-trivial three-forms is captured by the holomorphic one-forms of the Riemann surfaces $ R_\alpha $. Therefore we will now discuss the general theory of periods on Riemann surfaces $ R $.

A Riemann surface $ R $ is a compact K\"ahler manifold of complex dimension one and we are interested in its non-trivial cohomology, the non-trivial one-forms and their complex structure dependence. To do this, we introduce appropriate bases for the one-forms, first a topological and then a holomorphic basis.

The Riemann surface $ R $ we consider has genus $ g = h^{1,0}(R) $ and is equipped with a basis of integral one-cycles $ \hat A_a , \hat B^a \in H_1(R, \mathbb{Z}) $ with indices $ a = 1, \ldots, g $ and dual one-forms $ \hat \alpha_a , \hat \beta^a \in H^1(R, \mathbb{Z})$. We can choose this basis to satisfy canonically
\begin{equation} \label{oneCycleBase}
 \int_R \hat \alpha_a \wedge \hat \beta^b = \delta_a ^b \, , \quad \int_R \hat \alpha_a \wedge \hat \alpha_b = 0 \, , \quad 
\int_R \hat \beta^a \wedge \hat \beta^b = 0 \, . 
\end{equation}
For a $ n $-dimensional K\"ahler manifold we can choose the holomorphic $ n $-forms to vary holomorphically with the complex structure, \cite{Voisin03hodgetheory}. In particular, we can choose a basis of holomorphic one-forms on $ R $ denoted by $ \gamma_a \in H^0(R, \Omega^1 _R) $ depending holomorphically on the the complex structure of $ R $. The so called period matrices are obtained by integrating these one-forms over the basis of one-cycles as
\begin{equation}
 (\hat \Pi _a) ^b = \int_{\hat A_b} \gamma_a \, , \quad (\hat \Pi _a) _b = \int_{\hat B^b} \gamma_a \, .
\end{equation}
These $ g \times g $ matrices are holomorphic in the complex structure. The period vectors $ \hat \Pi^b $ and $ \hat \Pi_b $ are the column vectors of these matrices. These vectors are obtained by integrating the full basis of holomorphic one-forms over one fixed one-cycle. These $ 2g $ vecotrs are linearly indpendent over $ \mathbb{R} $ and therefore generate a lattice
\begin{equation}
 \hat \Lambda = \bigoplus_a \big( \hat \Pi_a \mathbb{Z} \oplus \hat \Pi^a \mathbb{Z} \big) \, .
\end{equation}
in $ \mathbb{C}^g \simeq H^{1,0}(R) $. The period matrices allow to expand the holomorphic one-forms into the topological basis as
\begin{equation}
 \gamma_a = (\hat \Pi _a) ^b \hat \alpha_b + (\hat \Pi _a) _b \hat \beta^b \, ,
\end{equation}
and define the projection of the integral cohomology, the lattice $ H^1(R, \mathbb{Z}) $, to the eigenspace of complex structure $ H^{1,0}(R) \subset H^1(R, \mathbb{C}) $. The object we want now to study is the Jacobian variety $ \cJ^1(R) $ of the Riemann surface $ R $ given by
\begin{equation} \label{RiemannIntermediateJacobian}
 \cJ^1(R) = \frac{H^{1,0}(R)}{H^1(R,\mathbb{Z})} \simeq \mathbb{C}^g / \hat \Lambda \, .
\end{equation}

We can normalize the basis $ \gamma_a $, since one of the two period matrices will be invertible, we choose $ (\hat \Pi _a) ^b $. This enables us to introduce the normalized basis $ \tilde \gamma_a \in H^{1,0}(R) $ of holomorphic one-forms on $ R $:
\begin{equation}
 \tilde \gamma_a = {(\hat \Pi^{-1} )_a} ^b \gamma_b \, , \quad \int_{\hat A^b} \tilde \gamma_a = \delta_a ^b \, , 
\end{equation}
which now depends meromorphically on the complex structure captured by the normalized period matrix
\begin{equation}
 i \hat f_{ab} = {(\hat \Pi^{-1} )_a} ^c (\hat \Pi_c )_b \quad \Rightarrow \quad \tilde \gamma_a = \hat \alpha_a + i \hat f_{ab} \hat \beta^b \, .
\end{equation}
In the following we will assume our basis of holomorphic one-forms to always be of this form and hence we drop the tilde and write for the normalized basis $ \gamma_a \in H^{1,0}(R) $.

It can now be shown that this holomorphic (in general meromorphic) normalized period matrix $ \hat f_{ab} $ of the Riemann surface $ R $ satisfies the relations
\begin{equation}
 \hat f_{ab} = \hat f_{ba} \, , \quad \text{Re} \, \hat f_{ab} > 0 \, ,
\end{equation}
the period matrix is symmetric and its real part is positive definite. Therefore, we can find a positive definite quadratic form on $ H^{1,0}(R) $ that is given in the normalized basis as
\begin{equation} \label{fhat}
 -i \int_{R} \gamma_a \wedge \bar \gamma_b = 2 \, \text{Re} \, \hat f_{ab} \, .
\end{equation}
This normalized period matrix $ \hat f_{ab} $ will be the physical quantity we are interested in. Especially its complex structure dependence will be of interest and we will study this by giving an explicit representation of the holomorphic one-forms for Riemann surfaces as hypersurfaces $ R $ in simplicial complete toric varieties $ \cA_2 $.

Here we specialize to the more general construction as discussed in \autoref{PoincareSection} of mapping rational two-forms on $ \cA_2 $ with poles along the hypersurface $ R $ of order $ r $ denoted by $ \Omega^2_{\cA_2}(rR) $ to the middle cohomology of $ R $ via the Poincar\'e residue. The general description of these rational forms that arise from the restriction of the anti-canonical hypersurface on a ambient space $ \cA_5 $ to $ \cA_2 $ reads
\begin{align}
 H^0(\cA_2, \Omega^2 _{\cA_2}(r R)) &= \{ \frac{g \, d \omega_{\cA_2}}{p^r _\theta} \, : \, g \in S_2 ( -r K_{\cA_5}|_{\cA_2} + K_{\cA_2}) \}  \\
 &\simeq S_2 ( -r K_{\cA_5}|_{\cA_2} + K_{\cA_2}) \, . \nonumber
\end{align}
We denoted here by $ - K_{\cA_5}|_{\cA_2} \in A_1(\cA_2) $ the divisor class of the restriction of the anti-canonical divisor class $ -K_{\cA_5} $ of $ \cA_5 $ to $ \cA_2 $. The representative of this divisor is given by  $ R $ which is the zero set of $ p_\theta = p_\Delta |_{\cA_2} \in S_2 (- K_{\cA_5}|_{\cA_2} ) $. Similarly we denoted by $ -K_{\cA_2} $ the anti-canonical divisor class of $ \cA_2 $ and as we already seen in \eqref{hol_volume_form} the holomorphic volume form $ d \omega_{\cA_2} $ of $ \cA_2 $ has the same degree as the anti-canonical divisor class $ -K_{\cA_2} $.

Using now the theory outlined in \autoref{PoincareSection} we can express the (anti-)holomorphic one-forms via the residue construction as
\begin{align}
 \cR_\theta ( -r K_{\cA_5}|_{\cA_2} + K_{\cA_2}) \quad &\rightarrow \quad H^{2-r,r-1}(R) \, , \quad r = 1,2 \nonumber \\
 q \quad &\mapsto \quad \int_{\Gamma} \frac{q}{p_\theta ^{r}} d \omega_{\cA_2} \, .
\end{align}
Here $ \Gamma $ is a small one-dimensional curve winding around the Riemann surface $ R $ in $ \cA_2 $. 

Moving on to the study of the complex structure dependence of these forms, we see that due to the fact that $ R $ is an ample hypersurface its holomorphic one-forms are entirely determined by the polynomial $ p_\theta $ which is a restriction of $ p_\Delta $ to $ \cA_2 $. This implies that the complex structure of $ H^1(R,\mathbb{C}) $ can only depend on algebraic deformations of the Calabi-Yau fourfold and only the very few surviving the projection of the full polynomial $ p_\Delta $ to $ p_\theta $. Recall that the family of Calabi-Yau hypersurfaces in the toric simplicial complete ambient space $ \cA_5 $ was given as the zero set of
\begin{align}
 p_\Delta &= \sum_{\nu_j \in \Delta \cap M} a_j \prod_{\nu^\ast _i \in \Delta^\ast \cap N} X_i ^{\langle \nu_j, \nu^\ast _i\rangle + 1} \, \\
 & \in S_5(-K_{\cA_5}) \simeq H^0(\cA_5, \cO_{\cA_5}(-K_{\cA_5})) \, . \nonumber
\end{align}
The restriction to $ \cA_2 $ is simply given by
\begin{align}
 p_\Delta |_{\cA_2} &= p_\theta = \sum_{\nu_j \in \theta \cap M_2} a_j \prod_{\nu^\ast _i \in \theta^\ast \cap N_2} X_i ^{\langle \nu_j, \nu^\ast _i\rangle + 1} \, \\
 &\in S_2(-K_{\cA_5}|_{\cA_2}) \simeq H^0(\cA_2, \cO_{\cA_2}(-K_{\cA_5}|_{\cA_2})) \, . \nonumber
\end{align}
Therefore, the complex structure of $ R $ is fixed by the vector of the prefactors $ a = (a_j) $ of the monomials deformations $ p_j $ corresponding to integral points $ \nu_j $ of $ \Delta $. Therefore, we can denote the Riemann surface with complex structure at a point $ a $ in complex structure moduli space by $ R_a $, this is the induced complex structure of the full Calabi-Yau fourfold $ Y_{4,a} $. From the previous analysis, we deduce that a one-form $ \gamma \in H^1(R_a,\mathbb{C}) $ on $ R_a $ will only depend on complex structure moduli corresponding to integral points in the interior of $ \theta $:
\begin{equation}
 \frac{\partial}{\partial a_j} \, \gamma (a) = 0 \, , \quad \forall \, \gamma \in H^1(R_a,\mathbb{C}) \, , \, \nu_j \notin \text{int} ( \theta) \, .
\end{equation}
In the following we will therefore consider only deformations of $ p_\theta $ by monomials $ p_b $ corresponding to the integral points $ \nu_b \in \text{int}(\theta) $ and similar the corresponding complex structure coordinates $ a_b $. As we have seen the explicit representations of $(1,0)$-forms $ \hat \gamma_b (a) \in H^{1,0}(R_a) $ depend holomorphic on the coordinates $ a_b $. This makes the normalized period matrix $ \hat f_{ab}(a) $ a holomorphic function of the complex structure moduli $ a_b $. Note that we will find $ \hat f_{ab}(a) $ to be a rational function of the moduli $ a $, but we will only consider the normalized period matrix in a region of moduli space where it is holomorphic.

In more mathematical language we can interpret the residue representation of the one-forms as local trivializations of the Hodge bundles with base the complex structure moduli space $ \cM_c $ of $ Y_4 $ and fibers given by $ H^{1,0}(R_a) $ and $ H^1(R_a,\mathbb{C}) $, respectively. These are holomorphic bundles, for details we refer to \cite{Voisin03hodgetheory}. This will enable us to derive the complex structure dependence of the holomorphic $(1,0)$-forms
\begin{equation}
 \gamma_b (a) = \int_{\Gamma} \frac{p^\prime _b}{p_\theta} \, d \omega_{\cA_2} \, \in H^{1,0}(R_a) \, , \quad \nu_b \in \text{int}(\theta) \cap M \, .
\end{equation}
Here we denoted by $ p^\prime _b = p_b / \prod_{\nu^\ast _i \in \theta^\ast} X_i \in S_2(-K_{\cA_5} |_{\cA_2} + K_{\cA_2}) $. This is still a monomial, since the corresponding integral point $ \nu_b \in \text{int}(\theta) \cap M $. We have
\begin{equation}
 p_b ^\prime = \prod_{\nu^\ast _i \in \theta^\ast \cap N_2} X_i ^{\langle  \nu^\ast _i , \nu_b \rangle}  \in S_2(-K_{\cA_5} |_{\cA_2} + K_{\cA_2}) \, , \quad \nu_b \in \text{int}(\theta) \cap M \, .
\end{equation}
We can now see easily what happens if we vary the complex structure by taking derivative with respect to the moduli $ a_b $. The first derivative is given by
\begin{equation}
\frac{\partial}{\partial a_c} \, \gamma_b (a) = \frac{\partial}{\partial a_b} \, \gamma_c (a) =- \int_{\Gamma} \frac{p^\prime _b p_c}{p^2 _\theta} \, d \omega_{\cA_2} \, \in H^1(R_a,\mathbb{C}) \, .
\end{equation}
The ring structure of the chiral ring $ \cR_{\theta} = S_2 / \langle \partial_i p_\theta \rangle $ determines the Hodge type of this form. Note that although $ S_2 $ does not depend on $ a $, $ \cR_\theta $ does depend on $ a $ through $ p_\theta $. We have
\begin{align}
 \frac{\partial}{\partial a_c} \, \gamma_b (a) &=- \int_{\Gamma} \frac{p^\prime _b p_c}{p^2 _\theta} \, d \omega_{\cA_2} \in H^{1,0}(R_a) \, , \quad p^\prime _b p_c \in \langle \partial_i p_\theta \rangle \, , \\
 \frac{\partial}{\partial a_c} \, \gamma_b (a) &=- \int_{\Gamma} \frac{p^\prime _b p_c}{p^2 _\theta} \, d \omega_{\cA_2} \in H^1(R_a,\mathbb{C}) \, , \quad p^\prime _b p_c \not \in \langle \partial_i p_\theta \rangle \, .
\end{align}
It can be shown in general that the one-forms $ \gamma_b $ and its first derivatives are sufficient to generate all of $ H^1(R_a,\mathbb{C}) $ as a $ \mathbb{C} $-module. This implies in particular, that we can express the second derivatives of the $ \gamma_b $ in terms of lower derivatives. This can be seen from
\begin{align}
\frac{\partial}{\partial a_d}\frac{\partial}{\partial a_c} \, \gamma_b (a) &= 2 \int_{\Gamma} \frac{p^\prime _b p_c p_d}{p^3 _\theta} \, d \omega_{\cA_2} \, \in H^1(R_a,\mathbb{C}) \,\\
&\Rightarrow \quad p^\prime _b p_c p_d \in \langle \partial_i p_\theta \rangle \, .  \nonumber 
\end{align}
Expressing the monomials of the second derivatives $ p^\prime _b p_c p_d $ in terms of lower degree monomials and partial derivatives of $ p_\theta $ introduced coefficients depending rationally on the complex structure moduli $ a_b $. These coefficients are structure constants of the chiral ring $ \cR_{\theta} $. Therefore, we can write
\begin{equation} \label{RiemannSurfacePF}
 \frac{\partial}{\partial a_c}\frac{\partial}{\partial a_d} \, \gamma_b (a) = \big( {c ^{(1)}(a)_{cd b} }^{e f} \frac{\partial}{\partial a_e} + {c^{(0)} (a)_{cd b} }^f \big) \, \gamma_f (a) \, .
\end{equation}
Here we denoted the structure constants of $ \cR_{\theta} $ as $ {c ^{(1)}(a)_{cd b} }^{e f}  $ and $ {c^{(0)} (a)_{cd b} }^f $, respectively. These are rational functions in the complex structure moduli $ a_b $ determined by modding out the partial derivatives of $ p_\theta $ from the homogeneous coordinate ring $ S_2 $ of the ambient space $ \cA_2 $. From the previous considerations, it is easy to see that these structure constant are symmetric in their lower and upper indices. The relations between the second derivative and the lower derivatives are Picard-Fuchs equations which are central in the derivation of the complex structure dependence of the holomorphic one-forms $ \gamma_b $ and the normalized period matrix $ \hat f_{ab} $.

Due to the fact the Picard-Fuchs equations are determined by the structure constants of the chiral ring $ \cR_\theta $ which is a quotient of the full Jacobian $ \cR_\Delta $ of the Calabi-Yau fourfold $ Y_4 $, the flat complex structure coordinates $ z^\cK (a) $ for which the structure constants of $ \cR_\Delta $ trivialize, as described in \cite{Hosono:1993qy,Mayr:1996sh}, are also flat coordinates for the Hodge bundles of $ R $ trivializing the structure constants in \eqref{RiemannSurfacePF}, and we find
\begin{equation}
 \frac{\partial}{\partial z^\cL}\frac{\partial}{\partial z^\cK} \, \gamma_b (a(z)) = 0 \, .
\end{equation}
This implies that in the flat coordinates $ \gamma_a(z) $ depend at most linearly on the complex structure moduli $ z^\cK $. Integrating these over a basis of topological one-cycles as introduced in \eqref{oneCycleBase} we obtain the period matrices $ (\hat \Pi^b)_a $ and $ (\hat \Pi_b)_a $ also depending at most linearly on the flat coordinates $ z^\cK $. Therefore we find that the normalized period matrix $ \hat f_{ab}(z) $ can be expanded around the large complex structure point $ z \gg 1 $ as
\begin{equation} \label{fexpand}
 \hat f_{ab} (z)= z^\cK \hat M_{\cK ab} + \hat C_{ab} + \cO(z^{-1}) \, .
\end{equation}
The constants $ \hat M_{\cK ab}, \hat C_{ab} \in \mathbb{C} $ can be determined from boundary conditions as found in \cite{Greiner:2015mdm} for the large complex structure point. There we found that the numbers $ \hat M_{\cK ab} $ correspond to certain intersection numbers of the mirror Calabi-Yau fourfold. The coefficients of \eqref{fexpand} are likely to be further restricted by shift-symmetries of the intermediate Jacobian \eqref{RiemannIntermediateJacobian}.

In the next section we will lift these holomorphic one-forms to the full Calabi-Yau fourfold realized as a semiample hypersurface in a complete simplicial toric variety using the Gysin-map.

\section{The intermediate Jacobian of a Calabi-Yau fourfold hypersurface} \label{FourFoldJacobianSection}

In this section we will lift the previously established theory for Riemann surfaces and their intermediate Jacobian to the three-form cohomology of the full Calabi-Yau fourfold realized as a semiample hypersurface in a simplicial complete toric variety. This will enable us to define the intermediate Jacobian of the Calabi-Yau fourfold hypersurface and express its metric in terms of the normalized period matrices of Riemann surfaces and certain intersection numbers of divisors inducing the three-form cohomology via the Gysin-map.

In the following we consider a smooth and semiample Calabi-Yau fourfold hypersurface $ Y_4 $ embedded in a complete and simplicial toric variety $ \cA_5 $. As we have already seen, in this case the non-trivial three-form cohomology of $ Y_4 $ is induced by the Gysin-map, \eqref{Gysin-square}. This is a topological mapping induced by the inclusion $ \iota_i $ of toric divisors $ D^\prime _i $ into $ Y_4 $ and hence respects the complex structure induced on the cohomology. We have the isomorphism
\begin{equation}
 H^{2,1}(Y_4) \simeq \bigoplus_{\nu^\ast _i \in \Delta^\ast \cap N} H^{1,0}(D^\prime _i) \, .
\end{equation}
In \autoref{divisorSection} we have seen that a toric divisor $ D^\prime _i $ of $ Y_4 $ can only have non-trivial one-forms if it is two-semiample, i.e.~ the integral point $ \nu^\ast _{l_\alpha} $ of $ \Delta^\ast $ defining the toric divisor $ D^\prime _{l_\alpha ^\ast} $ is contained in a two-dimensional face $ \theta_\alpha ^\ast $ of $ \Delta^\ast $, $ \nu^\ast _{l_\alpha} \in \text{int}(\theta^\ast _\alpha) $. These divisors are fibration over Riemann surfaces $ R_\alpha $ defined by the dual face $ \theta_\alpha $ of $ \theta^\ast _\alpha $. The fibers $ E_{l_\alpha} $ of these divisors are connected toric surfaces and can be assumed to not depend on the complex structure of $ Y_4 $, since they are toric. Therefore, we argued that
\begin{equation} \label{ThreeFormSplit}
 H^{2,1}(Y_4) \simeq \bigoplus_{\text{dim}(\theta^\ast _\alpha) =2} \bigoplus_{\nu^\ast _{l_\alpha} \in \text{int}(\theta_\alpha ^\ast)} H^{1,0}(R_\alpha) \otimes_{\mathbb{C}} H^{0,0}(E_{l_\alpha}) \, .
\end{equation}
Since we are now dealing with several Riemann surfaces $ R_\alpha $ with genus $ g_\alpha $ we will denote their respective holomorphic one-forms for which we constructed representatives in the previous section by
\begin{equation}
 \gamma_{a_\alpha} \in H^{1,0}(R_\alpha) \, , \quad a_\alpha = 1, \ldots, g_\alpha = \ell^\prime(\theta_\alpha) \, .
\end{equation}
These lift to holomorphic $ (2,1) $-forms $ \psi_\cA $ of $ Y_4 $ by first pulling them back via $ \pi^\ast _\alpha $ to the full toric divisor $ D^\prime _{l_\alpha} $ with the projection $ \pi_{l_\alpha} : D^\prime _\alpha \rightarrow R_\alpha $ and then pushing forward via the Gysin-map $ \iota_{l_\alpha \ast} $ induced by the inclusion $ \iota_{l_\ast} : D^\prime _{l_\alpha} \rightarrow Y_4 $. Explicitly this is given by
\begin{align} \label{psiAnsatz}
 \psi_\cA &= \iota_{l_\alpha \ast} (\pi^\ast _{l_\alpha} \gamma_{a_\alpha}) \, , \\ 
 \cA &= (\alpha, l_\alpha, a_\alpha)= (1,1,1) , \ldots, (k_2, \ell^\prime(\theta^\ast_\alpha), \ell^\prime(\theta_\alpha) ) \, , \nonumber
\end{align}
where we made use of a multi-index $ \cA $ keeping track of the base Riemann surface $ R_\alpha $ the toric fiber $ E_{l_\alpha} $ and the holomorphic one-form $ \gamma_{a_\alpha} \in H^{1,0}(R_\alpha) $. Recall that $ k_2 $ is the number of two-dimensional faces $ \theta^\ast _\alpha $ of $ \Delta^\ast $ and hence counts the Riemann surfaces $ R_\alpha $. The projection $ \pi_{l_\alpha} $ as well as the inclusion $ \iota_{l_\alpha} $ are both topological maps independent of the complex structure or the metric of $ Y_4 $. Therefore, also a basis of topological integral one-forms $(\hat \alpha_{a_\alpha}, \hat \beta^{a_\alpha})$ as introduced in \eqref{oneCycleBase} for each $ R_\alpha $ will be pushed forward to a basis of topological three-forms $ (\alpha_\cA, \beta^\cA) $ of $ H^3(Y_4, \mathbb{Z}) $, via the same construction
\begin{equation} \label{ThreeFormAlphaBeta}
 \big( \alpha_\cA =  \iota_{l_\alpha \ast} (\pi^\ast _{l_\alpha} \hat \alpha_{a_\alpha}) \, , \,  \beta^\cA =  \iota_{l_\alpha \ast} (\pi^\ast _{l_\alpha} \hat \beta^{a_\alpha}) \big) \in H^3(Y_4, \mathbb{Z}) \, .
\end{equation}
These also satisfy the canonical properties of \eqref{oneCycleBase} which will simplify the upcoming discussion. Note that we do only consider here integral cohomology without torsion. This implies that the intermediate Jacobian $ \cJ^3(Y_4) $ also splits as a topological space using \eqref{ThreeFormSplit} into a direct product of intermediate Jacobian of Riemann surfaces as
\begin{equation}
 \cJ^3(Y_4) = \frac{H^{2,1}(Y_4)}{H^3(Y_4, \mathbb{Z})} \simeq \prod_{\alpha = 1} ^{k_2} (\cJ^1(R_\alpha)) ^{\ell^\prime (\theta^\ast_\alpha)} \, . 
\end{equation}
This is the intermediate Jacobian we already introduced in \autoref{threeFormAnsatz}. In the upcoming discussion we will derive the normalized period matrix $ f_{\cA \cB}(z) $ of $ Y_4 $, which can be seen to be a block-diagonal matrix with the blocks being the normalized period matrices of the Riemann surfaces $ R_\alpha $. Due to the direct sum in \eqref{ThreeFormSplit} these blocks are independent. At special points in complex structure, however, the lattice in $ H^{2,1}(Y_4) $ induced by $ H^3(Y_4,\mathbb{Z}) $ may degenerate. This will require an extension of the diagonal ansatz we consider here and we hope to come back to this situation in the future. In this work we will restrict ourselves to the non-degenerate case. As we have already seen in \autoref{threeFormAnsatz}, we can find a positive quadratic form $ Q $ on the intermediate Jacobian $ \cJ^3(Y_4) $ given by
\begin{align} \label{Qdef}
 Q(\psi_\cA, \psi_\cB) &= \int_{Y_4} \psi_\cA \wedge \ast \bar \psi_\cB \\ 
 &= -i v^\Sigma \int_{Y_4} \omega_\Sigma \wedge \psi_\cA \wedge \bar \psi_\cB \, , \quad \psi_\cA, \psi_\cB \in H^{2,1}(Y_4) \, . \nonumber 
\end{align}
Here we inserted the expansion of the K\"ahlerform $ J = v^\Sigma \omega_\Sigma $ into real harmonic two-forms $ \omega_\Sigma \in H^{1,1} (Y_4) $ as already introduced in \eqref{KaehlerFormExpand}. The two-forms $ \omega_\Sigma $ can be chosen to be Poincar\'e dual to a set of homologically independent divisors $ D^\prime _\Sigma $ of $ Y_4 $. As we have seen around \eqref{nonToricDivisors} we can assume these divisors to be induced by toric divisors of the ambient space and for simplicity we will assume that all the toric divisors of $ Y_4 $ are connected. Therefore, we can also obtain the non-trivial two-forms as a push-forward of the generator $ 1_\Sigma $ of the constant functions on the toric divisors $ D^\prime _\Sigma $ via the Gysin map as
\begin{equation}
 \omega_\Sigma = \iota_{\Sigma \ast} (1_\Sigma) \in H^{1,1} (Y_4) \, , \quad 1_\Sigma \in H^{0,0}(D^\prime _\Sigma) \, .
\end{equation}
In the case where the toric divisor $ D^\prime _\Sigma $ has several components, $ H^{0,0}(D^\prime _\Sigma) $ has several generators, one for each component of $ D^\prime _\Sigma $. The generalization to non-connected toric divisors is straightforward, but will just clutter up the notation. Let us now evaluate $ Q $ for the constructed $ (2,1) $-forms of \eqref{psiAnsatz}.

\begin{figure}
\begin{center}
\setlength{\unitlength}{0.7cm}
\begin{picture}(12,12)
\put(-1,0){\includegraphics[height=8.4cm]{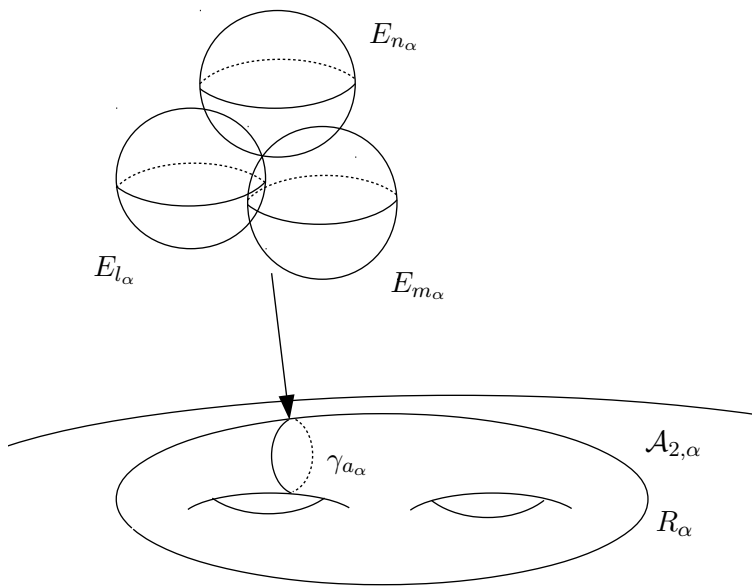} }
\put(11.2,1.5){$ R_\alpha $}
\put(11,3.0){$ \cA_{2,\alpha} $}
\put(5.0,2.7){$ \gamma_{a_\alpha} $}
\put(0.6,6.3){$ E_{l_\alpha} $}
\put(6.2,6.0){$ E_{m_\alpha} $}
\put(5.8,10.7){$ E_{n_\alpha} $}
\end{picture}
\caption{\textit{Intersection structure of the divisors $ D^\prime _{l_\alpha} $ that are fibration over $ R_\alpha $ with fiber $ E_{l_\alpha} $ and holomorphic one-forms $ \gamma_{a_\alpha} $.}}
\label{intersection_structure_picture}
\end{center}
\end{figure}

The key insight that will allow us to reduce to $ Q $ to quantities accessible to calculation is to analyze the intersection structure implied by the integral in \eqref{Qdef}. Since all three of the appearing forms $ \omega_\Sigma, \psi_\cA, \psi_\cB $ arise from toric divisors $ D^\prime _\Sigma, D^\prime _{l_\alpha}, D^\prime _{l_\beta} $ it is natural to consider the intersection of these divisors which will result in a curve $ \cC $ in $ Y_4 $
\begin{equation}
 \cC = D^\prime _\Sigma \cap D^\prime _{l_\alpha} \cap D^\prime _{l_\beta} \subset Y_4 \, .
\end{equation}
Evaluating the integral of \eqref{Qdef} will hence only have a non-vanishing result if the homology class of $ \cC $ is the same as $ R_\alpha $ and $ R_\beta $, respectively. This can be deduced from the fact that the three-forms $ \psi_\cA $ are induced by one-forms $ \gamma_{a_\alpha} $ that have support on $ R_\alpha $. Another way to see this is that $ \cC $ is again a hypersurface in the toric ambient space $ D_\Sigma \cap D_{l_\alpha} \cap D_{l_\beta} $. If the one-forms on this $ \cC $ lift to $ Y_4 $ the hypersurface has to be two-semiample which requires $ \cC $ to be homologous to one of the Riemann surfaces $ R_\alpha $. Note that this also implies $ D^\prime _\Sigma $ to be two-semiample which is a severe restriction on the number of K\"ahler moduli $ Q $ can depend on. Consequently, all three divisors are fibrations over the same $ R_\alpha $, but with possibly different fibers $ E_{l_\alpha} $ corresponding to integral points of $ \text{int}(\theta^\ast _\alpha) $. Therefore the intersections have the form
\begin{equation}
 D^\prime _{l_\alpha} \cap D^\prime _{n_\alpha} \cap D^\prime _{m_\alpha} = \hat M_{l_\alpha n_\alpha m_\alpha} \cdot R_\alpha \, .
\end{equation}
The intersection numbers $ \hat M_{l_\alpha n_\alpha m_\alpha} $ account for possible multi-components of the intersection. These intersection numbers can be computed in the so called generalized Hirzebruch-Jung Sphere-Tree, $ \bigcup_{l_\alpha} E_{l_\alpha} $, which is the union over all fibers $ E_{l_\alpha} $ at a point of $ R_\alpha $. Two components of this sphere-tree only intersect in codimension one subvarieties. The intersection numbers $  \hat M_{l_\alpha n_\alpha m_\alpha}  $ are given by
\begin{equation}
 E _{l_\alpha} \cap E _{n_\alpha} \cap E_{m_\beta} = \hat M_{l_\alpha n_\alpha m_\alpha} \, .
\end{equation}
We depicted the intersection structure in figure \ref{intersection_structure_picture}.
In practice we can calulate these intersection numbers from the intersection numbers of $ Y_4 $ by taking an intersection of four toric divisors with three of them as above fibrations over $ R_\alpha $ and the fourth one being transverse to $ R_\alpha $ intersecting it in a single point (or several points, but then we have to take multiplicities into account). The easiest choice for the fourth divisor is the dual of $ R_\alpha $, we can call it $ D_\alpha $, which satisfies $ D_\alpha \cdot R_\alpha = 1 $. Therefore, we find that
\begin{equation} \label{MDivisorDef}
  \hat M_{l_\alpha n_\alpha m_\alpha} = D^\prime _{l_\alpha} \cap D^\prime _{n_\alpha} \cap D^\prime _{m_\alpha} \cap D_\alpha \,  , \quad D_\alpha \cdot R_\alpha = 1 \, .
\end{equation}
This implies that we can compute the intersection numbers $ \hat M_{l_\alpha n_\alpha m_\alpha} $ by standard techniques, for example by calculating the intersection numbers of five divisors of the ambient space $ \cA_5 $ and then choosing the fifth divisor the anti-canonical divisor class of the Calabi-Yau fourfold $ Y_4 $. From this analysis we see that there is a convenient expansion of the K\"ahlerform suited to calculate $ Q $.
\begin{equation} \label{JforThreeForm}
 J = v^\Sigma \omega_\Sigma = \sum_{\alpha = 1} ^{n_2} \sum_{l_\alpha} v^{\l^\alpha} \omega_{l_\alpha} + \ldots \, , \quad  \omega_{l_\alpha}  = \text{PD} [D_{l_\alpha}] \, .
\end{equation}
Here we only displayed the expansion in two-forms that will contribute to $ Q $ and used Poincar\'e duality to relate the divisor classes $ [D_{l_\alpha}] $ and two-forms $ \omega_{l_\alpha} \in H^{1,1}(Y_4) $ in the natural way. Combining now the insights on the intersection structure, we can express the positive bilinear form $ Q $ in a part depending on the intersection pattern independent of the complex structure and a second part only depending on the positive bilinear form on a Riemann surface
\begin{equation} \label{QFromDivisors}
 Q(\psi_\cA, \psi_\cB) = -i \delta_{\alpha \beta} v^{l_\alpha} \hat M_{l_\alpha m_\alpha n_\beta} \int_{R_\alpha} \gamma_{a_\alpha} \wedge \bar \gamma_{b_\beta} \, ,
\end{equation}
where we used the multi-indices $ \cA = (\alpha, m_\alpha, a_\alpha) $ and $ \cB = (\beta, n_\beta, b_\beta) $. Geometrically this can be interpreted in the following picture. For fixed Riemann surface $ R_\alpha $, the $ E_{l} $-fibers form a generalized Hirzebruch-Jung sphere-tree, which is usually used for the resolution of codimension two orbifold singularities, where a chain of $ \mathbb{P}^1 $'s (the spheres of the tree) are fibered over the singularity locus. The intersection matrix of these spheres is then related to the symmetry group of the orbifold. We use here the resolution of codimension three singularities requiring complex two-dimensional resolution fibers $ E_l $ that intersect in a more complicated pattern which is captured by the fully symmetric three-tensor $ M_{lmn} $. The intersection number $ M_{lmn} $ are independent of the hypersurface and can also be computed directly in the ambient space geometry $ \cA_5 $, but depends well on the triangulation of $ \Delta^\ast $ as well as the K\"ahler moduli. The codimension three-singularity we also assume to be an orbifold singularity and it would be interesting to find a group theoretic interpretation of the intersection matrices $ M_{lmn} $.

Let us now connect the result for $ Q $ in \eqref{QFromDivisors} to the general formula of \autoref{threeFormAnsatz} given by \eqref{HpsiDef}. Therefore, we need to identify the intersection numbers \eqref{Mdef} with \eqref{MDivisorDef} as
\beq \label{McAcB}
    {M_{\Sigma \cA}}^\cB = \left\{ \begin{array}{ccc} \hat M_{l_\alpha m_\alpha n_\alpha} \delta_{a_\alpha}^{b_\alpha} & & \text{for} \, \, \alpha =\beta\ \text{and} \ \Sigma= l_\alpha \\[.1cm]
                                                                                      0 & & \text{otherwise}\, ,\end{array} \right. 
\eeq
with mulit-indices $\cA=(\alpha,m_\alpha,a_\alpha)$ and $\cB=(\beta,n_\beta,b_\beta)$. The second intersection number $ {M_\Sigma}^{\cA \cB} $ vanishes,
\begin{equation} \label{McBetaBeta}
  {M_\Sigma}^{\cA \cB} = 0 \, ,
\end{equation}
since we showed that we can choose a canonically normalized basis of topological three-forms as in \eqref{ThreeFormAlphaBeta} induced by the canonical basis of one-cycles \eqref{oneCycleBase} on the Riemann surfaces $ R_\alpha $. This implies that the bilinear form $ Q $ will only depend on the real part of the normalized period matrix $ f_{\cA \cB} $ as in \eqref{QFromDivisors} in contrast to the general form \eqref{HpsiDef}.

As already mentioned, the normalized period matrix of \eqref{fThreeFormDef} has on its diagonal the normalized period matrices $ \hat f^{(\alpha)} _{a_\alpha b_\alpha} $ of the Riemann surfaces $ R_\alpha $.
This can be read off by inserting \eqref{fhat} into \eqref{QFromDivisors}
\begin{equation} \label{QpsiFinal}
 Q(\psi_\cA, \psi_\cB) = 2 \, \delta_{\alpha \beta} v^{l_\alpha} \hat M_{l_\alpha m_\alpha n_\beta} \text{Re} \, \hat f^{(\alpha)} _{a_\alpha b_\alpha} \, ,
\end{equation}
and comparing to \eqref{HpsiDef}. The precise identification is given by
\beq \label{f_final}
     f_{\cA\cB} = \left\{ \begin{array}{ccc} \hat f_{a_\alpha b_\alpha}^{(\alpha)} \delta_{m_\alpha n_\alpha} & & \text{for} \, \, \alpha =\beta\ \\[.1cm]
                                                                                      0 & & \text{otherwise}\, , \end{array} \right.
\eeq
where we used again the multi-indices $\cA=(\alpha,m_\alpha,a_\alpha)$ and $\cB=(\beta,n_\beta,b_\beta)$.

Combining the identifications \eqref{McAcB}, \eqref{McBetaBeta} and \eqref{f_final} with the results on Picard-Fuchs equations of \eqref{RiemannSurfacePF} we found the quantities of \autoref{threeFormAnsatz} relevant for the three-form moduli dynamics on Calabi-Yau fourfold hypersurfaces in toric varieties. The normalized period matrix $ f_{\cA \cB} $ shows a block structure, one block for each Riemann surface that serves as a basis of several toric divisors giving rise to the non-trivial three-forms on the hypersurface. The relations of the divisors forming a generalized sphere tree fibered over a fixed Riemann surface are encoded by the topological intersection number $ {M_{\Sigma \cA}} ^{\cB} $ that obtain a similar block-structure. Some of the important physical quantities relevant for effective theories can be computed with this information as was done in \cite{Grimm:2014vva,Carta:2016ynn}. In contrast, some applications require a more general class of geometries. The non-Abelian structures discussed in \cite{Corvilain:2016kwe,Grimm:2015ona,Grimm:2015ska} are not covered by the block-diagonal structures appearing in the hypersurface scenario and are likely to make an extension of our considerations to complete intersections necessary. This is an interesting topic that will provide exciting opportunities for the next generation of students.

In the next section we will illustrate the previously encountered structures on the simplest non-trivial examples of Calabi-Yau fourfold hypersurfaces in simplicial complete toric ambient geometries. We will consider Fermat hypersurfaces in weighted projective spaces.

\section{Calabi-Yau fourfold hypersurfaces in weighted projective spaces} \label{WeightedProjectiveSpaceSection}

We end our discussion of the construction of three-form periods on Calabi-Yau fourfold hypersurfaces by giving simple examples. A particular well-suited class of geometries are Fermat  hypersurfaces in weighted projective spaces providing both simplicity as well as non-trivial features illustrated in the previous sections. These are examples of hypersurfaces in toric varieties and hence we can apply what we learned before directly in this context. The discussion here lays the groundwork for our examples we will discuss in the following section that also fit into this scheme.

The focus of our discussion will be on the calculation of the normalized period matrix $ f_{\cA \cB} $ of \eqref{f_final} that was shown to only depend on Riemann surfaces that serve as bases for fibrations of blow-up divisors resolving singularities along the Riemann surfaces. Therefore it is not necessary to blow-up the singularities and explicitly resolve them for the calculation of $ f_{\cA \cB} $, which is very practical to simplify our analysis. For the derivation of the intersection numbers $ {M_{\Sigma \cA}} ^{\cB} $ as in \eqref{McAcB} the structure of these blow-ups is crucial. We close this section by specifying to geometries whose three-forms are all induced by a single toric divisor and hence by only one Riemann surface.

A weighted projective space $ \cA_D = \mathbb{P}^D(w_1, \ldots, w_{D+1}) $ is a simplicial complete toric space whose geometry is determined by its weights $ w_i \in \mathbb{N} $. These spaces are in general singular and require blow-ups to resolve these singularities. For our ambient space $ \cA_5 $ we will consider a weighted projective space with one weight $ w_6 = 1 $ whose defining polyhedron $ \Delta^\ast \subset \mathbb{Q}^5 = N_{\mathbb{Q}} $ can be represented by the simplex with the six vertices
\begin{equation}
 \nu^\ast _i = e_i  \in \mathbb{Z}^5 \, , \quad i = 1, \ldots, 5 \, , \quad \nu^\ast _6 = (-w_1, -w_2, -w_3, -w_4, -w_5) \in \mathbb{Z}^5 \, .
\end{equation}
These vertices are integral, i.e.~ elements of the lattice $ N = \mathbb{Z}^5 $ with generators the unit vectors $ e_i $. Choosing $ w_6 = 1 $ allows us to relate the toric divisor classes $ D_i $ corresponding to the vertices to each other, as all of these divisors are rational equivalent to multiples of the divisor $ D_6 = H $ as
\begin{equation}
 [D_i] = w_i [H] \, .
\end{equation}
The divisor class $ [H] $ can be interpreted as a generalization of the hyperplane class of classical projective spaces. The homogeneous coordinate ring of $ \cA_5 $
\begin{equation}
 S_5 = \mathbb{C}[X_1, \ldots, X_6]
\end{equation}
has therefore the usual grading of a monomial by a positive number
\begin{equation}
 \text{deg}_{S_5} \big( \prod_i X_i ^{k_i}\big) = \sum_i w_i k_i \in \mathbb{Z}_{\geq 0} \, .
\end{equation}
In particular, we can choose globally quasi-homogeneous coordinates, denoted by
\begin{equation}
 [X_1: \ldots : X_6] =  [\lambda^{w_1} X_1: \ldots : \lambda^{w_6} X_6] 
\end{equation}
that are invariant under a rescaling by a non-zero factor $ \lambda \in \mathbb{C} - \{ 0 \}  $.

The polynomial $ p_\Delta \in S_5 (-K_{\cA_5}) $ defining the anti-canonical hypersurface $ Y_4 ^{sing} \subset \cA_5 $ has therefore degree $ d $ determined by
\begin{equation}
 -K_{\cA_5} = \sum_i [D_i] = \big(\sum_i w_i \big) [H] \, , \quad d = \sum_i w_i \, .
\end{equation}
A polynomial $ p $ is called of Fermat type, if it is the sum of monomials containing only one variable $ X_i $ raised to certain power $ k_i $. Schematically this reads
\begin{equation} \label{Fermat}
 p_{Fermat} = \sum_i X_i ^{k_i} \, , \quad \Rightarrow \quad k_i = d / w_i \in \mathbb{N} \, ,
\end{equation}
where we indicated already the condition on the degree $ d $ of our anti-canonical hypersurface $ p_\Delta $. Choosing $ p_\Delta $ a deformation of a Fermat type polyhedron, we can represent the dual polyhedron $ \Delta $ as a simplex in $ \mathbb{Q}^5 = M_\mathbb{Q} $ with integral vertices $ \nu_i \in \mathbb{Z}^5 = M $ corresponding to the monomials of \eqref{Fermat}. They are given by
\begin{equation}
 \nu_i = - \sum_j e_j + \frac{d}{w_i} e_i \in \mathbb{Z}^5 \, , \quad i = 1, \ldots, 5 \, , \quad \nu_6 = - \sum_j e_j \in \mathbb{Z}^5 \, .
\end{equation}
We denoted again by $ e_i $, $ i= 1, \ldots, 5 $ the unit vectors generating the lattice $ \mathbb{Z}^5 \subset \mathbb{Q}^5 $. Note that the assumptions that we can find a anti-canonical hypersurface that allows for a Fermat representative of its class is a very restrictive assumption that will, however, simplify our calculations considerably. In general $ \Delta $ will not be a simplex and have more vertices. We will in the following consider small deformations of Fermat hypersurfaces that will make them non-degenerate. Pure Fermat hypersurfaces have a high degree of symmetry and hence will introduce orbifold singularities that do not stem from the ambient toric space. Small deformations by monomials in $ p_\Delta $ will resolve these singularities. Therefore, we consider $ p_\Delta $ of the form
\begin{equation} \label{FermatPDelta}
 p_\Delta = \sum_i X_i ^{k_i} + \sum_{\nu \in \theta, \text{codim}(\theta) > 1} a_\nu p_\nu
\end{equation}
where we have chosen a set of inequivalent deformations by restricting to monomials that correspond to integral points $ \nu $ not contained in vertices or edges of $ \Delta $. These monomials can be reabsorbed by linear coordinate redefinitions and hence are equivalent to the remaining monomials up to derivative $ \partial_i p_\Delta $. We will denote a degree $ d $ hypersurface in the weighted projective space by $ \mathbb{P}^D (w_1, \ldots, w_{D+1})[d] $. In particular we will consider
\begin{equation}
 Y_4 ^{sing} = \mathbb{P}^D (w_1, \ldots, w_5, w_6 = 1)[d] \, , \quad w_i \, | \, d \, , \quad d = \sum_i w_i \, .
\end{equation}

For the resolved smooth fourfold $ Y_4 $ to have non-trivial three-form cohomology, we need $ Y_4 ^{sing} $ and hence also $ \cA_5 $ to have codimension three orbifold singularities. A toric surface $ \cA_2 $ of $ \mathbb{C}^3/\mathbb{Z}_n $ singularities in the ambient space $ \cA_5 $ exists if and only if exactly three of the six weights $ w_i $ have a common divisor $ n $. Up to renaming the coordinates we can assume that
\begin{equation} 
 n \, | \, w_3, w_4, w_5 \, , \quad n \not | \, w_1, w_2, w_6 \, .
\end{equation}
Therefore $ \cA_2 $ is the subspace of $ \cA_5 $ given by $ X_1 = X_2 = X_6 = 0 $ which is the intersection of three toric divisors
\begin{equation} \label{A2def}
 \cA_2 = D_1 \cap D_2 \cap D_6 \subset \cA_5 \, .
\end{equation}
The hypersurface $ Y_4 ^{sing} $ will intersect $ \cA_2 $ in general transversely and hence inherit the $ \mathbb{C}^3/\mathbb{Z}_n $ singularities along a Riemann surface $ R $. To resolve the hypersurface $ Y_4 ^{sing} $ we will need to blow-up the ambient space several times in general. This will introduce the toric divisors that give rise to the three-forms of $ Y_4 $, their complex structure dependence is, however, completely captured by $ R \subset Y_4 ^{sing} $ which justifies to work with the singular geometry in order to obtain the normalized period matrix $ f_{\cA \cB} $.

In the special situation of Fermat hypersurfaces of degree $ d $ in the weighted projective space $ \cA_5 $, it is easy to see that $ n $ divides $ d $, $ n \, | \, d $ for $ n $ the order of the cyclic orbifold group $ \mathbb{Z}_n $. From \eqref{A2def} it is easy to see that $ \cA_2 $ is also a weighted projective space
\begin{equation}
 \cA_2 = \mathbb{P}^2(w_3, w_4, w_5) \, , \quad S_2 = \mathbb{C}[X_3,X_4,X_5] \, .
\end{equation}
From the homogeneous coordinate ring $ S_2 $ and the fact that $ X_3, X_4, X_5 $ have a common divisor $ n $, it is easy to see that we can represent $ \cA_2 $ also as
\begin{equation}
 \cA_2 \simeq \mathbb{P}^2(w_3 / n, w_4 / n, w_5 / n) \, ,
\end{equation}
since they have the same homogeneous coordinate ring $ S_2 = \mathbb{C}[X_3,X_4,X_5] $. Restricting $ p_\Delta $ to $ \cA_2 $ produces $ p_\theta $ where $ \theta $ is the face with vertices $ \nu_3, \nu_4, \nu_5 $. The restricted polynomial $ p_\theta $ has the same degree $ d $ and therefore, we can represent the Riemann surface $ R $ of singulalarites in $ Y_4 ^{sing} $ by
\begin{equation}
 R = \mathbb{P}^2(w_3 / n, w_4 / n, w_5 / n) [d/n] \, .
 \end{equation}
 This is still a Fermat hypersurface, since we obtain it from \eqref{FermatPDelta} by setting $ X_1 = X_2 = X_6 = 0 $ and this can be written as
 \begin{align}
  p_\theta &= X_3 ^{k_3} + X_4 ^{k_4} + X_5 ^{k_5}  \\
  &\quad + X_3 X_4 X_5 \big( \sum_{\text{deg}(p^\prime _b) = w_1 + w_2 + w_6} a_b \, p^\prime _b (X_3, X_4, X_5)\big) \, , \nonumber 
 \end{align}
 where we made contact with the notation introduced in \autoref{RiemannSection} using the monomials
\begin{equation}
 p^\prime _a \in S_2(-K_{\cA_5}|_{\cA_2} + \cA_2) = S_2 (w_1 + w_2 + w_3) \, , \quad \nu_a \in \text{int}(\theta) \cap M \, ,
\end{equation}
whose equivalence classes in $ \cR_\theta = S_2 / \langle \partial_i p_\theta \rangle $ form a basis of the monomials of degree $ w_1 + w_2 + w_6 $ up to linear transformations. These monomials correspond to the integral interior points $ \nu_a \in \text{int}(\theta) \cap M $. Recall that the number of these interior points $ \ell^\prime (\theta) $ is the genus $ g $ of the Riemann surface $ R $.

With construction of the holomorphic one-forms of $ R $ outlined in \autoref{RiemannSection} in mind, we will now proceed with the holomorphic volume form of $ \cA_2 $ defined in \eqref{hol_volume_form}. This simplifies in the situation of $ \cA_2 = \mathbb{P}^2(w_3, w_4, w_5) $ drastically to
\begin{equation} \label{weightedHolomorphicVolumeForm}
 d \omega_{\cA_2} = w_3 X_3 dX_4 \wedge dX_5 - w_4 X_4 dX_3 \wedge dX_5 + w_5 X_5 dX_3 \wedge dX_4 \, \in \Omega^2 _{\cA_2} \, ,
\end{equation}
which has degree $ w_3 + w_4 + w_5 $ or in terms of divisors classes $ -K_{\cA_2} = (w_3 + w_4 + w_5) [H] $. This holomorphic volume-form enables us to define the global meromorphic two-forms of $ \cA_2 $ with first order poles along $ R $ as
\begin{equation}
 \frac{p^\prime _a}{p_\theta} \, d \omega_{\cA_2} \, \in H^0 (\cA_2, \Omega_{\cA_2} ^2 (R)) \, .
\end{equation}
These forms have degree zero, therefore they are independent of the quasi-projective equivalences of the weighted projective space and hence they are globally well-defined on $ \cA_2 $. The first order pole facilitates the residue construction which enables us to associate to the monomials $ p^\prime _a $ the holomorphic $ (1,0) $-forms
\begin{equation}
 \gamma_a = \int_\Gamma \frac{p^\prime _a}{p_\theta} \, d \omega_{\cA_2} \, \in H^{1,0}(R) \, , \quad  \nu_a \in \text{int}(\theta) \cap M \, ,
\end{equation}
where $ \Gamma $ is as usual a small one-dimensional curve winding around $ R $ in $ \cA_2 $. To derive the Picard-Fuchs equations of \eqref{RiemannSurfacePF}, we need to apply the relations in the Jacobian ring $ \cR_\theta $ of the Riemann surface to relate the second derivatives of $ \gamma_a $ with the respect to the complex structure moduli $ a_b $ to its lower derivatives. This is, however, connected with a tremendous amount of work (at least $ \cO(g^2) $) for which an adapted algorithm needs to be found and implemented into a computer program. We will outline the simplest case $ g=1 $ in the upcoming section.

For general orbifold singularities along a curve $ R $ in a toric Calabi-Yau fourfold hypersurface $ Y_4 $, the toric blow-up divisors will intersect in complicated patterns, we need to understand, how to calculate the intersection numbers $ {M_{\Sigma \cA}}^\cB $ of \eqref{McAcB}. The number of three-tensors grows with $ \cO(\ell^\prime (\theta^\ast) ^3) $, where $ \ell^\prime (\theta^\ast) $ is the number of toric divisors necessary to resolve the orbifold singularity along $ R $.

\begin{figure}
\begin{center}
\setlength{\unitlength}{0.7cm}
\begin{picture}(12,12)
\put(0,0){\includegraphics[height=8.4cm]{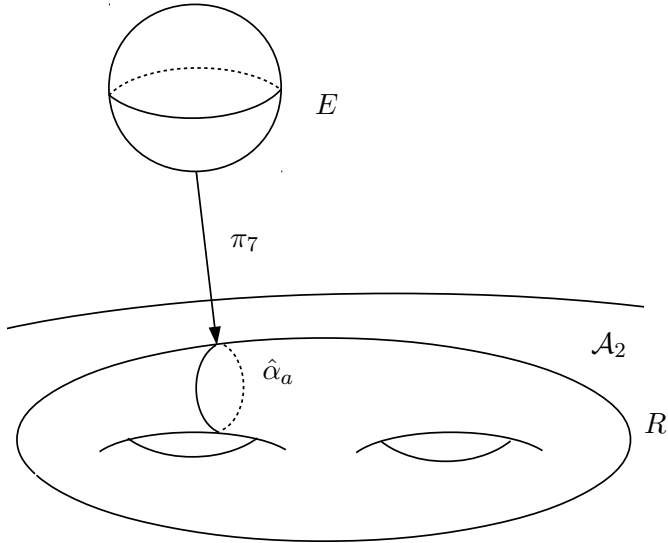} }
\put(12.0,3){$ R $}
\put(11,4.5){$ \cA_2 $}
\put(4.2,6.5){$\pi_{7} $}
\put(4.8,4.0){$\hat \alpha_a $}
\put(5.8,9.0){$ E $}
\end{picture}
\caption{\textit{Fibration structure of $ D^\prime _7 $. The Riemann surface $ R $ is a hypersurface of the toric space $ \cA_2 $ over which the toric surface $ E $ is fibered.}}
\label{fibration_structure_picture}
\end{center}
\end{figure}

To illustrate the concept, let us consider the simplest codimension-three orbifold singularity $ \mathbb{C}^3/\mathbb{Z}_3 $ leading to $ n = 3 $. This simple singularity can be resolved by a single toric blow-up, with coordinate $ X_7 $ and divisor $ D^\prime _7 = \{ X_7 = 0 \} $. This will be a fibration over a Riemann surface $ R $ as before, but with only a single exceptional fiber $ E $. The divisor $ D^\prime _7 $ corresponds to the integral interior point $ \nu^\ast _7 $ of the face $ \theta^\ast $ spanned by $ \nu^\ast _1 , \nu^\ast _2 , \nu^\ast _6 $ given by
\begin{equation}
 \nu^\ast _7 = \frac{1}{3} (\nu^\ast _1 + \nu^\ast _2 + \nu^\ast _6) \, .
\end{equation}
Following the construction of \autoref{divisorSection} one can show that the exceptional fiber $ E $ of $ D^\prime _7 $ is
\begin{equation}
 E = \mathbb{P}^2 (w_1, w_2, w_6) \, .
\end{equation}
The fiber $ E $ is for general $ w_1, w_2 $ not smooth, but resolving these singularities will not lead to new three-forms, as only $ H^{0,0}(E) $ contributes in the formula \eqref{ThreeFormSplit}, and this is unaffected by blow-ups.

In the special situation of a single divisor $ D^\prime _7 $ inducing the non-trivial three-forms $ \psi_\cA $ on the smooth fourfold $ Y_4 $ the discussion of \autoref{FourFoldJacobianSection} simplifies a lot. First, since we have only one divisor $ D^\prime _7 $ with fiber $ E $ the intersection numbers $ {M_{\Sigma \cA}}^\cB $ of \eqref{McAcB} reduce to a single number $ M $. This number $ M $ is for a smooth ambient space a positive integer, depending on normalization this can be set to one $ M=1 $. The holomorphic three-forms $ \psi_\cA $ correspond in this scenario exactly to the holomorphic one-forms $ \gamma_a $ of the Riemann surface $ R $ which is the base of the fibration of $ D^\prime _7 $. The multi-index $ \cA = (\alpha, l_\alpha, a_\alpha) = (1,7,1), \ldots, (1,7,g) $ of $ \psi_\cA $ can be replaced here by the index $ a = 1, \ldots, g $ labeling the holomorphic one-forms $ \gamma_a \in H^{1,0}(R) $. The formula \eqref{psiAnsatz} relating one-forms of $ R $ with three-forms of $ Y_4 $ reads here
\begin{equation}
 \psi_\cA = \iota_{7 \ast} (\pi^\ast _7 \gamma_a) \,  \in H^{2,1}(Y_4) \, , \quad a = 1, \ldots, g \, .
\end{equation}
For the positive bilinear form $ Q $ of \eqref{QpsiFinal} we calculate
\begin{equation}
 Q(\psi_\cA, \psi_\cB) = 2 \, v^7 M \cdot \text{Re} \, \hat f_{ab} \, ,
\end{equation}
where $ \hat f_{ab} $ is the normalized period matrix \eqref{fhat} of $ R $ and $ v^7 $ the volume modulus of $ D^\prime _7 $ in the expansion of the K\"ahler form $ J $ as we have seen in \eqref{JforThreeForm}. We conclude this section by a depiction of the encountered fibration structure of $ D^\prime _7 $ as seen in \autoref{fibration_structure_picture}. This should serve the reader as a guideline for the next section, where we will discuss explicit examples of Calabi-Yau fourfold hypersurfaces in weighted projected spaces of Fermat type.

\chapter{Calabi-Yau fourfold examples} \label{ExampleSection}

In the this section we will apply the previously introduced concepts to construct two simple examples of Calabi-Yau fourfold with non-trivial fourfold cohomology. These two geometries are elliptically fibered and hence allow to serve as a background for an effective description of F-theory in four dimensions. We will highlight the implications of non-trivial three-form cohomology in these F-theory geometries and focus especially on the weak coupling limit following Sen. Tracing the three-form moduli and their couplings through the F-theory and weak coupling limits we strengthen the case for the necessity to extend the weakly coupled Type IIB orientifold framework to strongly coupled regions in complex structure space provided by F-theory compactifications.

\section{General aspects}

In order to describe the F-theory examples in the upcoming sections, we first need to introduce some general aspects of F-theory on elliptically fibered Calabi-Yau fourfolds realized as hypersurfaces in weighted projective spaces as we discussed in \autoref{WeightedProjectiveSpaceSection}. In the four-dimensional effective theory we will find that the three-form modulie $ N_\cA $ yield complex scalar fields for general hypersurfaces in toric varieties. In a general F-theory compactifications these can have two possible origins. First, the NS-NS and R-R two-forms can have non-trivial zero-modes and for a second, the seven-branes may have continuous Wilson-line moduli. These two types of moduli are in general indistinguishable on a Calabi-Yau fourfold as they both arise in the same way as we have seen before. They can only be separated after performing the F-theory limit and the weak coupling limit which we will describe in detail. After introducing the general aspects we will discuss simple examples that exhibit both types of moduli.

\subsection{Weierstrass-form and non-trivial three-form cohomology}

We start our discussion be specifying to a possibly singular Calabi-yau hypersurface $ Y_4 ^{sing} $ in a weighted projective space
\begin{equation}
 Y_4 ^{sing} \subset \cA_5 = \mathbb{P}^5(w_1,\ldots, w_5, w_6 = 1) \, ,
\end{equation}
as already discussed in \autoref{WeightedProjectiveSpaceSection}. Furthermore, we will assume that we can find a resolution by toric blow-ups of the ambient space $ \cA_5 $ to a toric ambient space $ \hat \cA_5 $ with at most point-singularities that will have a anti-canonical hypersurface $ Y_4 $ that is in general smooth and has a non-trivial three-form cohomology. As already discussed in \cite{Batyrev:1994hm} such a resolution is not always possible for Calabi-Yau fourfolds, which is in contrast to the threefold situation where resolutions always exist. We made the complex structure dependence of the three-forms explicit in previous sections and we can derive this dependence from the induced complex structure variations of holomorphic one-forms on Riemann surfaces along which we have orbifold singularities in $ Y_4 ^{sing} $. To discuss the three-form cohomology it will only be necessary to be able to resolve the singularities along these Riemann surfaces.

In order to have a valid F-theory background we will restrict our considerations in the following to elliptically fibered Calabi-Yau fourfolds with a section. To do so, we use a so called Weierstrass-models with the elliptic fiber a hypersurface in the weighted projective space $ \cA_{fiber} = \mathbb{P}^2(2,3,1) $. The full five-dimensional space $ \cA_5 $ will be a fibration with fiber $ \cA_{fiber} $ over a toric basis $ B_3 $ that will be a (blow-up of a) weighted projective space.
\begin{equation}
 B_3 ^{sing} = \mathbb{P}^3(w_1,w_2,w_3,w_6 = 1) \, .
\end{equation}

Allowing $ B_3 $ to be a blow-up of a weighted projective space enables us to consider for example generalized Hirzebruch surfaces that are $ \mathbb{P}^1 $-fibrations over two-dimensional toric varieties. Due to the basis $ B_3 $ of the fibration being toric it cannot carry a non-trivial three-form cohomology itself. Furthermore do we not restrict the polyhedron $ \Delta^\ast _{base} $ of the toric base to be convex and hence it is in general non-Fano (the anti-canonical bundle is not semiample in previous terminology). This requires often toric resolutions of $ \cA_5 $ corresponding to adding integral vertices in the exterior or interior of $ \Delta^\ast $. In contrast to the crepant resolutions we discussed before such a resolution will alter the geometry of the Calabi-Yau fourfold and hence also change the properties of the effective field theory.

The weights of the full toric ambient space $ \cA_5 $ in which our elliptically fibered Calabi-Yau fourfold $ Y_4 ^{sing} $ will be embedded is completely determined by the geometry of the base $ B_3 ^{sing} $
\begin{align}
 \cA_5 &= \mathbb{P}^5(w_1,w_2,w_3,w_4 = 2w ,w_5 = 3w,w_6 = 1) \, ,\\
 w &= w_1 + w_2 + w_3 + w_6 \, .  \nonumber 
\end{align}
Here $ w $ is the degree of the anti-canonical divisor class of the base in its homogeneous coordinate ring $ S_3(-K_{B_3 ^{sing}}) = S_3 (w) $. As is common in the literature, we will denote the homogeneous coordinates of $ \cA_5 $ with weight $ 2w $ and $ 3w $ by $ X_4 = x $ and $ X_5 = y $, respectively. Imposing the Calabi-Yau condition on the hypersurface requires $ Y_4 ^{sing} $ to be a hypersurface with defining polynomial $ p_\Delta $ of degree $ d = 6 w $ that has so called Tate form
\begin{equation} \label{TateForm}
 y^2 + a_1 x y + a_3 y = x^3 + a_2 x^2 + a_4 x + a_6 \, ,
\end{equation}
with $ a_j $ global sections of various multiples of the anti-canonical bundle $ -K_{B_3} $ of the base $ B_3 $:
\begin{equation}
 a_j \in H_0 (B_3, -j K_{B_3}) \simeq S_3 (-j K_{B_3}) \, ,
\end{equation}
where $ S_3 $ is the homogeneous coordinate ring of the base $ B_3 $. On the singular base $ B_3 ^{sing} $, a weighted projective space, being a global section of $ -j K_{B_3^{sing}} $ is equivalent to being a quasi-homogeneous polynomial in the homogeneous base coordinates of degree $ \text{deg}(a_j) = w \cdot j $. After performing toric blow-ups the grading of $ S_3 $ becomes more complicated and we need to use the grading by divisor classes. Performing blow-ups of the base that do not preserve the anti-canonical divisor class will hence also change the global sections $ a_j $ and hence the structure of the full elliptically fibered Calabi-Yau fourfold $ Y_4 $.

\subsection{Sen's weak coupling limit} \label{SenLimitSection}

To make contact with the regular type IIB orientifold picture as for example discussed in \cite{Grimm:2005fa}, we will introduce next the weak string coupling limit by Sen \cite{Sen:1997bp,Sen:1997gv} for which also more refined versions exist \cite{Clingher:2012rg}. Going to weak string coupling amounts to moving to a special point in complex structure moduli space of the fourfold geometry as we will now describe. Redefining the variables of \eqref{TateForm}, we can always obtain the standard Weierstrass form given by
\begin{equation} \label{WeierstrassForm}
 y^2 = x^3 + fx + g \, .
\end{equation}
To analyze the underlying geometry we make this redefinition more explicit. First, we see that we can write the global sections $ f,g $ in this form as
\begin{align} \label{fgdef}
 f &= - \frac{1}{48} (b_2 ^2 - 24 \epsilon b_4) \, \in S_3(-4 K_{B_3}) \, , \\
 g &= - \frac{1}{864} (- b_2 ^3 + 36 \epsilon b_2 b_4 - 216 \epsilon^2 b_6) \, \in S_3(-6 K_{B_3}) \, ,
\end{align}
where also $ b_i \in S_3 (-i K_{B_3})$ are global sections of $ -i K_{B_3} $ that are related to the $ a_j $ of the Tate form \eqref{TateForm} as
\begin{equation} \label{bidef}
 b_2 = a_1 ^2 + 4 a_2 \, , \quad b_4 = a_1 a_3 + 2 a_4 \, , \quad b_6 = a_3 ^2 + 4 a_6 \, .
\end{equation}
We also introduced in \eqref{fgdef} the parameter $ \epsilon $ that can be thought of as the coordinate of the complex structure moduli space measuring the distance to the weak coupling region. Sending $ \epsilon $ to zero will provide us with the weak coupling description of the considered system whose details we consider next.

Starting from an F-theory compactification on the smooth elliptically fibered $ Y_4 $ we can find a weak string coupling configuration whose seven-branes are purely D7-branes or O7-planes. This configuration can be found by sending $ \epsilon \rightarrow 0 $ as we will now argue. The complex structure $ \tau $ of the elliptic fiber is a function that depends on the base coordinates and is related to the Weierstrass-form \eqref{WeierstrassForm} as
\begin{equation} \label{jDeltaDef}
 j(\tau) = \frac{4(24f)^3}{\Delta} \in S_3 (0)\, , \quad \Delta = 27 g^2 + 4 f^3 \in S_3 (-6 K_{B_3})\, .
\end{equation}
Here $ j $ is Klein's j-function and $ \Delta $ is the discriminant whose zero-set is the degeneration locus of the elliptic fiber. Combining \eqref{jDeltaDef} with \eqref{fgdef} we can expand $ \Delta $ and $ j $ to leading order in small $ \epsilon $ as
\begin{equation} \label{DeltajWeakCoupling}
 \Delta = \frac{1}{64} \epsilon^2 b_2 ^2(b_2 b_6 - b_4 ^2) \, , \quad j(\tau) = - \frac{32 b_2 ^4}{(b_2 b_6 - b_4 ^2) \epsilon^2} \, .
\end{equation}
Using the expansion $ j(\tau) = \text{exp}(- 2 \pi i \tau) + \ldots $ for $ j $ around its single pole, we see that in the limit $ \epsilon \rightarrow 0 $ we have $ \text{Im} \, \tau \propto - \text{log} \, \epsilon $ everywhere except at the vanishing locus of $ b_2 $. In Type IIB supergravity the axio-dilaton is given by $ \tau = C_0 + i e^{-\phi} $ with $ e^{\langle \phi \rangle} = g_s $ the string coupling. This $ \tau $ is geometrized in F-theory by the complex structure $ \tau $ of the elliptic fiber and therefore we conclude that the limit $ \epsilon \rightarrow 0 $ can be interpreted as the weak string coupling limit $ g_s \rightarrow 0 $.

As mentioned before, the non-perturbative seven-branes allow in this limit a global description by a configuration of D7-branes and O7-planes. The factorisation of the discriminant in \eqref{DeltajWeakCoupling} allows to identify the locations of these seven-branes as
\begin{equation} \label{O7D7locus}
 O7 \, :  \quad b_2 = 0 \, , \quad D7 \, : \quad b_2 b_6 - b_4 ^2 = 0 \, .
\end{equation}
The Calabi-Yau threefold $ Y_3 $ of the corresponding weakly coupled Type IIB orientifold set-up can be constructed as a double-cover of the toric base $ B_3 $ with branching locus the O7-planes. This can be done in the simplest fashion by representing $ Y_3 $ as a hypersurface in the anti-canonical line bundle $ -K_{B_3} $ of the base $ B_3 $ with fiber cooridinate $ \xi $ and general defining equation
\begin{equation}
 Y_3 \, : \quad Q = \xi^2 - b_2 = 0 \, .
\end{equation}
In this description we can easily identify the holomorphic orientifold involution $ \sigma : Y_3 \rightarrow Y_3 $ with $ \sigma^2 = id $ ($ id $ being the identity map) whose fixed point loci are the O7-planes. By construction $ \sigma $ acts only on the coordinate $ \xi $ as $ \sigma \, : \, \xi \rightarrow - \xi $ which implies that $ \sigma = b_2 = 0 $ is indeed the fixed-point set given by \eqref{O7D7locus}. These involutions were in detail discussed in \cite{Gao:2013pra}.

In the situation where the base $ B_3 $ is a blow-up of a weighted projective space $ B_3 ^{sing} = \mathbb{P}^3(w_1,w_2,w_3,w_6 = 1) $ the double-cover $ Y_3 $ can be embedded in a corresponding toric ambient space $ \hat \cA_4 $ which is a blow-up of
\begin{equation}
 \cA_4 = \mathbb{P}^4(w_1,w_2,w_3,w_6 = 1, w) \, , \quad w = w_1 + w_2 + w_3 + 1 \, ,
\end{equation}
as an anti-canonical hypersurface with $ Y_3 ^{sing} $ of degree $ 2 w $. After toric resolutions the ambient space $ \hat \cA_4 $ will be a $ \mathbb{P}^1 $-fibration over $ B_3 $ that can be interpreted as a compactification of the anti-canonical line-bundle $ - K_{B_3} $ of $ B_3 $. This hypersurface description enables us in particular to apply the previously developed techniques also for the Calabi-Yau threefold $ Y_3 $.

Let us now also discuss the fate of the toric divisors inducing the non-trivial three-form cohomology of $ Y_4 $ under the weak coupling limit. The geometric interpretation works best in the Poinc\'are dual picture of non-trivial five-cycles of the divisors that are non-trivial five-cycles on the full fourfold $ Y_4 $. Since the bases $ B_3 $ of the elliptic fibrations we consider are all toric, the non-trivial three-forms of $ Y_4 $ need to have at least one-leg in the fiber and therefore the dual five-cycles need to be circle fibrations over divisors in the base with its circle a cycle of the elliptic fiber. Expanding the M-theory three-form potential $ C_3 $ as in \cite{Denef:2008wq} leads in this case to
\begin{equation}
 C_3 = B_2 \wedge \alpha + C_2 \wedge \beta + \ldots
\end{equation}
where $ \alpha $ and $ \beta $ are a real basis of one-forms on the elliptic fiber dual to its two one-cycles $ A, B $ as we used for the general Riemann surfaces in \eqref{oneCycleBase}. Having a toric divisor $ D^\prime _{l_\alpha} $ of \autoref{divisorSection} that is an elliptic fibration and induces non-trivial three-forms immediately implies that it has the form $ D^\prime _{l_\alpha} = R_\alpha \times E_{l_{\alpha}} $ where $ R_\alpha $ is the elliptic fiber over a toric divisor $ E_{l_\alpha} $ of the toric $ B_3 $. Since $ R_\alpha $ is a fibration over $ E_\alpha $ but also vice versa, the divisor $ D^\prime _{l_\alpha} $ needs to be a direct product. From the direct product structure, we can deduce that $ R_\alpha $ the normalized period matrix $ \hat f^{(\alpha)} = i \tau $ which simplifies to a single number needs to be constant over $ E_{l_\alpha} $. This single number is by construction the axio-dilaton. In the weak coupling limit the three-form moduli $ \cN^\cA $ can be identified with the so called odd moduli $ \cN^\cA = G^{l_\alpha} $ or two-form scalars that arise from the expansion of the complexified two-form
\begin{equation} \label{Gexpand}
 G = G^{l_\alpha} \omega_{l_\alpha} = B_2 + i \tau C_2 \, \in \, H^{1,1} _-(Y_3) \, .
\end{equation}
Here we denoted by $ H^{1,1} _-(Y_3)  $ the subspace of two-forms $ \omega_{l_\alpha} $ of $ H^{1,1}(Y_3) $ that are odd under the orientifold involution $ \sigma $ as $ \sigma^\ast \omega_{l_\alpha} = - \omega_{l_\alpha} $. Geometrically this means that there are two four-cycles in $ Y_3 $ that get interchanged under $ \sigma $ and map under projection to the base $ B_3 $ to the same four-cycle $ E_{l_\alpha} $. For finite $ \tau $ in \eqref{Gexpand} the $ E_{l_\alpha} $ we have constructed will not intersect the O7-planes and hence their double-cover in $ Y_3 $ are two disjoint copies of $ E_{l_\alpha} \subset B_3 $ denoted by $ E_{l_\alpha} ^{\pm} \subset Y_3 $ and the two-forms $ \omega_{l_\alpha} \in H^{1,1} _-(Y_3) $ are the Poincar\'e dual two-forms to their difference
\begin{equation}
 \omega_{l_\alpha} = PD[E^+_{l_\alpha} - E^- _{l_\alpha}] \in H^{1,1} _- (Y_3) \, .
\end{equation}
For further details on the physics of odd moduli we refer to \cite{Grimm:2005fa}. This is just a different point of view on the situation we already encountered in \autoref{orientifold_limit}.

We can, however, also find a second kind of five-cycle. Starting with a four-chain in the base we can have a circle-fibration that degenerates at the boundaries of this four-chain which are three-cycles. An easy example of this phenomenon is to represent a two-sphere as a circle fibation over an interval, the one-chain, which degenerates at the boundaries of the interval. Taking the circle of the fibration as a cycle of the elliptic fiber, we see that at the loci in the base where this cycle and hence the elliptic fiber degenerates the five-cycle induces three-cycles. The degeneration loci are associated to seven-branes by constructios and hence this kind of five-cycle induces possibly non-trivial three-cycles on seven-branes. On these seven-branes, which are geometrically divisors in the base, we can dualize these three-cycles to one-forms. Taking into account the monodromy properties of the resulting cohomology classes we hence deduce that the three-forms on the elliptically Calabi-Yau fourfold hypersurface split in the weak coupling limit into two classes
\beq \label{WeakCouplingThreeForms}
    H^{2,1}(Y_4) \quad \longrightarrow \quad  \left\{ \begin{array}{c} H^{1,1} _- (Y_3) \, , \\[.1cm]
                                                                                      H^{1,0} _- (S) \, ,\end{array} \right. 
\eeq
where $ S $ denotes a divisor in $ Y_3 $ wrapped by a D7-brane. Both one-cycles of the elliptic fibration are odd under the orientifold involution, as explained in \cite{Denef:2008wq}, therefore also their linear combinations are odd and hence also the five-cycles with one leg in the fiber. There is a priori no reason why these arguments should not work for general elliptically fibered Calabi-Yau fourfolds $ Y_4 $ and therefore we conjecture that \eqref{WeakCouplingThreeForms} holds in general for the three-form splitting in the weak-coupling limit. In the two subsequent sections we will give examples for the two-form scalar moduli in \autoref{TwoFormScalarSection} and an example of the second kind will be presented in \autoref{Wilson_scalars}.

\section{Example One: An F-theory model with two-form scalars} \label{TwoFormScalarSection}

In this subsection we present the first example of an elliptic Calabi-Yau fourfold geometry with non-trival three-form cohomology. In this simple geometry there will only be one non-trival $ (2,1) $-form whose complex structure dependence is therefore induced by a genus one Riemann-surface, a two-torus. In the specific case we discussed before this two-torus can be identified with the elliptic fiber over a divisor in the base over which the elliptic fibration factors as a direct product as discussed at the end of the previous section. Due to the simple geometric structure of this example, we will be able to derive the Picard-Fuchs type equations for the periods explicitly and disucss the weak coupling limit in detail.

\subsection{Toric data and origin of non-trivial three-forms} \label{toric_data_example1}

A list of elliptically fibered Calabi-Yau fourfolds in weighted projective spaces was already presented in \cite{Klemm:1996ts}. Amongst them was also a particular simple example constructed from the weighted projective space
\begin{equation}
 \cA_5 = \mathbb{P}^5(1,1,1,3,12,18) \, .
\end{equation}
This space has two kinds of singularities that will be inherited by the anti-canonical hypersurface $ Y_4 ^{sing} $, that can be resolved by moving to a toric space $ \hat \cA_5 $ with at most point singularities. The singularity can easily be seen from the weights of $ \cA_5 $, since two of them have a common divisor $ 2 $, leading to $ \mathbb{C}^4/\mathbb{Z}_2 $ singularities along a curve in $ \cA_5 $ and three of the weights have a common divisor three leading to $ \mathbb{C}^3/\mathbb{Z}_3 $ along a toric surface $ \cA_2 $ in $ \cA_5 $. In the singular ambient space $ \cA_5 $ the anti-canonical hypersurface is given by a polynomial $ p_\Delta $ that has a quasi-homogeneous degree of $ 36 $. We can introduce complex homogeneous coordinates on $ \cA_5 $ denoted by 
$ [\underline u : w : x : y] $ with the abbreviation $ \underline u = (u_1, u_2, u_3) $ and identifications with the usual homogenous coordinates $ X_i $ given in the table below. We can always bring the most general hypersurface equation of the type of $ p_\Delta $ into the form
\begin{equation} \label{pDeltaSingExampleOne}
 p_\Delta ^{sing} = y^2 + x^3 + \hat a_1 \, xy + \hat a_2 \, x^2 + \hat a_3 \, y + \hat a_4 \, x + \hat a_6 \, ,
\end{equation}
highlighting the elliptic fibration structure and the coefficient functions $ \hat a_n $ that only depend on the coordinates $ \underline u = (u_1, u_2, u_3) $ and $ w $ which can be further split into
\begin{equation} \label{hatanExpandExampleOne}
 \hat a_n = \sum_{m=0} ^{2n} w^{2n-m} c_{n,m}(\underline u) \, ,
\end{equation}
where $ c_{n,m}(\underline u) $ are general homogeneous polynomials of degree $ 3m $ in $ \underline u $. In this description we can already find the Riemann surface $ R $ of $ \mathbb{C}^3 / \mathbb{Z}_3 $ singularities in $ Y_4 ^{sing} $ that will give rise to the non-trivial three-form cohomology after resolving the singularities. It is simply given by restricting  \eqref{pDeltaSingExampleOne} to $ \cA_2 \subset \cA_5 $ given by $ u_1 = u_2 = u_3 = 0 $ with equation
\begin{equation} \label{Requation}
 R \, : \quad p_\theta ^{sing} =  y^2 + x^3 + \hat c_1 \, xy + \hat c_2 \, w^4 x^2 + \hat c_3 w^6 \, y + \hat c_4 w^8 \, x + \hat c_6 w^{12} = 0 \, .
\end{equation}
Here we denoted by $ \hat c_n = c_{n,0} $ the constant non-zero coefficients that remain after restricting \eqref{hatanExpandExampleOne} to $ \cA_2 $.

The $ \mathbb{Z}_2 $-and $ \mathbb{Z}_3 $-singularities of the toric space $ \cA_5 $ can be resolved by moving to the toric space $ \hat \cA_5 $ whose fan is uniquely determined by the cones generated from the rays through the integral points $ \nu_i ^\ast $ of the following polyhedron $ \Delta^\ast $

\begin{equation} \label{poly_1}
\setlength{\tabcolsep}{5pt}
\begin{tabular}[0]{|c  r  r  r  r  r  r  | c | c | c | c|}\hline
   \multicolumn{7}{|c|}{\text{Example One: \ \ Toric data of $ \hat \cA_5 $}}    &\rule[-.2cm]{0cm}{.7cm} coords & $ \ell_1 $ & $ \ell_2 $ & $ \ell_3 $ \\
  \hline
    $ \nu^\ast_1 = $ & $ ( $ & $ 1$ & $ 0 $ & $ 0 $ & $ 0 $ & $ 0 ) $ & $ X_1 = u_1 $ & $ 0 $ & $ 1 $ & $ 0 $\\	
    $ \nu^\ast_2 = $ & $ ( $ & $ 0$ & $ 1 $ & $ 0 $ & $ 0 $ & $ 0 ) $ & $ X_2 = u_2 $ & $ 0 $ & $ 1 $ & $ 0 $\\
    $ \nu^\ast_3 = $ & $ ( $ & $ 0$ & $ 0 $ & $ 1 $ & $ 0 $ & $ 0 ) $ & $ X_3 = w   $ & $ 0 $ & $ 0 $ & $ 1 $\\
    $ \nu^\ast_4 = $ & $ ( $ & $ 0$ & $ 0 $ & $ 0 $ & $ 1 $ & $ 0 ) $ & $ X_4 = x   $ & $ 2 $ & $ 0 $ & $ 0 $\\
    $ \nu^\ast_5 = $ & $ ( $ & $ 0$ & $ 0 $ & $ 0 $ & $ 0 $ & $ 1 ) $ & $ X_5 = y   $ & $ 3 $ & $ 0 $ & $ 0 $\\
    $ \nu^\ast_6 = $ & $ ( $ & $-1$ & $-1 $ & $-3 $ & $-12$ & $-18) $ & $ X_6 = u_3 $ & $ 0 $ & $ 1 $ & $ 0 $\\	
    $ \nu^\ast_7 = $ & $ ( $ & $ 0$ & $ 0 $ & $-1 $ & $-4 $ & $-6 ) $ & $ X_7 = v   $ & $ 0 $ & $-3 $ & $ 1 $\\
    $ \nu^\ast_8 = $ & $ ( $ & $ 0$ & $ 0 $ & $ 0 $ & $-2 $ & $-3 ) $ & $ X_8 = z   $ & $ 1 $ & $ 0 $ & $-2 $\\
    \hline
\end{tabular} \, 
\, .
\end{equation}
In this table we also noted the three projective relations $ \ell_i $ between the homogeneous coordinates of $ \cA_5 $. These are chosen to highlight the fibration structure of the blown-up ambient space $ \hat \cA_5 $ and not necessarlily comprise the minimal set of generators, as is usual for the Mori-cone, the cone of all equivalence relations. This new ambient space $ \hat \cA_5 $ has only point singularities and therefore a general anti-canonical hypersurface in $ \hat \cA_5 $ will be smooth. This was already shown in \cite{Batyrev:1994hm}.

In the resolved ambient space $ \hat \cA_5 $ we can construct a smooth Calabi-Yau hypersurface $ Y_4 $ as the zero-locus of a polynomial $ p_\Delta \in S_5 (-K_{\cA_5}) $. A general polynomial of this kind can always be brought into Tate form via coordinate transformations
\begin{align} \label{pDeltaExampleOne}
 p_\Delta &=  y^2 + x^3 + a_1 \, xyz + a_2 \, x^2 z^2 + a_3 \, y z^3 + a_4 \, x z^4 + a_6 \, z^6 \, ,  \\
 p_\Delta ^{sing} &= p_\Delta |_{v=1,z=1} \, , \nonumber
\end{align}
where we used the homogeneous coordinates of \eqref{poly_1}. The coefficient functions $ \hat a_i $ depend on the remaining variables. The ambient space $ \hat \cA_5 $ can be seen to have a fibration structure with fiber given by $ \mathbb{P}^2(2,3,1) $ and coordinates $ [x:y:z] $ and coordinates for the base $ B_3 $ are $ [u_1:u_2:u_3:v:w] $. Due to the form of \eqref{pDeltaExampleOne} we find an elliptic curve embedded in $ \mathbb{P}^2(2,3,1) $ over every point in $ B_3 $ and hence we constructed an elliptically fibered Calabi-Yau fourfold $ Y_4 $. The coefficient function of \eqref{pDeltaExampleOne} are explicitly given by
\begin{equation} \label{a_example_1}
 a_n = \sum_{m=0} ^{2n} c_{n,m}(\underline u) w^{2n-m} v^m \, , \quad \hat a_n = a_n |_{v=1}
\end{equation}
where the coefficient functions $ c_{n,m}(\underline u) $ are the same homogeneous polynomials of degree $ 3m $ in the variables $ \underline u = (u_1, u_2, u_3) $ as in \eqref{hatanExpandExampleOne}. The toric base $ B_3 $ is itself a fibration over $ \mathbb{P}^2 $ with coordinates $ [\underline u] $ and fiber a $ \mathbb{P}^1 $ with coordinates $ [v:w] $. Therefore we can also interpret $ Y_4 $ as a fibration over $ \mathbb{P}^2 $ with fiber an elliptically fibered K3 surface.

The Hodge-numbers of $ Y_4 $ can be calculated as shown in \autoref{HodgeNumberSection} and are given by
\begin{align}
 h^{1,1}( Y_4) = 3, \quad
 h^{2,1}( Y_4) = 1, \quad 
 h^{3,1}( Y_4) = 4358\, . 
\end{align}
The Calabi-Yau fourfold $ Y_4 $ has therefore a single non-trivial $ (2,1) $-form whose origin we will describe in more detail in the following.

As we have seen in \autoref{JacobianSection} we need an integral point in the interior of a two-dimensional face $ \theta^\ast $  of $ \Delta^\ast $ to obtain a toric divisor with holomorphic one-forms. From the toric data in \eqref{poly_1} we infer that the only integral point satisfying this condition is
\begin{equation}
 \nu^\ast _7 = \frac{1}{3} (\nu^\ast _1 + \nu^\ast _2 + \nu^\ast _6) \, , \quad D_7 = \{ v= 0 \} \, ,
\end{equation}
and the corresponding toric divisor of $ \hat \cA_5 $ is $ D_7 $. The induced toric divisor $D^\prime _7 $ on $ Y_4 $ is hence a hypersurface in $ D_7 $ with equation $ p_\Delta |_{v=0} = p_\theta = 0 $, where $ \theta $ is the dual face of $ \theta^\ast $. Using the scaling relation $ \ell_3 $ to set $ w=1 $ we find the equation for $ D_7 ^\prime $ to be
\begin{equation} \label{D7primeEquation}
 D_7 ^\prime \, : \quad p_\theta =  y^2 + x^3 + \hat c_1 \, xy z + \hat c_2 \, x^2 z^2 + \hat c_3 \, y z^3 + \hat c_4 \, x z^4 + \hat c_6 z^6 = 0 \, ,
\end{equation}
which is of the same form the hypersurface equation of $ R $, \eqref{Requation}, along which we had $ \mathbb{Z}_3 $-singularities. Similarly, we have the same constant coefficients $ \hat c_n = a_n(\underline u, v= 0, w = 1) = c_{n,0} $ that are constant on all of $ Y_4 $, but depend on the complex structure moduli. This illustrates the fact that in order to determine the complex structure dependence of the non-trival three-forms it does not matter if we deal with the full divisor $ D^\prime _7 $ or only its base $ R $. The equation \eqref{D7primeEquation} is in Tate form\footnote{We will later see that it can also always be brought into Weierstrass form $ y^2 + x^3 + f x + g = 0 $.} and hence the Riemann surface is a torus and in particular the elliptic fiber over the base divisor $ v = 0 $. Due to the fact that the coefficients $ \hat c_n $ of this equation do not depend on the base coordinates, it is easy to see that
\begin{equation}
 D_7 ^\prime = R \times E \, , \quad R \simeq T^2 \, , \quad E \simeq \mathbb{P}^2 \, ,
\end{equation}
as already advocated in \autoref{SenLimitSection}. The divisor $ v = 0 $ in the base $ B_3 $ can easily be seen to be $ \mathbb{P}^2 $ with coordinates $ [u_1:u_2:u_3] $. The single holomorphic $(2,1)$-form is therefore induced by the single holomorphic one-form on $ R $ for which we will determine its Picard-Fuchs equation in the upcoming section.

\subsection{Picard-Fuchs equations on $ T^2 $} \label{subsec:PF_example1}

In this section we encounter the simplest example of theory we developed in \autoref{RiemannSection} and \autoref{WeightedProjectiveSpaceSection} to calculate the Picard-Fuchs equations from which we can learn about the behavior of the normalized period matrix $ f_{\cA \cB} $. This section serves as an illustration of the general toric conecpts we introduced before.

The form of $ p_\theta $ in \eqref{D7primeEquation} already implies that the toric ambient space $ \cA_2 = \mathbb{P}^2(2,3,1) $ of the Riemann surface $ R $ has coordinates $ [x:y:z] $ with weights $ 2,3,1 $, respectively. Therefore, its homogeneous coordinate ring $ S_2 $ and the Jacobian ring $ \cR_\theta $ are given by
\begin{equation}
 S_2 = \mathbb{C}[x,y,z] \, , \quad \cR_{\theta} = \frac{\mathbb{C}[x,y,z]}{\langle \partial_x p_\theta, \partial_y p_\theta, \partial_z p_\theta \rangle} \, ,
\end{equation}
and $ \text{deg}(p_\theta) = 6 $. From the Poincar\'e residue construction, we find therefore that
\begin{equation}
 H^{1,0}(R) \simeq \cR_\theta (0) \, , \quad H^{0,1}(R) = \cR_{\theta}(6) \, ,
\end{equation}
and it can be shown that both are one-dimensional. The holomorphic one-forms $ H^{1,0}(R) $ can be generated by
\begin{equation}
 \gamma = \int_{\Gamma} \frac{1}{p_\theta} \, d \omega_{\cA_2} \,  \in H^{1,0}(R) \, ,
\end{equation}
with $ \Gamma $ a small one-dimensional curve in $ \cA_2 - R $ winding around $ R $ and $ d \omega_{\cA_2} $ the holomorphic volume-form of $ \cA_2 $ as found in \eqref{weightedHolomorphicVolumeForm} for weighted projective spaces which can be represented by
\begin{equation}
 d \omega_{\cA_2} = z dx \wedge dy - 2 x dy \wedge dz + 3 y dx \wedge dz \, .
\end{equation}
We choose for $ p_\theta $ the representation in Weierstrass form \eqref{WeierstrassForm}, but other representations are possible using reparametrizations. This is equivalent to choosing an element of $ S_2 (6) $ to represent the generator of $ \cR_{\theta}(6) $. We use the Weierstrass form, since it allows for a comparison with the weak coupling description of the next section
\begin{equation} \label{ptheta}
 p_\theta = y^2 + x^3 + z^6 + a \, xz^4 \, ,
\end{equation}
with $ a = f $ the complex structure modulus. The function $ g $ is here just a constant that we can choose to be $ g = 1 $. The derivatives of the holomorphic one-form with respect to the complex structure modulus $ a $ are
\begin{align}
 \partial_a \gamma &= - \int_{\Gamma} \frac{x z^4}{p_\theta^2 } \, d \omega_{\cA_2} \,  \in H^{1,0}((R)_a) \, , \\
 \partial_a ^2 \gamma &=2 \int_{\Gamma} \frac{x^2 z^4}{p_\theta} \, d \omega_{\cA_2} \,  \in H^{1}(R,\mathbb{C}) \, .
\end{align}
Using the equivalence relations in $ \cR_\theta $ we can derive the identity
\begin{equation}
 (27 + 4 a^3) x^2 z^8 = 9 z^8 \partial_x p_\theta + (-\frac{3}{2} a z^7 + a^2 z^5 x) \partial_z p_\theta \, ,
\end{equation}
which we can use to find the Picard-Fuchs equation of $ \gamma $ at the vacuum configuration with $ a= 0 $
\begin{equation} \label{PFExampleOne}
 (27 + 4 a^3) \gamma^{\prime \prime} + 12 a^2 \gamma^\prime  + \frac{7}{4} a \gamma = 0 \, .
\end{equation}
This is a well known differential equation appearing in the literature for example in \cite{Lerche:1991wm} and combining these results with the boundary conditions derived in \autoref{mirror_section} allows to find a solution around $ a=0 $. Due to the fact that the Weierstrass-form is so well studied, as for example reviewed in \cite{Denef:2008wq} we know already that we can find the normalized period matrix $ \hat f(a) $ of the corresponding elliptic curve satisfies
\begin{equation}
 j(i \hat f(a)) = \frac{24(4a)^3}{\Delta} \, , \quad \Delta = 27 + 4 a^3 \, .
\end{equation}
Close to the three distinct zeroes of the discriminant $ \Delta $ given by $ a_i = 3 / 4^{1/3} \xi^i $, with $ \xi^3 = 1 $ the roots of $ a^3 = 1 $, we find
\begin{equation}
 i \hat f(a) \sim \frac{1}{2 \pi i} \, \text{log}(a - a_i)
\end{equation}
up to $ SL(2,\mathbb{Z}) $-transformations.  The boundary conditions derived in \cite{Greiner:2015mdm} are here trivially satisfied, $ \hat f = i \tau $, since the genus of the Riemann surface is one, and hence the coefficient of the linear term is the triple intersection number of the one blow-up divisor in the mirror geometry. Due to the fact that the mirror is also smooth, this number is one. 

Another way to interpret this result stems from Seiberg-Witten theory, like reviewed in \cite{Lerche:1996xu}. There the exact coupling of an $ SU(2) $ gauge theory was calculated using an elliptic curve  and we find here the same result as a coupling of scalars. The three singularities $ a_i $ can be used as points around which we can expand the period-matrix $ f $ and these three coordinate patches couple the full moduli space of the gauge theory. However, two of these $ a_i $ describe in SW language points of gauge enhancement. In contrast to this, we expand around the large complex structure point of the Calabi-Yau fourfold $ Y_4 $ after transforming to the proper complex structure coordinates $ z^\cK $. In the SW theory this corresponds to the solution at infinity in moduli space, i.e.~deep in the Coulomb branch of the gauge theory.

We have found that $ p_\theta $ is the equation for the elliptic fiber $ R$ over the divisor $ v = 0 $ in the base. This implies in particular, that $ p_\theta $ defines the complex structure $ \tau|_{v=0} $ of the elliptic fiber $ R$ over this divisor. This is defined such that up to 
$ SL(2,\mathbb{Z}) $-transformations we have a holomorphic one-form
\begin{equation} \label{elliptic_gamma}
 \gamma = \hat \alpha + \tau \hat \beta \, \in H^{1,0}(R) \, ,
\end{equation}
for $ \hat \alpha $, $ \hat \beta $ a canonical basis of $ H^1(R ,\mathbb{Z}) $ as introduced in \autoref{oneCycleBase}. This $ \tau $ is the axio-dilaton of Type IIB string theory varying over the base $ B_3$. The important observation here is that 
$ \tau|_{v=0} $ is constant along the divisor $v=0$ in $B_3$, i.e.~does not depend on the base coordinates, but does vary non-trivially 
with the complex structure moduli. To see this, we evaluate 
\begin{equation}
 j(\tau) \big|_{v=0} = \frac{4(24f)^3}{27g^2 + 4f^3} \Big|_{v=0} = C(\hat c_n) \, .
\end{equation}
In order to do that we determine $ f|_{v=0}$, $g|_{v=0} $ using \eqref{fgdef}, \eqref{bidef} with the $a_n|_{v=0}$ determined from 
$ p_\theta $ given in \eqref{ptheta}. The result is a non-trivial function of the coefficients $\hat c_n$ of $p_\theta$, these are constants on $Y_4$, but do depend on the complex structure moduli $z^\cK$ of $Y_4$. Note that there are $4358$ such complex structure 
moduli and we will not attempt to find the precise map to the five coefficients $\hat c_n$. Putting everything together, we can thus use $ \tau|_{v=0} $ as normalized period matrix of the curve $ R $ that induces the non-trivial three-forms in the fourfold $ Y_4$. Therefore, we have just shown that
\begin{equation}
 \hat f(z) = i \tau|_{v=0}(\hat c_n) \, ,
\end{equation}
on the full complex structure moduli space of the Calabi-Yau fourfold.

\subsection{Weak string coupling limit: a model with two-form scalars}

We next examine the weak string coupling limit of the geometry introduced in \autoref{toric_data_example1}. Using Sen's general procedure described in \autoref{SenLimitSection} we add an additional coordinate $ \xi $ to the homogeneous 
coordinate ring of the base $B_3$. The scaling weight of $\xi$ is the degree of the monomials associated to the 
anti-canonical bundle $-K_{B_3}$, i.e.~$ \xi $ has the degree of half the anti-canonical class in the homogeneous 
coordinate ring of $ \hat \cA_4$. Therefore, we find $ Y_3 \subset \cA_4 $ as the Calabi-Yau hypersurface obtained as the blow-up 
of the singular hypersurface $ Y_3 ^{\textrm  sing} = \mathbb{P}^4(1,1,1,3,6) [12]$. 
Recalling that $ B_3$ is a $ \mathbb{P}^1 $-fibration over $ \mathbb{P}^2 $, the double-cover $ Y_3 $ turns out to be the 
double-cover of $ \mathbb{P}^1 $ fibered over $ \mathbb{P}^2 $. The double-cover of the $\bbP^1$-fiber is a two-torus, or rather an elliptic curve,  $ \mathbb{P}^2(1,1,2)[4]$. 

To make this more explicit we again use a toric description. The fan of the ambient space for the three-fold is given by the cones generated by the rays through the points
\begin{equation}
\setlength{\tabcolsep}{5pt}
\begin{tabular}[0]{|c  r  r  r  r  r   | c | c | c |}
\hline
  \multicolumn{6}{|c|}{ \text{Example 1:\ \ Toric data of $ \cA_4 $ }}   &\rule[-.2cm]{0cm}{.7cm} coords & $ \ell_1 $ & $ \ell_2 $ \\
  \hline
    $ \nu^\ast_3 = $ & $ ( $ & $ 0$ & $ 0 $ & $ 1 $ &  $ 0 ) $ & $ z_3 = w $   & $ 1 $ & $ 0 $ \\	
    $ \nu^\ast_4 = $ & $ ( $ & $ 0$ & $ 0 $ & $ 0 $ &  $ 1 ) $ & $ z_4 = \xi $ & $ 2 $ & $ 0 $ \\
    $ \nu^\ast_6 = $ & $ ( $ & $ 0$ & $ 0 $ & $-1 $ &  $-2 ) $ & $ z_6 = v $   & $ 1 $ & $-3 $ \\
    \hline
    $ \nu^\ast_1 = $ & $ ( $ & $ 1$ & $ 0 $ & $ 0 $ &  $ 0 ) $ & $ z_1 = u_1 $ & $ 0 $ & $ 1 $ \\
    $ \nu^\ast_2 = $ & $ ( $ & $ 0$ & $ 1 $ & $ 0 $ &  $ 0 ) $ & $ z_2 = u_2 $ & $ 0 $ & $ 1 $ \\
    $ \nu^\ast_5 = $ & $ ( $ & $-1$ & $-1 $ & $-3 $ &  $-6 ) $ & $ z_5 = u_3 $ & $ 0 $ & $ 1 $	\\ 
    \hline
\end{tabular} 
\end{equation}
The hypersurface equation is then denoted by $Q=0$ and from subsection \ref{SenLimitSection} we can deduce that it has the form
\begin{equation}
 Q= \xi^2 - b_2(\underline u, v,w)
\end{equation}
in the fully blown-up ambient space with
\begin{equation}
 b_2 = a_1 ^2 + 4 a_2 
\end{equation}
specified by the Weierstrass-form of the corresponding fourfold in \eqref{a_example_1}.

One computes the Hodge-numbers to be
\beq
h^{1,1}( Y_3) = 3, \quad h^{2,1}( Y_3)  = 165 \, .
\eeq 
This example was already discussed in the context of mirror symmetry in \cite{Hosono:1993qy}. The resulting threefold is an elliptic fibration over $\bbP^2$ with two sections. It should be stressed that despite the fact that $h^{1,1}(Y_3)=3$ the toric ambient space only admits two non-trivial divisor classes. In fact, we will discuss in the 
following that this can be traced back to the fact that the divisor $v=0$ yields two disjoint $ \mathbb{P}^2 $ when intersected with the hypersurface constraint. These are the two sections, i.e.~two copies of the base. This is also noted in \cite{Gao:2013pra}, 
where a classification of orientifold involutions suitable for Type IIB orientifold compactifications is presented.

To make this more precise, let us analyze the singularities of $ Y_3 ^{ \textrm  sing} = \mathbb{P}^4(1,1,1,3,6) [12]$ and their resolutions via blow-ups further. 
The ambient space $ \cA_4 = \mathbb{P}^4(1,1,1,3,6) $ has $ \bbC^3 / \mathbb{Z}_3 $-singularities along a curve $ \mathbb{P}^1 $ given by $ [0:0:0:w:\xi] $. The hypersurface intersects this curve in two points, which are identified as double cover of the  
point of the not yet blown up base $ B_3 ^{ \textrm  sing} = \mathbb{P}^3(1,1,1,3) $, where 
we find $ \mathbb{C}^3/\mathbb{Z}^3 $-singularities. Blowing up this curve of singularities in the ambient space 
by adding $ \nu^\ast _6 $ leads to an exceptional divisor $v=0$, which is a $ \mathbb{P}^2 $ fibration over two points 
of the hypersurface. 
On the hypersurface $ Y_3$ we find that the ambient space divisor $v=0$ splits into two parts
\begin{equation}
 D_6 ^\prime = \{ v=0, \, Q^{(1)} = 0 \} \sim \mathbb{P}^2 \sqcup \mathbb{P}^2
\end{equation}
with coordinates $ [u_1,u_2,u_3,v=0,w, \pm \sqrt{c}w^2] $. Note that $ c $ is a constant, but depends on complex structure moduli. It is given by
\begin{equation} \label{c_def}
 c = b_2 |_{v=0} = c_{1,0}^2 + 4 c_{2,0} \, .
\end{equation}
The modulus $ c $ measures the separation between the two $ \mathbb{P}^2 $ in which $ D_6 $ splits when intersecting the threefold hypersurface. For $ \hat c_{2,0} = 0 $ we find that $ c $ is a perfect square.

We next investigate the action of the orientifold involution $\sigma: \xi \rightarrow -\xi$. 
From the coordinate description of $ D_6 ^\prime $ we find that the two disjoint $ \mathbb{P}^2 $ are interchanged by the involution $ \sigma $. Therefore, we introduce the two non-toric holomorphic divisors $ D_{6,1} ^\prime $ and $ D_{6,2} ^\prime $ that are the two disjoint $ \mathbb{P}^2 $ such that
$D_6 ^\prime = D_{6,1} ^\prime + D_{6,2} ^\prime$ and $\sigma^\ast (D_{6,1} ^\prime) = D_{6,2} ^\prime$.
It is now straightforward to define an eigenbasis for the involution $ \sigma $ as
\begin{align} \label{Kminusplus}
 K^+ _1 = D_4 ^\prime \, , \quad K^+ _2 = D_6 ^\prime\, , \quad K^- = D_{6,1} ^\prime - D_{6,2} ^\prime \, .
\end{align}
Therefore, we conclude that 
\beq 
h^{1,1} _+ (Y_3 ) = 2 , \qquad h^{1,1} _- (Y_3) = 1\ , 
\eeq
which shows that there is one negative two-from which yields zero-modes for the R-R and NS-NS two-forms
of Type IIB supergravity. Furthermore, we can evaluate the intersection ring to be
\begin{equation}
 I_{Y_3} = 18 (D_6 ^\prime )^3 + 144 (D_4 ^\prime ) ^3 = 18 (D_6 ^\prime) ^3 - 6 D_1 ^\prime (D_6 ^\prime )^2 + 2 (D_1 ^\prime )^2 D_6 ^\prime
\end{equation}
Note that $ D_{6,1} ^\prime \cap D_{6,2} ^\prime = \emptyset $. Due to the symmetry between the components of $ D_6 ^\prime$ and $ D_4 ^\prime$ being exactly the fixed point of this symmetry, we find that the intersections of $ K^- $ appearing linearly vanish. We learn that $ (D_{6,1} ^\prime)^3 = (D_{6,2} ^\prime) ^3 = 9 $, $ (D_{6,1} ^\prime )^2 D_4 ^\prime= (D_{6,1} ^\prime )^2 D_4 ^\prime = 0 $ and $ D_{6,1} ^\prime (D_4 ^\prime)^2 = D_{6,2} ^\prime (D_4 ^\prime)^2 = 0$.\\

From this analysis we see that  all toric divisors are invariant under the involution $ \sigma $. Therefore, we can choose the divisor basis of the base $ B_3 = \hat{\mathbb{P}}^3(1,1,1,3) $ obtained from $ \hat \cA _4 = \hat{\mathbb{P}}^4(1,1,1,3,6) $ by setting $ \xi = 0 $ . This corresponds on the lattice level to projecting to $ \mathbb{Z}^3 $, i.e.~dropping the fourth coordinate of every vertex.

\begin{equation}
\setlength{\tabcolsep}{5pt}
\begin{tabular}[0]{|c  r  r  r  r    | c | c | c |}
\hline
  \multicolumn{5}{|c|}{\text{Toric data of $ B_3 $}}  &\rule[-.2cm]{0cm}{.7cm} coords & $ \ell_1 $ & $ \ell_2 $ \\
  \hline
    $ \nu^\ast_3 = $ & $ ( $ & $ 0$ & $ 0 $ & $ 1 ) $ & $ z_3 = w $   & $ 1 $ & $ 0 $ \\	
    $ \nu^\ast_6 = $ & $ ( $ & $ 0$ & $ 0 $ & $-1 ) $ & $ z_6 = v $   & $ 1 $ & $-3 $ \\
    \hline
    $ \nu^\ast_1 = $ & $ ( $ & $ 1$ & $ 0 $ & $ 0 ) $ & $ z_1 = u_1 $ & $ 0 $ & $ 1 $ \\
    $ \nu^\ast_2 = $ & $ ( $ & $ 0$ & $ 1 $ & $ 0 ) $ & $ z_2 = u_2 $ & $ 0 $ & $ 1 $ \\
    $ \nu^\ast_5 = $ & $ ( $ & $-1$ & $-1 $ & $-3 ) $ & $ z_5 = u_3 $ & $ 0 $ & $ 1 $ \\
    \hline
\end{tabular}
\end{equation}
As a consequence, we can use $ D_6 $ and $ D_1 $ as a basis for the divisors on $ B_3 $. For $ Y_3 $ we can choose the corresponding basis via $ D_4 ^\prime = 2 D_6 ^\prime + 6 D_5 ^\prime$ and find
\begin{equation}
 I_{B_3} = 9 D_6 ^3 - 3 D_1 D_6 ^2 + D_1 ^2 D_6 = \frac{1}{2}(18 D_6 ^3 - 6 D_1 D_6 ^2 + 2 D_1 ^2 D_6) \sim \frac{1}{2}I_{Y_3} \, .
\end{equation}
This fits the fact that $ Y_3 $ double-covers $ B_3$ and $ D_{6,1} ^\prime $ and $ D_{6,2} ^\prime $ project down to the same $ \mathbb{P}^2 $ in $ B_3 $.

Let us now discuss what happens to the normalized period matrix $ \hat f = i \tau|_{v=0} $ that we have derived in subsection 
\ref{subsec:PF_example1}, in the weak coupling limit of complex structure space. In this orientifold limit 
the field $\tau_0 = C_0 + i e^{-\phi}$ is actually constant everywhere on $Y_3/\sigma$
and becomes an independent modulus. The identification $\hat f= i \tau_0$ then precisely yields 
the known moduli $N = c- \tau_0 b$ of the orientifold setting, where $c$, $b$ are the zero-modes 
of the R-R and NS-NS two-forms along $K_-$ introduced in \eqref{Kminusplus}.

We close by pointing out that it is important to have $c= \hat c_1 ^2 + 4 \hat c_2 \neq 0$ for this weak coupling analysis to apply. 
Indeed, if we go to the limit $c \rightarrow 0$ we find a spliting of the O7-plane located at $b_2=0$ into $v=0$ and 
$b'_2=0$. Not only would we find intersecting O7-planes, but also the simple identification $\hat f = i\tau_0$ would no longer hold.

\section{Example Two: An F-theory model with Wilson line scalars} \label{Wilson_scalars}

In this subsection we construct a second example geometry that we argue to admit Wilson line moduli 
when used as an F-theory background. In this example the three-forms of the Calabi-Yau fourfold stem from a 
genus seven Riemann surface. It turns out that this example features also other interesting properties, such as a 
non-Higgsable gauge group and terminal singularities corresponding to O3-planes.

\subsection{Toric data and origin of non-trivial three-forms}

Our starting point is the anti-canonical hypersurface in the weighted projective space $ \cA  _5 = \mathbb{P}^5(1,1,3,3,16,24) $ of degree $ d = 48 $. This space is highly singular, but admits an elliptic fibration necessary to serve as an F-theory background. 
It is easy to see that we have a curve $ R$ along which we find $ \mathbb{C}^3/ \mathbb{Z}_3 $-singularities.
In contrast to the first example this curve $R$ is not the elliptic fiber. It rather arises as a multi-branched cover over 
a $ \mathbb{P}^1 $ of the singular base $ B_3 ^{\textrm  sing} $.

We can resolve part of the  singularities of the ambient-space $\cA_5$ by moving to a toric space $ \hat \cA_5$ whose fan is obtained by the maximal subdivision of the polyhedron $ \Delta^\ast  $ of $\cA_5$:
\begin{equation} \label{poly_2}
\setlength{\tabcolsep}{5pt}
\begin{tabular}[0]{|c  r  r  r  r  r  r  | c | c | c | c | c |}
\hline
 \multicolumn{7}{|c|}{\text{Example 2:\ \ Toric data of $ \hat \cA_5 $}}   & \rule[-.2cm]{0cm}{.7cm} coords & $ F $ & $ \mathbb{P}^2 $ & $ B $ & $ E $\\
  \hline
    $ \nu^\ast_1 = $ & $ ( $ & $ 1$ & $ 0 $ & $ 0 $  & $ 0 $ & $ 0 ) $   & $ z_1 = w   	$ & $ 0 $ & $ 0 $  & $ 1 $ & $ 1 $ \\	
    $ \nu^\ast_2 = $ & $ ( $ & $ 0$ & $ 1 $ & $ 0 $  & $ 0 $ & $ 0 ) $   & $ z_2 = u_1 	$ & $ 0 $  & $ 1 $  & $ 0 $ & $ 0 $ \\
    $ \nu^\ast_3 = $ & $ ( $ & $ 0$ & $ 0 $ & $ 1 $  & $ 0 $ & $ 0 ) $   & $ z_3 = u_2 	$ & $ 0 $  & $ 1 $  & $ 0 $ & $ 0 $ \\
    $ \nu^\ast_4 = $ & $ ( $ & $ 0$ & $ 0 $ & $ 0 $  & $ 1 $ & $ 0 ) $   & $ z_4 = x   	$ & $ 2 $  & $ 0 $  & $ 0 $ & $ 1 $ \\
    $ \nu^\ast_5 = $ & $ ( $ & $ 0$ & $ 0 $ & $ 0 $  & $ 0 $ & $ 1 ) $   & $ z_5 = y   	$ & $ 3 $  & $ 0 $  & $ 0 $ & $ 0 $ \\
    $ \nu^\ast_6 = $ & $ ( $ & $-1$ & $-3 $ & $-3 $  & $-16$ & $-24) $   & $ z_6 = v   	$ & $ 0 $ & $ 0 $  & $ 1 $ & $ 1 $ \\
    $ \nu^\ast_7 = $ & $ ( $ & $ 0$ & $-1 $ & $-1 $ & $ -5 $ & $ -8) $   & $ z_7 = e 	$ & $ 0 $  & $ 0 $  & $ 0 $ & $ -3 $ \\ 
    $ \nu^\ast_8 = $ & $ ( $ & $ 0$ & $-1 $ & $-1 $  & $-6 $ & $ -9) $   & $ z_8 = u_3   $ & $ 0 $  & $ 1 $  & $-3 $ & $ 0 $ \\
    $ \nu^\ast_9 = $ & $ ( $ & $ 0$ & $ 0 $ & $ 0 $  & $-2 $ & $ -3) $   & $ z_9 = z   	$ & $ 1 $  & $-3 $  & $ 1 $ & $ 0 $ \\
    \hline
\end{tabular}
\end{equation}
Note already at this point, that the new ambient space $\hat  \cA_5  $ still contains singularities of the form
\begin{equation}
 \mathbb{C}^4 / \mathbb{Z}_2 \, : \quad (v,w,u_3,y)\quad \rightarrow \quad (-v,-w,-u_3,-y)
\end{equation}
and hence the hypersurface inherits singular points that do not allow for any crepant resolution as pointed out in \cite{Aspinwall:1994ev}. This can be related to the presence of O3-planes. \footnote{ Various aspects of O3-planes have been discussed recently for example in \cite{Collinucci:2010gz,Garcia-Etxebarria:2015wns}}

A number of intriguing features of this model arises due to the geometry of the base $ B_3$. It arises as a non-crepant blow-up of the weighted projective space $ B_3 ^{\textrm  sing} = \mathbb{P}^3(1,1,3,3) $ with toric data given by
\begin{equation}
\setlength{\tabcolsep}{5pt}
\begin{tabular}[0]{|c  r  r  r  r    | c | c | c |}
\hline
 \multicolumn{5}{|c|}{\text{Toric data of $ B_3 $}}    & \rule[-.2cm]{0cm}{.7cm} coords & $ \mathbb{P}^1 $ & $ \mathbb{P}^2 $ \\
  \hline
    $ \nu^\ast_1 = $ & $ ( $ & $ 1$ & $ 0 $ & $ 0 ) $ & $ z_1 = \tilde v $   & $ 1 $ & $ 0 $ \\
    $ \nu^\ast_2 = $ & $ ( $ & $ 0$ & $ 1 $ & $ 0 ) $ & $ z_2 = \tilde u_1 $ & $ 0 $ & $ 1 $ \\
    $ \nu^\ast_3 = $ & $ ( $ & $ 0$ & $ 0 $ & $ 1 ) $ & $ z_3 = \tilde u_2 $ & $ 0 $ & $ 0 $ \\
    $ \nu^\ast_4 = $ & $ ( $ & $-1$ & $-3 $ & $-3 ) $ & $ z_5 = \tilde w $   & $ 1 $ & $ 1 $ \\
    $ \nu^\ast_5 = $ & $ ( $ & $ 0$ & $-1 $ & $-1 ) $ & $ z_6 = \tilde u_3 $ & $-3 $ & $ 1 $ \\
    \hline
\end{tabular} \, .
\end{equation}
It can be interpreted as a generalization of a Hirzebruch surface, i.e.~a $ \mathbb{P}^2 $-fibration over $ \mathbb{P}^1 $. We note in particular, that the point $ \nu^\ast _5 $ does lie in the interior of the convex hull of the remaining points and correspondingly the new polyhedron is no longer convex. The consequence is that the anti-canonical bundle $ -K_{B_3 } $ of the base has only global sections that vanish over the locus $ \{ \tilde u_3 = 0 \} \simeq \mathbb{P}^1 \times \mathbb{P}^1 $, i.e.~$ -K_{B_3 } $ is not ample. In the F-theory picture this will lead to a non-Higgsable cluster as described in \cite{Grassi:2014zxa,Morrison:2014lca}, i.e.~to the generic existence of a non-Abelian gauge group in this setting. The base $ B_3 $ has been analyzed recently in detail in \cite{Berglund:2016nvh}.

The ambient space $\hat \cA _5 $ has the fibration structure given by the projection $ \pi: \, \hat \cA_5 \, \rightarrow \, B_3 $, which reads in homogeneous coordinates
\begin{align}
 \pi \, : \, &[v:w:u_1:u_2:u_3:x:y:z:e] \, \nonumber \\
 \mapsto \, &[\tilde v = v: \tilde w = w: \tilde u_1 = u_1: \tilde u_2 = u_2: \tilde u_3 = e u_3] \, .
\end{align}

Due to the non-Higgsable gauge group, $ Y_4 $ can only be written in Tate form after blowing down the exceptional divisor $e=0$, i.e.~setting $ e=1 $:
\beq \label{p2}
 p_{\Delta} = y^2 + e x^3 + \hat a_1\, x y + \hat a_2 \, x^2 +\hat a_3 \, y + \hat a_4 \, x  + \hat a_6 = 0\ ,
\eeq
with $ \hat a_i $ global sections of $ K_{B_3} ^{-i} $. 
Due to the properties of $ K_{B_3} ^{-1} $ these $ \hat a_n $ have common factors of $ u_3 e = \tilde u_3 $ 
independently of the point in complex structure space. This shows that the non-Higgsable cluster with the 
enhanced gauge group is located on the divisor $ \tilde u_3 = 0 $ in the base. The 
singularity type can be easily read of by translating \eqref{p2} into Weierstrass form using \eqref{fgdef}, \eqref{bidef}.
We then obtain a singularity of orders $ (2,2,4) = (f,g,\Delta) $, where $ \Delta $ is the discriminant as above. 
This leads to a type $ IV $ singularity and the exact gauge group, which is either $Sp(1)$ or $SU(3)$, 
can be derived from monodromy considerations as we recall below.
The generic anti-canonical hypersurface $ Y_4$ of the ambient space $ \hat \cA_5$ has Hodge numbers
\beq
   h^{1,1}( Y_4) = 4, \quad h^{2,1}( Y_4) = 7, \quad h^{3,1}( Y_4) = 3443 , \quad h^{2,2}( Y_4) = 13818 \, .
\eeq
This implies that $Y_4$ indeed has seven $(2,1)$-forms and we claim that these arise from a single Riemann surface of 
genus $ g = 7 $.

There is only one two-dimensional face $ \theta^\ast $ of the polyhedron spanned by $ \nu^\ast _1, \nu^\ast _4, \nu^\ast _6 $ that contains an interior integral point. This interior point is $ \nu^\ast _7 $ and we add this point to resolve the $ \mathbb{C}^3 /\mathbb{Z}_3 $-singularity along the surface $ \cA_2 = \mathbb{P}^2(1,1,8) $ given as the subspace of $ \cA_5 $ with $ w=v=x=0 $. The anti-canonical hypersurface $ Y_4$ intersects $ \cA_2$ in a Riemann surface $ R $ given by
\begin{equation}
 R = \mathbb{P}^2(1,1,8)[16],  \quad g = 7 \, .
\end{equation}
This can also be seen from the dual face $ \theta $ whose inner points correspond to the monomials
\begin{equation}
 p^\prime _a = u_1 ^a u_2 ^{6-a} \, \in \cR_\theta (6) , \quad a = 0, \ldots, 6
\end{equation}
where we already divided out the common factor $ u_1 u_2 y $ as described in \autoref{RiemannSection}. The exceptional divisor resolving this singularity is a fibration over $ R$ with fiber $ E = \mathbb{P}^2(1,1,16) $.

Expanding the Weiserstrass form \eqref{WeierstrassForm} of $Y_4$ around the singular divisor $ D_e = \{ e = 0 \} $, we find
\begin{equation}
 g = g_2 e ^2 + \cO(e ^3) \, , \quad g_2 = g_2(u_1,u_2)
\end{equation}
and this $ g_2 $ is precisely the degree $ 16 $ polynomial in $ u_1, u_2 $ defining the Riemann surface $ R  $ by
\begin{equation} \label{p_theta_ex2}
 R : \quad p _\theta = y^2 - g_2  = 0\, .
\end{equation}
The resulting gauge group over $ D_3 = \{ \tilde u_3 = 0 \} $ in $ B_3$ is $ Sp(1) $ for general $ g_2 $ and if $ g_2 = \gamma^2 $, i.e.~for $ g_2 $ a perfect square, we have an enhancement to $ SU(3) $.

\subsection{Comments on the weak string coupling limit}

So what happens to this curve in the weak coupling limit? 
For a $ IV $ singularity, there should be no straightforward perturbative limit in which 
$ \tau $ can be made constant and $\text{Im}\,\tau$ can be made very large over the base.
The general hypersurface equation derived from the naive Sen limit is
\begin{equation}
 Q=\xi^2 - b_2 = \xi^2 - \tilde u_3 \cdot b^\prime _2 = 0\, ,
\end{equation}
implying that the O7-plane splits in two intersecting branches, $ \tilde u_3 = 0 $ and $ b^\prime _2 = 0 $. At the intersection of the 
O7-planes perturbative string theory breaks down and hence there is no weak coupling description. However, we can still try to learn 
some of the aspects of the D7-branes in this setting. 

In fact, in the following we want to connect the curve \eqref{p_theta_ex2}
and Wilson line moduli located on D7-branes. As explained in \cite{Jockers:2004yj} 
the number of Wilson line moduli arising from a D7-brane image-D7-brane on a divisor $S \cup \sigma(S)$ 
of the threefold $ Y_3 $ is given by
\begin{equation}
 \text{Number of Wilson line moduli on } S \, : \quad h^{1,0} _- (S\cup \sigma(S)) \, .
\end{equation}
These are the $(1,0)$-forms on the union of $ S $ and its image that get 
projected out when considering the orientifold quotient. Therefore, we suggest that the Wilson lines 
arise in $ S \cup \sigma(S) $ as arcs in $ S $ that connect two components of $ S \cap \sigma(S)$. These arcs close to one-cycles in $ S \cup \sigma(S) $, but get projected out when we take the quotient $ Y_3/ \sigma = B_3$. Note here that $ S \cap \sigma(S) $ is equal to $ \text{O7} \cap S $. In our situation $Y_3$ is still a fibration over $ \mathbb{P}^1 $ with coordinates $ [v:w] $ and hence this will also hold for $ S \cap \sigma(S) $, 
i.e.~we suggest that $ S \cap \sigma(S) $ is a covering space of the base $ \mathbb{P}^1 $ given by 
\begin{equation}
 S \cap \sigma(S) = \{ \xi=0, \tilde u_3=0, g_2 = 0 \} \subset Y_3 \, ,
\end{equation}
where $ \xi = \tilde u_3 = 0 $ is the location of one branch of the O7-plane in $ Y_3 $. We also note that the divisor inducing the three-forms in the fourfold projects down to the $ \tilde u_3 =0 $ divisor of $B_3$. Recall that the locations of the seven-branes in a 
general F-theory model are given by the zeroes of the discriminant $ \Delta $. We can expand $\Delta$ around $ \tilde u_3 = 0 $ to
\begin{equation}
 \Delta \approx b_2 ^2 (b_2b_6 - b_4 ^2) = \tilde u_3 ^5 (b^\prime _2 )^3 g_2 + \cO(\tilde u_3 ^6) \, .
\end{equation}
This implies that in the weak coupling limit $ g_2 $ describes the intersection of the D7-brane  
in the form of a Whitney-Umbrella explained in \cite{Collinucci:2008pf} with the O7-branch given by $ \tilde u_3 = 0 $. 
For our considerations, it is just important that a D7-brane is path connected, 
but the shape away from the O7-plane is irrelevant for our analysis of Wilson lines. 
Therefore, we find that
\begin{equation}
 S \cap \sigma(S) = \bigcup_{i=1} ^{16} \big( \{p_i\} \times \mathbb{P}^1 \big) \, , \quad g_2 (p_i) = 0 \, .
\end{equation}

The points $ p_i $ can be interpreted as branching loci of the auxiliary hyperelliptic curve which is given by \eqref{p_theta_ex2}. Hence we find
\begin{equation}
 h^{1,0} _- (S\cup \sigma(S)) = 7 \, .
\end{equation}
Choosing a normalized basis $ \hat \alpha_a, \hat \beta^a $ for the cocycles arising from this procedure we can give a basis for $ H^{1,0} _- (S \cup \sigma(S)) $ as
\begin{equation}
 \gamma_a = \hat \alpha_a + i \hat f_{ab} \hat \beta^b \, \in H^{1,0} _- (S \cup \sigma(S)) \, ,
\end{equation}
with $ \hat f_{ab} $ the normalized period matrix of the curve $ R $ discussed in \autoref{RiemannSection}.
The coupling of the corresponding fields, the Wilson moduli $ N_\cA = N_a $, is given by the the normalized period matrix $ f_{\cA \cB} = \hat f_{ab} $ of $ R $.

Let us close by making one final observation for this example geometry. We can also resolve the $ \mathbb{Z}_2 $-singular points of the fourfold by blowing-up the ambient space $ \cA _5 $. This requires adding the exterior point
\begin{equation}
 \nu^\ast _{10} = (0,-2,-2,-10,-15) \, .
\end{equation}
This has, however, drastic consequences. As already mentioned before, there is no way to resolve the $ \mathbb{Z}_2 $-singular 
points in a crepant way, i.e.~preserving the anti-canonical bundle of the ambient-space. 
Closer inspection of the blow-up tells us that this blow-up is not crepant, but leads to a Calabi-Yau hypersurface in a new ambient-space 
that has a different triangulation not compatible with the old triangulation structure. 
This leads to a change in topology, which can be seen from the Hodge-numbers
\begin{equation}
 h^{1,1} _{new} = 5, \quad h^{2,1}_{new} = 0 , \quad h^{3,1}_{new} = 3435, \quad \chi = \chi_{old} = 20688\, ,
\end{equation}
with the Euler number $ \chi $ being preserved. This extremal transition between the two fourfolds follows a similar pattern 
as the conifold transition along curves described in \cite{Intriligator:2012ue}. The relations to the non-trivial three-form cohomology 
can also be made precise: the blow-up obstructs precisely the complex structure deformations described by $ g_2 $ setting it to zero and hence also obstructing the three-form cohomology. This obstruction leads to a further gauge-enhancement to $ G_2 $ along $ D_3 $ and also the weak coupling limit is no longer singular, i.e.~the O7-plane does no longer branch.

\newpage

\chapter{Conclusions and Outlook} \label{ConclusionSection}

In the first part of this work we studied 
the two-dimensional low-energy effective action obtained from Type IIA string theory 
on a Calabi-Yau fourfold with non-trivial 
three-form cohomology. 
The couplings of the three-forms were shown to 
be encoded by two holomorphic functions $f_{\cA \cB}$ and 
${h_\cA} ^\cB$, where the former depends on the complex structure moduli 
and the latter on the complexified K\"ahler structure moduli. 
Performing a large volume dimensional reduction of Type IIA supergravity, we 
were able to derive $ {h_\cA} ^\cB $ explicitly as a linear function. 
We argued that $ f_{\cA \cB}$ and $ {h_\cA} ^\cB$ computed 
on mirror pairs of Calabi-Yau manifolds will be exchanged, at least,
if one considers the theories at large volume and large complex structure.
In order to show this, we investigated the non-trivial map 
between the three-form moduli arising from mirror geometries 
and argued that it involves a scalar field dualization together 
with a Legendre transformation.
This can be also motivated by the fact that chiral and 
twisted-chiral multiplets are expected to be exchanged 
by mirror symmetry. We thus established a linear dependence of  
the function $f_{\cA \cB}$ on the complex structure moduli near 
the large complex structure point and determined the constant topological
pre-factor. 

In this part we also included a  discussion of 
the superymmetry properties of the two-dimensional 
low-energy effective action. This action is expected to be an 
$\cN=(2,2)$ supergravity theory, which we showed to 
extend the dilaton supergravity action of \cite{Gates:2000fj}.
The bosonic action was brought to an elegant 
form with all kinetic and topological terms determined 
by derivatives of a single function $\tilde K = \cK + e^{2 \tilde \varphi} \cS$,
where $\cK$ and $\cS$ can depend on the scalars in chiral and 
twisted-chiral multiplets, but are independent of the two-dimensional  dilaton $\tilde \varphi$. 
In the Type IIA supergravity reduction the three-form scalars only 
appeared in the function $\cS$ and are thus suppressed by 
$e^{2 \tilde \varphi} = e^{2 \phi_{\textrm{IIA}}}$. 
In this analysis the complex structure moduli and the 
three-form moduli were argued to fall into chiral multiplets, while the  
complexified K\"ahler moduli are in twisted-chiral multiplets. 
However, due to apparent shift symmetries of the three-form moduli 
and complexified K\"ahler moduli a scalar dualization accompanied by 
a Legendre transformation can be performed in 
two dimensions. This lead to dual descriptions in which certain 
chiral multiplets are replaced by twisted-chiral multiplets and vice versa. 
Remarkably, if one dualizes a subset of scalars appearing in $\cK$, 
we found that the requirement 
to bring the dual action back to the standard $\cN=(2,2)$ dilaton supergravity 
form imposes conditions on viable $\cK$.
These constraints include a no-scale type condition on $\cK$. 
The emergence of such restrictions arose from general arguments 
about two-dimensional theories coupled to an overall
$e^{-2\tilde \varphi}$ factor. For Calabi-Yau fourfold reductions we checked that these conditions are
indeed satisfied. 
It would be interesting to investigate this further and to get a  
deeper understanding of this result. 

Having shown that in the large complex structure limit
the function $f_{\cA \cB}$ is linear in the complex structure 
moduli, we discussed the 
application of this result in an F-theory compactification. 
By assuming that the Calabi-Yau fourfold is elliptically fibered and that
the three-forms exclusively arise from the base of this fibration, 
we recalled that $f_{\cA \cB}$ is actually the gauge-coupling 
function of four-dimensional R-R vector fields. 
This gauge-coupling function was already evaluated 
in the weak string coupling limit in the orientifold literature. 
In this orientifold limit 
one can double-cover the base with a Calabi-Yau threefold. 
We found compatibility of the fourfold result with the expectation
from mirror symmetry for Calabi-Yau threefold orientifolds. 
In this analysis we only included closed string moduli in 
the orientifold setting. Clearly, the results obtained from the 
Calabi-Yau fourfold analysis are more powerful and it would be 
interesting to further investigate the open string dependence in 
orientifolds using our results. Additionally we commented briefly 
on the case in which the three-forms have legs in the fiber 
of the elliptic fibration. In this situation the inverse of $\R f_{\cA \cB}$
sets the value of decay constants of four-dimensional axions \cite{Grimm:2014vva}. 
Again we found compatibility in the closed string sector at weak string 
coupling in which $f_{\cA \cB} \propto i \tau$. It would be interesting 
to include the open string moduli in the orientifold 
setting and derive corrections to $f_{\cA \cB}$ without restricting to the 
weak string coupling limit. The latter task requires to compute $f_{\cA \cB}$ away from the 
large complex structure limit for elliptically fibered Calabi-Yau fourfolds.

In the second part of this thesis we introduced a framework to explicitly derive the moduli dependence of non-trivial three-forms on 
Calabi-Yau fourfolds. Our focus was on geometries realized as hypersurfaces in toric ambient spaces for which we argued that properties of the non-trivial cohomology groups can be split into two parts, one arising from the ambient space 
and one part from the toric divisors using the Gysin-sequence. We have explicitly shown, how to obtain algebraic and non-algebraic complex structure deformations
and also K\"ahler deformations that arise from toric and non-toric divisors. The special tool we used to do so was the homogeneous coordinate ring of the ambient space
that enabled us to derive explicit expressions of non-trivial holomorphic forms using the chiral ring of the hypersurface and the Poincar\'e residue. We recovered in particular the
well known formulas to calculate the spectrum of the effective theories from toric data.

After the general considerations we focused on the three-form cohomology, essentially inherited from one-forms on Riemann surfaces along which we have orbifold singularities
supplemented by topological information about the corresponding resolution divisors. The three-form scalars were argued to parametrize the intermediate Jacobian of the Calabi-Yau fourfold which was shown to arise as a product of Jacobian varieties of Riemann surfaces. We derived how to obtain the Picard-Fuchs operators from the data of the chiral ring of a Riemann surface embedded in a toric variety which then lift to the Picard-Fuchs operators of the three-form cohomology of the fourfold. From this and the boundary conditions derived in the first part of this work a calculation of the normalized period matrix of three-forms is possible. This normalized period matrix combined with topological intersection numbers of a generalized sphere-tree used to resolve the singularities along the Riemann surface comprise the metric on the intermediate Jacobian and hence determine the couplings of the three-form scalars.

In the following we discussed the three-form cohomology in hypersurfaces that are elliptic fibrations in weighted projective spaces and explained the fate of these three-forms under Sen's weak coupling limit. The three-forms give rise to two-form scalars or Wilson-line moduli in the weakly coupled IIB theory. We concluded with two explicit examples for the two types of three-forms on elliptically fibered Calabi-Yau fourfolds and discussed in detail their weak coupling limit. In the first example we found that the Riemann surface inducing a non-trivial three-form is the elliptic fiber over a base divisor. In this case the normalized period matrix of the three-form cohomology maps to the axio-dilaton which is constant over the same base divisor. In the second example the Riemann surface is a double cover over a $ \mathbb{P}^1 $ in the base and the normalized period matrix contains the information about the location of the branching points on this $ \mathbb{P}^1 $. In the Calabi-Yau threefold of the weakly coupled description this normalized period matrix can be interpreted as as the coupling of Wilson-line moduli on D7 branes. This intricate second example has many non-trivial features like O3-planes and a non-Higgsable gauge-group and deserves further study. In the following we would like to point out several directions for future research.

A first interesting direction is to further extend and interpret the calculations outlined in \autoref{JacobianSection} in the context of mirror symmetry for Calabi-Yau fourfolds \cite{Greene:1993vm,Mayr:1996sh,Alim:2012gq}. In particular, it would be desirable to derive a general expression 
for the Picard-Fuchs equations for three-form periods in terms of the toric data of the ambient space in analogy to the discussion 
of \cite{Hosono:1993qy}. Furthermore, one striking observation to exploit mirror symmetry can be made by recalling 
the construction of the period matrix of the intermediate Jacobian. We note that mirror symmetry exchanges the two-dimensional faces 
$\theta_\alpha$ with their duals $\theta^\ast _\alpha$ and hence maps the one-forms on the Riemann surface $R_\alpha$ 
to the resolution divisors $D'_{l^\alpha}$. Indeed the number of $(1,0)$-forms, given by $\ell^\prime (\theta_\alpha)$ in \eqref{h21count},  
and the number of resolution divisors, given by $\ell'(\theta^\ast _\alpha)$ in \eqref{h21count}, are exchanged. 
This implies that the relevant intersection data for the $D'_{l^\alpha}$ must be captured by the period matrix of three-forms on the mirror 
geometry, at least at certain points in complex structure moduli space, as we have seen in the first part of this work. This observation is further supported by 
basic facts from Landau-Ginzburg orbifolds \cite{Intriligator:1990ua,Vafa:1989xc,Lerche:1989uy}, since in these 
constructions both the intersection data and periods are determined by the structure of the chiral rings of the fourfold and its mirror. 
One can thus conjecture that the complex structure dependent three-form periods calculate on the mirror geometry the K\"ahler moduli dependent quantum corrections to the intersection numbers between integral three-forms and two-forms.
It is then evident to suggest that these K\"ahler moduli corrections already cover world-sheet instanton corrections to the three-form couplings, when using the Calabi-Yau fourfold as a string theory background. It would be very interesting to access these corrections directly on 
the K\"ahler moduli side and establish their physical interpretation. As the three-form cohomology was shown to be localized on divisors a consideration of local Calabi-Yau fourfolds and local mirror symmetry would be sufficient to do so.

A second promising direction for future research is to apply our results in the duality between F-theory and the heterotic string theories. 
The relevance of three-forms in this duality was already pointed out, for example, in \cite{Friedman:1997yq,Curio:1998bva,Diaconescu:1999it}. Indeed, in heterotic compactifications on elliptically fibered Calabi-Yau threefolds with stable vector bundles, the moduli space of certain vector bundle moduli also admits the structure of a Jacobian variety. By duality this Jacobian turns out to be isomorphic to the intermediate Jacobian of the corresponding $ K3 $-fibered Calabi-Yau fourfold. The described powerful techniques 
available for analyzing the three-form periods on fourfolds might help to shed new light on the derivations required in the dual heterotic setting. 
Our first example describes a simple case of such an F-theory compactification with non-trivial intermediate Jacobian for which the comparison to its heterotic dual geometry can be performed explicitly. It is an interesting task to analyze several such dual settings in detail. 

The possibility of a direct calculation of the three-form metric also has immediate applications in string phenomenology. The scalars arising 
from the three-form modes can correspond to scalar fields in an F-theory compactification to four space-time dimensions.  These scalars are naturally axions, since the shift-symmetry is inherited from the forms of the higher-dimensional theory. The axion decay constants are thus given by 
the three-form metric and determines the coupling to the K\"ahler and complex structure moduli and thus can be 
derived explicitly for a given fourfold geometry. Since these geometries might not be at the weak string coupling limit of F-theory, one 
might be lead to uncovered new possibilities for F-theory model building. 
For example, our second example is admitting, if at all, a very complicated weak string coupling limit, but can be 
analyzed nevertheless using the presented geometric techniques. In this example also 
non-Higgsable clusters and O3-planes are present and it is interesting to investigate the 
physics of these objects in the presents of a non-trivial three-form cohomology. 
It is important to stress that consistency of Calabi-Yau fourfold compactifications generically require the inclusion 
of background fluxes \cite{Sethi:1996es}. It is well-known that these are also relevant in 
most phenomenological applications. Therefore, it is of immediate interest to generalize 
our discussion to include background fluxes. 
This will be particularly interesting in singular limits of the geometry, which are relevant in the 
construction of F-theory vacua. In particular, the intermediate 
Jacobian plays an important role in the computation of the spectrum of the effective theory as, 
for example, suggested by the constructions of \cite{Bies:2014sra,Bies:2017fam}.
The generalization 
to include fluxes will also be relevant in discussing extremal transitions in Calabi-Yau 
fourfolds that change the number of three-forms \cite{Intriligator:2012ue}.

To conclude this list of potential future directions, let us also mention the probably most obvious generalization 
of the discussions presented in this work and its immediate relevance for F-theory compactification. 
In fact, in this thesis we have only considered hypersurfaces in 
toric ambient spaces. A generalization to complete intersections \cite{Gray:2013mja,Gray:2014fla}, i.e.~Calabi-Yau manifolds described by 
more then one equation, would be desirable. This is particularly evident when recalling that in F-theory compactifications 
on elliptically fibered fourfolds, the non-trivial three-form cohomology of the base yields U(1)-gauge fields in 
the four-dimensional effective theory \cite{Grimm:2010ks} as we discussed in the first part. Since the function $f_{\cA \cB}$ then corresponds to the 
gauge coupling function, it is an interesting task to use geometric techniques for Calabi-Yau fourfolds to study 
setups away from weak coupling. For bases that are toric hypersurfaces the same techniques as we developed apply directly and enable a calcualtion of the Picard-Fuchs equations which we outlined in \autoref{PoincareSection}.

\newpage

\newpage


\begin{appendices}
{
\addtocontents{toc}{\protect\setcounter{tocdepth}{1}}
\makeatletter
\addtocontents{toc}{%
  \begingroup
  \let\protect\l@chapter\protect\l@section
  \let\protect\l@section\protect\l@subsection
}
\makeatother


\chapter{Three-dimensional $ \cN=2 $ supergravity on a circle}  \label{3d-2dreduction}

In this appendix we consider $\cN=2$ supergravity compactified on 
a circle of radius $r$. Our goal is to derive the resulting $\cN=(2,2)$ 
action. We also briefly discuss the dualization of vector multiplets in 
three dimensions and point out the relation to \autoref{detailed_dual}.
  
We start with a three-dimensional $ \cN=2 $ supergravity theory 
coupled to chiral multiplets with complex scalars $ \phi^\kappa $ and vector
multiplets with bosonic fields $ (L^\Sigma, A^\Sigma) $. Here$ L^\Sigma $ is a real scalar and $ A^\Sigma $ a vector of
an $ U(1) $ gauge theory. The bosonic part of the ungauged $\cN=2$ action takes the form
\begin{align} \label{3Dsugra}
S^{(3)} &= \int \frac{1}{2} R^{(3)} \ast 1 
      - \tilde{K}_{\phi^\kappa \bar \phi^\lambda} d\phi^\kappa \wedge \ast d\bar{\phi}^\lambda  
          + \frac{1}{4} \tilde{K}_{L^\Sigma L^\Lambda} dL^\Sigma \wedge \ast dL^\Lambda \nonumber \\
	&\qquad + \frac{1}{4} \tilde{K}_{L^\Sigma L^\Lambda} \ dA^\Sigma \wedge \ast dA^\Lambda + dA^\Sigma \wedge \I(\tilde{K}_{L^\Sigma \phi^\kappa} d\phi^\kappa) 
\end{align}
where the kinetic terms of the vectors and scalars are determined by the single real kinetic potential $\tilde{K}$.

We want to put this on a circle of radius $ r $ and period one, i.e. the background metric is of the form
\beq \label{circle_ansatz}
ds^2_{(3)} = g_{\mu \nu} dx^\mu dx^\nu + r^2 dy^2
\eeq
where we already drop vectors, since in an un-gauged theory they do not carry degrees of freedom
 in two dimensions. Similarly, the vectors $ A^\Sigma $ are only reduced to real scalars $dA^\Sigma =  d b^\Sigma \wedge dy $.
The resulting two-dimensional action thus reads
\begin{align} \label{actcc}
S^{(2)} &= \int \frac{1}{2} rR  \ast 1 - r\tilde{K}_{\phi^\kappa \bar \phi^\lambda } d\phi^\kappa \wedge \ast d\bar{\phi}^\lambda  
       + \frac{1}{4} r \tilde{K}_{L^\Sigma L^\Lambda} dL^\Sigma \wedge \ast dL^\Lambda \nonumber \\
	&\qquad \quad+ \frac{1}{4 r} \tilde{K}_{L^\Sigma L^\Lambda} db^\Sigma \wedge \ast db^\Lambda 
	- db^\Sigma \wedge \I (\tilde{K}_{L^\Sigma \phi^\kappa} d\phi^\kappa)\ , 
\end{align}
with a two-dimensional $R$ and Hodge star $*$. Note that the last term is topological and does 
not couple to the radius $ r $ of the circle.
We can perform Weyl rescaling of the two-dimensional  metric setting  $\tilde g_{\mu \nu} = e^{2 \omega} g_{\mu \nu}$.
This transforms the Einstein-Hilbert term as 
\beq \label{WeylEH}
   \int   \frac{1}{2} r \tilde R \ \tilde \ast1    = 
    \int \frac{1}{2} r R \  \ast1 + d\omega \wedge \ast dr \ ,
\eeq
while leaving all other terms in the action \eqref{actcc} invariant. 
We then find the action
\begin{align}  \label{action_new1}
S^{(2)} = \int r &\left( \frac{1}{2} R \ast 1 + d\log r \wedge \ast d\omega - \tilde{K}_{\phi^\kappa\bar \phi^\lambda } d\phi^\kappa \wedge \ast d\bar{\phi}^\lambda \right.  \nonumber \\
	&\quad \left.   + \frac{1}{4} \tilde{K}_{L^\Sigma L^\Lambda} dL^\Sigma \wedge \ast dL^\Lambda + \frac{1}{4r^2} \tilde{K}_{L^\Sigma L^\Lambda} db^\Sigma \wedge \ast db^\Lambda \right) \nonumber \\
	&\quad - db^\Sigma \wedge \I(\tilde{K}_{L^\Sigma \phi^\kappa} d\phi^\kappa) 
\end{align}
To make contact with the $ \cN=(2,2) $ dilaton supergravity action \eqref{bosonic_dilatonsugra_extended}
we set
\bea
 L^\Sigma  &=&  r^{-1}  v^\Sigma\ ,\qquad  r= e^{-2\tilde{\varphi}}  \ ,\\
 \sigma^\Sigma &\equiv& b^\Sigma + i v^\Sigma\ .
\eea
Inserted into \eqref{action_new1} we then obtain
\begin{align} \label{action_new2}
S^{(2)} &= \int e^{-2 \tilde \varphi} \left( \frac{1}{2} R  \ast 1 \right. \nonumber \\ 
&\qquad - 2 d \tilde \varphi \wedge \ast 
        \big(d\omega - \frac{1}{2}  \tilde{K}_{v^\Sigma v^\Lambda}v^\Sigma dv^\Lambda - \frac{1}{2}\tilde{K}_{v^\Sigma v^\Lambda} v^\Sigma  v^\Lambda d\tilde \varphi \big)  \nonumber \\
&\qquad - \tilde{K}_{\phi^\kappa\bar \phi^\lambda } d\phi^\kappa \wedge \ast d\bar{\phi}^\lambda 
 +  \tilde{K}_{\sigma^\Sigma \bar \sigma^\Lambda} d\sigma^\Sigma \wedge \ast d \bar \sigma^\Lambda  \nonumber \\
 &\qquad \left.- d \, \R\, \sigma^\Sigma \wedge \I(\tilde{K}_{v^\Sigma \phi^\kappa} d\phi^\kappa) \right)  \, . 
\end{align}
In order to match the action \eqref{bosonic_dilatonsugra_extended} one 
therefore has to find an $\omega$ such that
\beq \label{dw=phi}
      d\omega = -d \tilde \varphi + \frac{1}{2}  \tilde{K}_{v^\Sigma v^\Lambda}v^\Sigma dv^\Lambda + \frac{1}{2}\tilde{K}_{v^\Sigma v^\Lambda} v^\Sigma  v^\Lambda d\tilde \varphi \, .
\eeq
To solve this condition, we first notice that any term in $\tilde K$ that is linear in $v^\Sigma$ drops out from this 
relation, i.e.~$\tilde K$ can take the form
\beq \label{SAsplit}
   \tilde K = \cK + v^\Sigma \cS_\Sigma \, ,
\eeq
with an arbitrary function $\cS_\Sigma (\phi,\bar \phi)$. Furthermore, we can solve \eqref{dw=phi} by 
assuming that $\cK= \cK_1 + \cK_2$ splits into a $v^\Sigma$-independent term $\cK_1(\phi,\bar \phi)$ and a
term $\cK_2(v)$ that only depends on $v^\Sigma$. Then \eqref{dw=phi} is satisfied if
\beq \label{no-scaleappM}
  v^\Sigma \cK_{v^\Sigma} = -k \ , \qquad \omega = - \tilde \varphi + \frac{k}{2} \tilde \varphi -\frac{\cK_2(v)}{2} \, , 
\eeq
It is easy to check that the conditions \eqref{SAsplit} and
\eqref{no-scaleappM} are actually satisfied for the M-theory example \eqref{def-tildeKM}
of $\tilde K$. One finds 
\begin{align}
   \cK_1(z) &= - \log \int_{Y_4} \Omega \wedge \bar \Omega \, ,\quad   \cK_2(v) = \log \cV \, , \nonumber \\
   \cS_\Sigma &= e^{2\varphi} {d_\Sigma}^{\cA \bar \cB}\,  \R \, N_\cA \,  \R \, N_\cB \, ,
\end{align}
such that $k = -4$. Finally, in order to show that \eqref{action_new2} is indeed identical
to the action \eqref{bosonic_dilatonsugra_extended}, we still have to complete the last term
in \eqref{action_new2} to $\I (d \sigma^\Sigma \wedge \tilde{K}_{v^\Sigma \phi^\kappa} d\phi^\kappa)$. 
In order to do that we  use 
\beq
    d\I \, \sigma^\Sigma \wedge \R(\tilde{K}_{v^\Sigma \phi^\kappa} d\phi^\kappa)=    \frac{1}{2} d\I \, \sigma^\Sigma \wedge d \tilde K_{v^\Sigma} \, ,
\eeq
which follows from the fact that $d \tilde K_{v^\Sigma} = 2 \R( \tilde K_{v^\Sigma \phi^\kappa} d \phi^\kappa )  + \tilde K_{v^\Sigma v^\Lambda} d v^\Lambda$.
This implies that these terms simply yield a total derivative 
and shows that the reduction of $\cN=2$ supergravity of the form \eqref{3Dsugra} indeed 
yields the  extended form of 
$ \cN=(2,2) $  dilaton supergravity suggested in \eqref{bosonic_dilatonsugra_extended} 
coupled to the chiral multiplets with scalars $ \phi^\kappa $ and twisted-chiral multiplets with 
scalars $ \sigma^\Sigma $. Interestingly, we had to employ the conditions \eqref{SAsplit} and \eqref{no-scaleappM}, which hints
to the fact that the action \eqref{bosonic_dilatonsugra_extended} might admit further interesting
extensions. 

Let us end this appendix by pointing out that we could also have first dualized the vectors $ A^\Sigma $
to real scalars in three dimensions and then performed the circle reduction. 
The dual multiplets to the vector multiplets $ (L^\Sigma, A^\Sigma) $ 
are three-dimensional chiral multiplets with bosonic parts being complex scalars $ T_\Sigma $ 
given by
\beq \label{3Dchiral}
T_\Sigma =  \partial_{L^\Sigma} \tilde K  + i \rho_\Sigma\, .
\eeq
The metric is determined now from a proper K\"ahler potential given by 
\beq
\textbf{K}(T+\bar T, M) = K - \R\, T_\Sigma \, L^\Sigma \, ,
\eeq
such that the final action reads
\begin{align}   
S^{(3)} &= \int \frac{1}{2} R^{(3)} \ast 1 - \textbf{K}_{{M}^I \bar{ { M}}^J } d {M}^I \wedge \ast d\bar{{M}}^J \, ,
\end{align}
with $ {M}^I= (T_\Sigma, \phi^\kappa )$. We can again reduce this theory on a circle \eqref{circle_ansatz}
and perform a Weyl-rescaling \eqref{WeylEH} to find
\begin{align}
S^{(2)} &= \int \frac{1}{2} rR \ast 1 + dr \wedge \ast d\omega - r\textbf{K}_{{M}^I \bar{ { M}}^J } d{M}^I \wedge \ast d\bar{{M}}^J \, .
\end{align}
With the choices $ r = e^{-2\tilde{\varphi}}$ and $ \omega = - \tilde{\varphi} $ this reads
\begin{align}
S^{(2)} &= \int  e^{-2\tilde{\varphi}}\left(\frac{1}{2} R\ast 1 + 2d\tilde{\varphi} \wedge \ast d\tilde{\varphi} 
- \mathbf{K}_{{M}^I \bar{ { M}}^J } d{M}^I \wedge \ast d\bar{{M}}^J \right) .
\end{align}
This result should also be obtainable from \eqref{action_new2} by dualizing the chiral multiplets
with scalars $\sigma^\Sigma$. This is possible since $ b^\Sigma $ appears only with its field-strength $db^\Sigma $. 
The details of this dualization in two dimensions will be discussed in \autoref{detailed_dual}.

\chapter{Twisted-chiral to chiral dualization in two dimensions} \label{detailed_dual}

In this appendix we present the details of the dualization discussed 
in \autoref{sec_Legendre} of 
a twisted-chiral multiplet to a chiral multiplet in two dimensions. 
The starting point is the action 
\begin{align}  \label{action1app}
 S^{(2)}_{\text{C-TC}} = \int e^{-2\tilde{\varphi}} & \left( \frac{1}{2}R \, \ast 1 + 2d \tilde{\varphi} \wedge * d\tilde{\varphi} 
 -  \tilde K_{\phi^\kappa \bar \phi^\lambda} \, d\phi^\kappa \wedge * d \bar {\phi}^\lambda
	   \right.\nonumber \\
   & \quad +  \tilde  K_{ \sigma^\Sigma \bar \sigma^\Lambda}\, d \sigma^\Sigma \wedge * d \bar {\sigma}^\Lambda -\tilde K_{ \phi^\kappa \bar \sigma^\Lambda} \, d \phi^\kappa \wedge d \bar \sigma^\Lambda \nonumber \\
   & \quad \left. - \tilde K_{ \sigma^\Sigma \bar \phi^\lambda} \, d \bar {\phi}^\lambda \wedge d\sigma^\Sigma \right)  ,
\end{align}
where $\tilde K$ is given by 
\beq 
    \tilde K = \cK + e^{2 \tilde \varphi} \cS \, . 
\eeq
In the following we 
use sub-scripts to indicate derivatives with respect to fields, e.g.~$\tilde K_{\phi^\kappa} \equiv \partial_{ \phi^\kappa} \tilde K $.
$\tilde K$ depends on a number of chiral multiplets with complex
scalars $\phi^\kappa$ and a number of twisted-chiral 
multiplets with complex scalars $\sigma^\Sigma$.

In order to perform a dualization, we assume that 
$\R\, \sigma^\Sigma$ has a shift symmetry and only appears 
via $d \,\R\, \sigma^\Sigma$ in \eqref{action1app}. 
This implies that $\R\, \sigma^\Sigma$ can be dualized 
into a scalar $\rho_\Sigma$ by the standard procedure. One 
first replaces $d\R\, \sigma^\Sigma \rightarrow F^\Sigma$ in \eqref{action1app} 
and then adds a Lagrange multiplier term promotional to 
$F^\Sigma \wedge d\rho_\Sigma$. Then $F^\Sigma$ can be consistently eliminated 
from \eqref{action1app}. 
Denoting the imaginary part of $\sigma^\Sigma$ by $v^\Sigma = \I \, \sigma^\Sigma$
the resulting action reads
\begingroup
\allowdisplaybreaks
\begin{align}  \label{action2app}
 S^{(2)}_{\textrm  C} &= \int e^{-2\tilde{\varphi}} \left( \frac{1}{2}R \, \ast 1 + 2d \tilde{\varphi} \wedge * d\tilde{\varphi}
 -  \tilde K_{\phi^\kappa \bar \phi^\lambda} \, d\phi^\kappa \wedge * d \bar {\phi}^\lambda \right. \nonumber  \\
& \quad + \tilde K ^{v^\Sigma v^\Lambda} \big(e^{2\tilde \varphi} d\rho_\Sigma - \I \, (\tilde K_{v^\Sigma \phi^\kappa} d\phi^\kappa) \big) \wedge 
        \ast \big(e^{2\tilde \varphi} d\rho_\Lambda - \I \, (\tilde K_{v^\Lambda \phi^\lambda} d\phi^\lambda) \big)   \nonumber \\
& \quad \left.  + \frac{1}{4} \tilde  K_{ v^\Sigma  v^\Lambda}\, d v^\Sigma \wedge * d v^\Lambda  \right)
\end{align}
\endgroup
To compute the dualized action we make the following
ansatz for the Legendre transformed variables $T_\Sigma$ 
\beq \label{TA_app}
   T_\Sigma =  e^{-2\tilde \varphi} \frac{\partial \tilde K}{\partial v^\Sigma} + i \rho_\Sigma=
    e^{-2\tilde \varphi} \frac{\partial \cK}{\partial v^\Sigma}+ \frac{\partial \cS}{\partial v^\Sigma} + i \rho_\Sigma \ , 
\eeq
and the dual potential $\mathbf{K}$ 
\beq \label{bfK_app}
    \mathbf{K} = \tilde K - e^{2\tilde \varphi}\, \R \, T_\Sigma \, v^\Sigma\ .  
\eeq
We want to derive the conditions on $\tilde K$ under which the action \eqref{action2app} can be 
brought to the form
\beq \label{S2_app}
   S^{(2)}_{\text{C}} = \int e^{-2\tilde{\varphi}} \left( \frac{1}{2}R *1 + 2d \tilde{\varphi} \wedge * d\tilde{\varphi} 
 -  \mathbf{K}_{M^I \bar M^J} \, dM^I\wedge * d \bar M^J \right)\ ,
\eeq
with $M^I = (\phi^\kappa, T_\Sigma)$.

We first determine from \eqref{TA_app} and \eqref{bfK_app} that
\begin{align} \label{usefull}
   \frac{\partial v^\Sigma}{\partial T_\Lambda} &= \frac{1}{2} e^{2 \tilde \varphi} \tilde K^{v^\Sigma v^\Lambda} \ ,  
     &\frac{\partial v^\Sigma}{\partial \phi^\kappa} &= -  \tilde K^{v^\Sigma v^\Lambda} \tilde K_{v^\Lambda  \phi^\kappa}\ , \\
     \mathbf{K}_{T_\Sigma} &= - \frac{1}{2} e^{2 \tilde \varphi} v^\Sigma\ ,   & \mathbf{K}_{ \phi^\kappa} &= \tilde K_{ \phi^\kappa}\ , \nn
\end{align}
where $\tilde K^{v^\Sigma v^\Lambda}$ is the inverse of $\tilde K_{v^\Sigma v^\Lambda} \equiv \partial_{v^\Sigma} \partial_{v^\Lambda} \tilde K 
= 4 \tilde K_{\sigma^\Sigma \bar \sigma^\Lambda}$. Crucially, one also derives from \eqref{TA_app} that
\beq \label{dRT}
  d \R T_\Sigma = e^{-2 \tilde \varphi} \big( \tilde K_{v^\Sigma v^\Lambda} d v^\Sigma + 2 \R ( \tilde K_{v^\Sigma  \phi^\kappa} d \phi^\kappa ) - 2 \cK_{v^\Sigma} d\tilde \varphi \big)\ .
\eeq
Note that there is the additional $d\tilde \varphi $-term, which is absent in the standard dualization procedure. 
The conditions on $\tilde K$ arise from demanding that the dual action can be brought to 
the form \eqref{action1app} and no additional mixed terms involving $d\tilde \varphi $ appear. 
To evaluate \eqref{action1app} one uses \eqref{usefull} to derive the identities
\bea  \label{metric1}
  \mathbf{K}_{T_\Sigma \bar T_\Lambda} &=& - \frac{1}{4} e^{4 \tilde \varphi} \tilde K^{v^\Sigma v^\Lambda}\ , \qquad 
  \mathbf{K}_{T_\Sigma \bar \phi^\kappa} =  \frac{1}{2} e^{2 \tilde \varphi} \tilde K^{v^\Sigma v^\Lambda} \tilde K_{v^\Lambda \bar \phi^\kappa}\ , \\
  \mathbf{K}_{\phi^\kappa \bar \phi^\lambda} &=&   
  \tilde {K}_{\phi^\kappa \bar \phi^\lambda} -  \tilde K_{\phi^\kappa v^\Sigma}  \tilde K^{v^\Sigma v^\Lambda} \tilde K_{v^\Lambda \bar \phi^\lambda}\ .\nn
\eea
Inserting \eqref{dRT}, \eqref{metric1} into \eqref{S2_app} one finds the following terms involving $d \tilde \varphi$
\beq \label{extradphi}
  S_{d \tilde \varphi}^{(2)} = \int e^{-2 \tilde \varphi} \Big( \big(2 + \cK_{v^\Sigma}  \cK^{v^\Sigma v B} \cK_{v^\Lambda} \big)
  d\tilde \varphi \wedge * d \tilde \varphi + \cK_{v^\Sigma} dv^\Sigma \wedge * d \tilde \varphi \Big)\ .
\eeq
These terms can be removed by a Weyl rescaling of the three-dimensional metric 
if certain conditions on $\cK$ are satisfied. To see this, we perform a Weil rescaling 
\beq \label{Weyl-change}
      \tilde g_{\mu \nu} = e^{2 \omega} g_{\mu \nu}
\eeq
which transforms the Einstein-Hilbert term as 
\beq \label{change_R}
   \int  e^{-2 \tilde \varphi}  \frac{1}{2} \tilde R \ \tilde *1    = 
    \int e^{-2 \tilde \varphi}\left( \frac{1}{2} R \  *1 - 2 d\omega \wedge * d\tilde \varphi \right) \ ,
\eeq
while leaving all other terms invariant. Hence we can absorb the extra terms 
in \eqref{extradphi} by a Weyl rescaling iff
\beq
  - 2 d\omega = \cK_{v^\Sigma}  \cK^{v^\Sigma v B} \cK_{v^\Lambda} d\tilde \varphi + \cK_{v^\Sigma} dv^\Sigma\ .
\eeq
Clearly, a simple solution to this equation is found if $\cK$ satisfies 
\beq \label{nsc}
   \cK_{v^\Sigma}  \cK^{v^\Sigma v B} \cK_{v^\Lambda}  = k\ , \qquad  \cK = \cK_1(\phi,\bar \phi) + \cK_2(v)\ , 
\eeq
for a constant $k$, a function $\cK_1(\phi,\bar \phi) $ independent of $v^\Sigma$, and a function 
$\cK_2(v)$ independent of $\phi^\kappa$. 
In this case one can chose
\beq
    \omega = - \frac{k}{2} \tilde \varphi - \frac{1}{2} \cK_2(v)\ . 
\eeq
Note that \eqref{nsc} is satisfied for the result found in a Calabi-Yau fourfold reduction \eqref{idhatKS}, 
i.e.~$k=-4$ and $\cK_2 = \log \cV$. 

\addtocontents{toc}{\endgroup}
}
\end{appendices}

\newpage

\backmatter

\chapter{Summary}

In this summary we will give a short overview of the content of this thesis. The introduction \autoref{Introduction} explains the role of string theory in modern particle physics as a candidate for quantum gravity. We explain how to derive the effective low-energy supergravity description of string theory and how to make contact with the observable world via dimensional reduction. A special class of such theories is given by F-theory that allows for an inclusion of non-perturbative effects of stringy physics. F-theory is here understood as a decompactification limit of M-theory on elliptically fibered Calabi-Yau fourfolds. This elegant description of string theory is a powerful branch of string phenomenology and motivates the study of the geometry of Calabi-Yau fourfolds.\\
We start the main body of this thesis by introducing the geometric properties of Calabi-Yau fourfolds and their harmonic forms in \autoref{FourfoldBasics}. Here we focus especially on the harmonic three-forms whose physical properties are determined by the complex structure dependence of their so called normalized period matrix $ f_{\cA \cB} $, which was not well understood before. The next part of the thesis focuses on effective theories of Calabi-Yau fourfolds.\\
In \autoref{IIAreduction} we perform the dimensional reduction of Type IIA supergravity on a general Calabi-Yau fourfold and find a $ \cN =(2,2) $ dilaton-supergravity in two dimensions. Here we determine the massless spectrum and the kinetic potential of the resulting supergravity. We conjecture an extension of the usual K\"ahlerpotential of this dilaton-supergravity by a term depending on the dilaton of the theory to account for the kinetic coupling of the novel three-form scalars.\\
Subsequently we apply mirror symmetry in \autoref{mirror_section}, a duality of the same supergravity theory resulting from two different Calabi-Yau fourfolds, to determine the structure of the normalized period matrix at the large complex structure point in moduli space. We see that it is mirror to a function $ {h_\cA} ^\cB $ holomorphic in the mirror K\"ahler moduli at the large volume point. The linear leading order behavior of $ f_{\cA \cB} $ at the large complex structure point is specified by topological intersection numbers $ {M_{\Sigma \cA}} ^\cB $ of the mirror.\\
Following this discussion we extend our analysis to the dimensional reduction of eleven-dimensional supergravity in \autoref{F-theoryapp}. Afterwards we lift the resulting three-dimensional $ \cN=2 $ supergravity to the effective theory of F-theory on elliptically fibered Calabi-Yau fourfolds leading to an effective $ \cN=1 $ supergravity description in four dimensions. We close this part by discussing the implications of non-trivial three-form cohomology and defer some technical details of the first part into two appendices in \autoref{3d-2dreduction},\autoref{detailed_dual}. Three-forms located on the base of the elliptically fibered fourfold lead to a $ U(1) $-gauge theory in the effective four-dimensional theory whose coupling is determined by $ f_{\cA \cB} $. If the harmonic three-forms have a leg in the fiber they lead to scalars in the effective theory whose coupling is determined by their normalized period matrix $ f_{\cA \cB} $. We also show consistency of our results with the known weakly coupled IIB orientifold description.\\
In the second part of this thesis we construct explicit Calabi-Yau fourfold examples via hypersurfaces in toric varieties. We begin in \autoref{ToricSection} with the general construction of toric spaces and their hypersurfaces. We then focus on Calabi-Yau fourfolds realized as so called semiample hypersurfaces avoiding the Lefschetz-hyperplane theorem that forbids non-trivial three-form cohomology on ample hypersurfaces. This is followed by a discussion of the origin of non-trivial harmonic forms on the hypersurface and we state well known formulas for the number of the various harmonic forms in terms of toric data. We also include in our discussion so called non-toric and non-algebraic deformations of the hypersurface that are easy to describe in the same framework. The derived description via chiral rings and Poincar\'e residues allows not only to determine the spectrum of the effective theory upon compactification, but also sets the stage for a calculation of the couplings.\\
In \autoref{JacobianSection} we discuss the moduli space of the three-form scalars which is called intermediate Jacobian of a Calabi-Yau fourfold whose complex structure dependence reduces to the complex structure dependence of Riemann surfaces. The normalized period matrices of these Riemann surfaces are the building blocks of the three-form normalized period matrix $ f_{\cA \cB} $ of the Calabi-Yau fourfold. The K\"ahler dependence is captured by a so called generalized sphere-tree that can be computed in terms of the ambient space intersection theory and provides us with the aforementioned intersection numbers $ {M_{\Sigma \cA}} ^\cB $. We determine in this situation the map between the geometrical quantities, necessary to derive the effective theories discussed before.\\ 
This part is completed by a lengthy discussion of two simple examples of Calabi-Yau fourfold geometries with non-trival three-form cohomology in \autoref{ExampleSection}. Here we introduce the mathematical basics of elliptic fibrations and discuss the corresponding F-theory physics including the weak coupling limit of Sen. The first example contains one non-trivial three-form that gives rise to a so called two-form scalar in the weak coupling limit and in this case the normalized period matrix $ f_{\cA \cB} $ reduces to the axio-dilaton $ \tau $ of type IIB supergravity. In the second example we have a geometry with multiple non-trivial three-forms that we argue to give rise to so called Wilson-line scalars on a seven-brane. This example is in particular interesting as it includes many non-trivial features, like O3-planes and non-Higgsable clusters and deserves further investigation.\\
We conclude the thesis in \autoref{ConclusionSection} giving an outlook of possible future research directions. After recalling the results of our work, we point out that a further study of the $ \cN=(2,2) $-supergravity theory is necessary to fully settle the inclusion of three-form moduli in type IIA reductions on Calabi-Yau fourfolds, as we only derived a dilaton-dependent correction to the kinetic potential of the effective theory which is therefore not K\"ahler in an obvious way. Furthermore, we argue that a better understanding of mirror symmetry of Calabi-Yau fourfold hypersurfaces with non-trivial three-form cohomology is desirable. This may enable a direct expression of the normalized period-matrix $ f_{\cA \cB} $ in toric terms. Furthermore, we argue that this matrix determines the couplings of axions in effective theories and hence is relevant for string phenomenology. Therefore it is in particular necessary to determine $ f_{\cA \cB} $ in more complicated geometries, for instance complete intersections in toric varieties.

\newpage

\chapter{Samenvatting}

In deze samenvatting geven we een kort overzicht van de inhoud van
dit proefschrift. In the introductie, sectie 1, leggen we uit wat
de rol is van snaartheorie in de moderne deeltjesfysica als de kandidaat
voor een theorie van kwantumzwaartekracht. We leggen uit hoe een effectieve
lage-energie supergravitatie beschrijving van snaartheorie kan worden
afgeleid en hoe contact kan worden gemaakt met de wereld om ons heen
via dimensie reductie. Een speciale klasse van zulke theorieën wordt
gegeven door F-theorie dat ook niet-perturbatieve effectieve effecten
van snaartheorie beschrijft. We zien F-theorie hier als een decompactificatie
limiet van M-theorie op elliptisch gevezelde Calabi-Yau viervariëteiten.
Deze elegante beschrijving van snaartheorie is een krachtige tak van
snaarfenomenologie en geeft een motivatie voor de studie van de geometrie
van Calabi-Yau viervariëteiten. \\
Daarna introduceren we in sectie 2 de geometrische eigenschappen van
Calabi-Yau viervariëteiten en hun harmonische differentiaalvormen.
Hier focussen we ons op de harmonische drie-vormen waarvan de fysische
eigenschappen bepaald worden door de afhankelijkheid van hun zogenoemde
genormaliseerde periode matrix $f_{\mathcal{AB}}$ van de complexe
structuur. Deze afhankelijkheid is niet voldoende begrepen in de eerdere
literatuur. Het volgende deel van het proefschrift richt zich op de
effectieve theorieën van Calabi-Yau viervariëteiten. \\
In sectie 3 doen we de dimensionale reductie van Type IIA supergravitatie
op een algemene Calabi-Yau viervariëteit en vinden we een $\mathcal{N}=(2,2)$
dilaton-supergravitatie in twee dimensies. We bepalen het spectrum
van massaloze deeltjes en de kinetische potentiaal van de resulterende
supergravitatie theorie. Om de kinetische koppeling van de nieuwe
drie-vorm scalairen te verklaren, doen we een voorstel voor een uitbreiding
van de gebruikelijke Kählerpotentiaal van deze dilaton-supergravitatie
door een term toe te voegen die afhangt van de dilaton. \\
Daarna, in sectie 4, passen we spiegelsymmetrie toe, dat een dualiteit
is van dezelfde theorie van supergravitatie op twee verschillende
Calabi-Yau viervariëteiten, om de structuur van de genormaliseerde
periode matrix te bepalen op het grote complexe structuur punt (Engels:
large complex structure point) in de moduliruimte. We zullen zien
dat het gespiegeld wordt naar een holomorfe functie ${h_{\mathcal{A}}}^{\mathcal{B}}$
in de gespiegelde Kähler moduli op het grote volume punt (Engels:
large volume point). Het lineaire hoogste orde gedrag van $f_{\mathcal{AB}}$
op het grote complexe structuur punt wordt gespecificeerd door de
topologische intersectie getallen ${M_{\Sigma\mathcal{A}}}^{\mathcal{B}}$
van de spiegelvariëteit.\\
Na deze discussie breiden we in sectie 5 onze analyse uit naar de
de dimensionale reductie van elf-dimensionale supergravitatie. Daarna
liften we de resulterende drie-dimensionale $\mathcal{N}=2$ supergravitatie
op naar de effectieve theorie van F-theorie op elliptisch gevezelde
vier-variëteiten. Dit geeft een effectieve $\mathcal{N}=1$ supergravitatie
beschrijving in vier dimensies. We sluiten dit deel af met een discussie
van de implicaties van niet-triviale  drie-vorm cohomologieën. We
verwijzen naar twee addendums in sectie 9 voor een aantal technische
details van het eerste deel van de thesis. Drie-vormen die zich zich
in de basis van de elliptisch gevezelde Calabi-Yau viervariëteit bevinden,
geven een $U(1)$-ijktheorie in de effectieve vier-dimensionale theorie.
De koppeling wordt bepaald door $f_{\mathcal{AB}}.$ Als de harmonische
drie-vormen zich ook deels op de vezel bevinden, geven ze scalairen
in de effectieve theorie waarvan de koppeling bepaald is door hun
genormaliseerde periode matrix $f_{\mathcal{AB}.}$ We laten ook zien
dat onze resultaten consistent zijn met de bekende zwak-gekoppelde
IIB orientifold beschrijving.\\
In het tweede deel van dit proefschrift construeren we expliciete
Calabi-Yau viervariëteit voorbeelden via hyperoppervlakken in torische
variëteiten. We beginnen in sectie 6 met een algemene constructie
van torische ruimten en hun hyperoppervlakken. We richten ons daarna
op Calabi-Yau viervariëteiten die gerealiseerd worden als zogenoemde
semiampel (Engels: semiample) hyperoppervlakken. Deze oppervlakken
vermijden de Lefschetz-hypervlak stelling dat stelt dat het onmogelijk
is een niet-triviale  drie-vorm cohomologie op ampel (Engels: ample)
hyperoppervlakken te hebben. Dit wordt gevolgd door een discussie
over de oorsprong van niet-triviale harmonische vormen op het hyperoppervlak
en we geven bekende formules waarin het aantal harmonische vormen
uitgedrukt wordt in torische data. In onze discussie behandelen we
ook de zogenoemde niet-torische en niet-algebraïsche deformaties van
het hyperoppervlak die gemakkelijk te beschrijven zijn in hetzelfde
kader. De beschrijving die we afleiden via chirale ringen en Poincaré
residuen, geeft ons niet alleen het spectrum van de effectieve theorie
dat we krijgen na compactificatie, maar maakt ook de weg vrij voor
een berekening van de koppelingen.\\
In sectie 7 behandelen we de moduli ruimte van de  drie-vorm scalairen
dat de tussen-Jacobiaan van de Calabi-Yau viervariëteit genoemd wordt.
De complexe structuur afhankelijkheid van de tussen-Jacobiaan reduceert
naar de complexe structuur afhankelijkheid van Riemann-oppervlakken.
De genormaliseerde periode matrices van deze Riemann oppervlakken
zijn de bouwstenen van de  drie-vorm genormaliseerde periode matrix
$f_{\mathcal{AB}}$ van de Calabi-Yau viervariëteit. De Kähler-afhankelijkheid
wordt beschreven door een zogenoemde gegeneraliseerde sfeer-boom die
uitgedrukt kan worden in de intersectietheorie van de omgevende ruimte.
Dit geeft ons bovendien de eerder genoemde intersectiegetallen ${M_{\Sigma\mathcal{A}}}^{\mathcal{B}}$.
We bepalen ook de afbeelding tussen de geometrische grootheden die
nodig is om de effectieve theorieën, waar we het eerder over gehad
hebben, af te leiden. \\
We eindigen dit deel in sectie 8 met een lange discussie van twee
simpele voorbeelden van Calabi-Yau viervariëteit geometrieën die een
niet-triviale  drie-vorm cohomologie hebben. Hier introduceren we
de wiskundige basiskennis van elliptische vezelingen en bediscussiëren
we de corresponderende F-theorie natuurkunde, inclusief de zwakke
koppelings limiet van Sen. Het eerste voorbeeld heeft een niet-triviale
 drie-vorm dat een zogenoemde twee-vorm scalair geeft in de zwakke-koppelinglimiet.
In dit geval reduceert de genormaliseerde periode matrix $f_{\mathcal{AB}}$
naar de axio-dilaton $\tau$ van type IIB supergravitatie. In het
tweede voorbeeld hebben we een geometrie met meerdere niet-triviale
 drie-vormen en we beargumenteren dat ze zogenoemde Wilson-lijn scalairen
geven op een zeven-braan. Dit voorbeeld is interessant omdat het heel
veel niet-triviale eigenschappen heeft, zoals O3-vlakken en clusters
die je niet kunt Higgsen. Daarom zou het goed zijn als er meer onderzoek
naar gedaan wordt. \\
We eindigen dit proefschrift in sectie 9 door vooruit te kijken naar
mogelijke toekomstige onderzoeksrichtingen. Na het samenvatten van
de resultaten van dit werk, beargumenteren we dat een verdere studie
van de $\mathcal{N}=(2,2)$-supergravitatie theorie nodig is om de
inclusie van drie-vorm moduli in type IIA reducties op Calabi-Yau
viervariëteiten volledig te begrijpen. Dit omdat we alleen een dilaton-afhankelijke
correctie voor de kinetische potentiaal van de effectieve theorie
afgeleid hebben, waardoor het niet duidelijk is of deze nog Kähler
is. We beargumenteren ook dat het wenselijk is een beter begrip te
krijgen van de spiegelsymmetrie van Calabi-Yau viervariëteit hyperoppervlakken
met niet-triviale drie-vorm cohomologie. Dit zou het mogelijk kunnen
maken om de genormaliseerde periode matrix $f_{\mathcal{AB}}$ direct
uit te drukken in torische data. Bovendien beargumenteren we dat deze
matrix de koppelingen van axions in effectieve theorieën bepaalt en
daarom relevant is voor snaarfenomenologie. Daarom is het in het bijzonder
nodig om $f_{\mathcal{AB}}$ te bepalen in gecompliceerdere geometrieën,
bijvoorbeeld complete intersecties in torische variëteiten.

\newpage

\chapter{About the Author}

I was born on September 9th 1990 in G\"unzburg, Germany, where I graduated from the Dossenberger Gymnasium in 2010. Afterwards I studied Mathematics at the University of Augsburg where I obtained a Bachelor degree in 2013. In parallel I completed the Elite-Bachelor "TopMath" of the TU Munich with honors. This program enables students to start their own research during the last year of their Bachelor studies. In my case I focused on algebraic topology under the supervision of Prof. Dr. Bernhard Hanke. My Bachelor thesis provided a proof of the Jordan separation theorem on smooth manifolds.

After my undergraduate studies I went in 2013 to the LMU in Munich for the Elite-Master "Theoretical and Mathematical Physics" where I studied the geometry and physics of string theory. At the Max-Planck-Institute for Physics in Munich I wrote my Master thesis on three-forms on Calabi-Yau fourfolds and their effective theories under the supervision of Dr. Thomas Grimm who later became the supervisor of my Ph.D. studies. 

After completion of my Master's degree with honors in 2015, I started my research at MPI Munich building on previous work. Since Thomas became Associate Professor at Utrecht University in 2016, I followed to Utrecht in 2017 where I completed my studies towards a doctoral degree in 2018. This work is an account of the work completed during these studies.

\newpage

\chapter{Aknowledgements}

Most importantly I like to thank my supervisor Thomas Grimm for countless discussions and the freedom to follow my interests. Furthermore I thank Stefan Vandoren for making my studies possible and promoting me. I also like to thank my collaborators Ralph Blumenhagen and Irene Valenzuela from whom I learned a lot.\\
The many great people that made my studies a wonderful experience I would also like to thank. My office-mates Andreas Kapfer and Michael Fuchs in Munich and Pierre Corvilain and Raffaele Batillomo in Utrecht. Florian Wolf, Daniel Kl\"awer, Daniela Herschmann, Nina Miekley, Kilian Mayer, Huibert het Lam, Aron Jansen, Markus Dierigl and Miguel Montero for many deep and not so deep conversations. In addition, I would like to thank all the exceptional people that I met during my travels to conferences and seminars of whom there are too many to thank individually.\\

I also like to thank the Max-Planck-Institute for Physics in Munich where the first half of my studies was completed. The International Munich Particle Research School (IMPRS) provided an inspirational environment for my research. The same holds for Utrecht University where the second half of my studies took place and I was welcomed with hospitality.\\

Special thanks goes to my family for their never-ending support and understanding.

\newpage

\bibliography{bibmaster}
\bibliographystyle{utcaps}

\end{document}